\documentclass[a4paper,12pt,oneside,openright,parindent,headsepline,cleardoubleplain,chapterprefix,pointlessnumbers]{thesis_style}

\usepackage[latin1]{inputenc}
\usepackage[T1]{fontenc}
\usepackage{graphicx}
\usepackage{hyperref}
\usepackage[numbers,sort&compress]{natbib} 
\usepackage{amsmath}
\usepackage{amssymb}
\usepackage{color}
\usepackage{scrpage2}			
\usepackage{slashed}

\automark[section]{chapter}

\allowdisplaybreaks 

\begin{document}


\newcommand{\mhalfo}{\frac{1}{2}}	
\newcommand{\mhalf}[1]{\frac{#1}{2}}
\newcommand{\ka}{\kappa}
\newcommand{\al}{\alpha}
\newcommand{\be}{\beta}
\newcommand{\ga}{\gamma}
\newcommand{\la}{\lambda}
\newcommand{\de}{\delta} 
\newcommand{\vp}[0]{\varphi} 
\newcommand{\vpb}[0]{\bar{\varphi}} 
\newcommand{\equ}[1]{\begin{equation} #1 \end{equation}}
\newcommand{\ba}{\begin{align}}
\newcommand{\ea}{\end{align}}	
\newcommand{\eref}[1]{eq.~(\ref{#1})}
\newcommand{\fref}[1]{fig.~\ref{#1}}
\newcommand{\ddotp}[1]{\frac{d^d #1}{(2\pi)^d}}	
\newcommand{\nnnl}{\nonumber\\}	
\newcommand{\G}[1]{\Gamma(#1)}
\newcommand{\nq}{\nu_1}	
\newcommand{\nw}{\nu_2}	
\newcommand{\nd}{\nu_3}	
\newcommand{\dhalf}{\frac{d}{2}} 
\newcommand{\fig}[5]{\begin{figure}[#1]\centering\includegraphics[#5]{#3}\caption{#4}\label{#2}\end{figure}}
\newcommand{\cb}{$\bigstar$} 
\newcommand{\ce}{$\blacksquare$} 

\definecolor{grey}{gray}{0.95}

\def\bcommandlist #1 \ecommandlist{
\begin{table}[t]
 \fcolorbox{black}{grey}
	{\begin{tabular}{p{0.37\textwidth}p{0.54\textwidth}}
		#1
	\end{tabular}}
\end{table}
}

\begin{titlepage}

\begin{center}
\Huge
\textbf{On gauge fixing aspects of the infrared behavior of Yang-Mills Green functions}\\

\vskip1cm
\LARGE
Markus Q. Huber\\

\vskip1cm
\large
\textit{Institut f\"ur Physik, Karl-Franzens-Universit\"at Graz,\\Universit\"atsplatz 5, 8010 Graz, Austria}

\end{center}

\vskip1cm
\normalsize
The infrared behavior of propagators and vertices is derived for the maximally Abelian gauge and the Gribov-Zwanziger action relying on functional equations. The derivation and analysis of Dyson-Schwinger equations increase considerably in complexity when going beyond the standard Landau gauge fixing and the available tools have to be improved. For the derivation of the equations a computer program (\textit{DoDSE}) was developed in order to handle the plethora of terms. The process of determining possible infrared solutions is abstracted to obtain further insight into the structure of the so-called infrared scaling solutions. It is found that a few simple steps suffice to determine possible infrared scaling relations directly from the action. This makes also complicated actions easily accessible.

For the maximally Abelian gauge an infrared enhanced diagonal gluon propagator is found, while the off-diagonal degrees of freedom are infrared suppressed. This is in agreement with the idea of Abelian infrared dominance. Furthermore, it is proven that $SU(2)$ and higher $SU(N)$ have the same infrared behavior, although the corresponding actions differ. Under a suitable truncation the Dyson-Schwinger equations are solved in the deep infrared to obtain values for the exponents of the power laws. The maximally Abelian gauge constitutes the first instance where a consistent scaling solution was found besides the Landau gauge.

Restricting the integration in field configuration space to the Gribov region of the Landau gauge with the Gribov-Zwanziger action leads to two solutions qualitatively equivalent to the one obtained with the standard Faddeev-Popov gauge fixing. In both cases the gluon propagator is infrared suppressed and the infrared enhanced ghost propagator dominates the Dyson-Schwinger equations. This result corroborates the conjecture by Zwanziger that the functional integration can be cut at the first Gribov horizon and only the applied boundary conditions are important. For one solution the Dyson-Schwinger equations reduce in the deep infrared exactly to those obtained with the usual Faddeev-Popov gauge fixing.

For further investigations of Yang-Mills theory the possibility to derive Dyson-Schwinger equations with the computer will prove useful. The presented method for determining possible infrared scaling solutions will also alleviate their analyses.

\vskip5mm
\textit{Keywords:} Confinement, Yang-Mills theory, infrared behavior, Dyson-Schwinger equations, maximally Abelian gauge, Gribov-Zwanziger action

\end{titlepage}

\thispagestyle{empty}
\Large
\textbf{Note added to the eprint version:}
\vskip5mm
\normalsize
This doctoral thesis has been submitted to the Karl-Franzens-University Graz and successfully defended on Apr. 23, 2010.
The part about Dyson-Schwinger equations in Chapter \ref{chp:FunctionalEquations} is based on ref. [108] (arXiv:0808.2939). Chapters \ref{chp:scalingSolutions}, \ref{chp:MAG} and \ref{chp:GZ} contain reordered and extended material from refs. [67,179] (arXiv:0904.1873, arXiv:0910.5604).

\begin{titlepage}

\setcounter{page}{1}

\vspace*{1cm}
\begin{center}
\LARGE
 Markus Q. Huber\\
\vspace{1cm}
\Huge

\textbf{On gauge fixing aspects of the infrared behavior of Yang-Mills Green functions}\\

\vfill
\large
Dissertation\\
\normalsize
zur Erlangung des Doktorgrades der Naturwissenschaften

\vspace{0.5cm}
verfasst am Institut f\"ur Physik, Fachbereich Theoretische Physik\\
der Karl-Franzens-Universit\"at Graz

\vspace{0.5cm}
Betreuer: Univ.-Prof. Dr. rer. nat. Reinhard Alkofer

\vspace{0.5cm}
Graz, M\"arz 2010
\end{center}

\end{titlepage}





\tableofcontents

\chapter{Introduction}

Symmetries play an important role in physics especially in the theories of the elementary forces observed in nature: gravitation, the electromagnetic force, the weak force and the strong force. They are described by gauge theories, which are invariant under certain \textit{local} symmetry transformations. The force of interest for this thesis is the strong force whose theory is quantum chromodynamics (QCD) \cite{Fritzsch:1973pi,Itzykson:1980ft,Pokorski:1990gf,Pascual:1984zb,Peskin:1995,Ryder:1996qf,Muta:1998}. The elementary particles of QCD are gluons and quarks, which build up hadrons like protons and neutrons. To be precise this thesis is about the gluonic part of QCD, which is called Yang-Mills theory \cite{Yang:1954ek}.

The overall structure of the Lagrangian of QCD is similar to that of quantum electrodynamics (QED) which describes electrons and photons: The electrons/quarks have mass and interact via the massless photons/gluons. But while electrons and photons can be observed as free particles, as attested by the human eye, quarks and gluons only appear in bound states like protons or neutrons. This phenomenon of QCD is called confinement \cite{Alkofer:2006fu,Greensite:2003bk} and will be of special interest in this thesis.

For the investigation of QCD many different tools are available, each with their own disadvantages and advantages.
The approach chosen for the present work are functional equations. They form an infinite tower of coupled equations, which naturally has to be truncated for calculations. However, in some asymptotic cases it is possible to obtain general results for the whole system. This provides then constraints for the solutions of the equations and can determine the usefulness of a given truncation. Hence such information constitutes a reliable starting point for numerical calculations. The main topic of this thesis are systems where such solutions can be obtained.

Functional methods, mainly Dyson-Schwinger equations \cite{Dyson:1949ha,Schwinger:1951ex,Schwinger:1951hq} and functional renormalization group equations \cite{Wetterich:1992yh}, have by now a long tradition in the investigation of Yang-Mills theory and also QCD. As it is the most convenient choice in this context, the Landau gauge was usually employed \cite{vonSmekal:1997is,vonSmekal:1997vx,Alkofer:2000wg,Fischer:2002hn} with a few exceptions. During the last decade the methods have improved considerably and while at the beginning several assumptions had been required, many of them could be put aside and the results got more and more rigorous. Unfortunately the situation is not as clear as one would wish, because two different solutions are known, which only differ in the deep infrared \cite{Fischer:2008uz}. The connection between these two solutions has been elucidated only partially and it is not settled if both or only one solutions are correct. It has even been conjectured that in the former case they correspond to two different non-perturbative Landau gauges that coincide perturbatively \cite{Maas:2009se}.

In Chapter \ref{chp:IRYM} I shortly review these two solutions for the Landau gauge. In the remainder of the thesis I focus then on one of them, the so-called scaling solution. Chapter \ref{chp:IRYM} contains also an overview over confinement scenarios and aspects of the infrared regime of Yang-Mills theory as well as of the gauge fixing procedure.

In the Landau gauge the methods for the investigation of its scaling solution have proven to be very reliable and we have by now a fairly good understanding of many aspects of QCD like dynamical chiral symmetry breaking \cite{Alkofer:2006gz,Alkofer:2008tt} and confinement in this gauge. Transferring this knowledge to other gauges seems like a natural next step to understand the complete picture of QCD. So it is promising to focus on a gauge with different features than Landau gauge. The maximally Abelian gauge \cite{'tHooft:1981ht} provides a perfect first candidate for such an endeavour: It is a covariant and renormalizable gauge that is closely related to the dual superconductor picture of confinement \cite{Mandelstam:1974pi,'tHooft:1975pu}. The meaning of this confinement scenario in Landau gauge is somewhat unclear, as there are no chromomagnetic monopoles which seem to play a decisive role. The findings of this thesis about the maximally Abelian gauge, presented in Chapter \ref{chp:MAG}, further elucidate the connection to Landau gauge confinement scenarios.

The second issue investigated is the influence of the Gribov horizon on the Landau gauge scaling solution. It arises when attempting a complete gauge fixing, which is required for non-perturbative calculations, as the standard Landau gauge definition is insufficient. As a consequence the integration in the path integral should be restricted to the interior of the Gribov horizon \cite{Gribov:1977wm}. This is implemented in the Gribov-Zwanziger action by additional terms \cite{Gribov:1977wm,Zwanziger:1989mf}. Interestingly the gluons are then confined already at tree-level as their propagator vanishes at zero momentum. The qualitative picture of an infrared vanishing gluon propagator and an infrared enhanced ghost propagator is also found by Dyson-Schwinger and functional renormalization group equations \cite{vonSmekal:1997is,vonSmekal:1997vx,Alkofer:2000wg,Fischer:2002hn,Pawlowski:2003hq}. However, these functional equations did up to now not explicitly take into account the restriction to the interior of the Gribov horizon as it was argued that one can formally cut off the integration at the Gribov horizon without changing the equations \cite{Zwanziger:2001kw}. This conjecture is explicitly tested in Chapter \ref{chp:GZ} and confirmed by analyzing the Dyson-Schwinger equations of the Gribov-Zwanziger action.

As both the maximally Abelian gauge and the Landau gauge with the Gribov-Zwan\-ziger action are considerably more complicated than the standard Landau gauge, the available methods had to be improved and extended for their investigations. Chapter \ref{chp:scalingSolutions} describes this improved method as general as possible so that it can be employed for a broad range of different actions and is not restricted to the cases investigated in this thesis. To demonstrate its usefulness I give the examples of the Landau gauge, linear covariant gauges and ghost anti-ghost symmetric gauges. Under certain assumptions this method provides the means to derive all possible scaling solutions for a theory from the structure of its interactions.

Additionally, the basic derivation of the Dyson-Schwinger equations proved to be quite time consuming so that I developed a computer program dedicated to this task. The underlying algorithm, which is suited for manual derivations as well, is explained in Chapter \ref{chp:FunctionalEquations}, where also an introduction to Dyson-Schwinger and functional renormalization group equations can be found. The program is named \textit{DoDSE} ("Derivation of Dyson-Schwinger equations") and freely available on the \textit{Computer Physics Communications Program Library}\footnote{\href{http://cpc.cs.qub.ac.uk/summaries/AECT_v1_0.html}{http://cpc.cs.qub.ac.uk/summaries/AECT\_v1\_0.html}}. Given the fields and interactions of a theory it derives the Dyson-Schwinger equations in symbolic notation and even the algebraic expressions when Feynman rules are provided.

Finally, I present in several appendices details of some calculations. The derivations of formulae used in Section \ref{chp:scalingSolutions} can be found in Appendix \ref{chp:details-scaling} together with the generalization to $d$ dimensions. The employed conventions for Grassmann fields are described in Appendix \ref{chp:GrassmannFields}. Appendix \ref{chp:MAGFeynmanRules} contains the Feynman rules of the maximally Abelian gauge and Appendix \ref{chp:sunset} a formula for the calculation of the sunset diagram. An overview over the use of \textit{DoDSE} is given in Appendix \ref{chp:DoDSE} including the calculation of the two-point function Dyson-Schwinger equations in the maximally Abelian gauge and for the Gribov-Zwanziger action.

\chapter{Yang-Mills theory and its infrared behavior}
\label{chp:IRYM}

Yang-Mills theory \cite{Yang:1954ek} describes the gluonic part of the strong interaction. It is a gauge theory and consequently the choice of a gauge is required in functional approaches. This amounts to choosing one representative among all the physically equivalent gauge field configurations. In Section \ref{sec:YMAction} I will describe the standard procedure for gauge fixing, as employed in perturbation theory, as well as the problems arising when going to the non-perturbative regime. An improved gauge fixing taking these into account will be presented in Chapter~\ref{chp:GZ}.

As in this thesis the aspects of the infrared (IR) behavior of correlation functions is investigated, I will describe in Section \ref{sec:aspectsIR} two well known ways of connecting the IR behavior of propagators to confinement: First I explain the basic idea behind the Kugo-Ojima scenario \cite{Kugo:1995km,Kugo:1979gm} and then I introduce the Gribov-Zwanziger scenario \cite{Gribov:1977wm,Zwanziger:1993dh,Zwanziger:1992qr,Zwanziger:1991gz}. The idea behind the latter will be important for Chapter \ref{chp:GZ}.

In the final section of this chapter I will shortly review the development of our understanding of the IR regime in the Landau gauge. This gauge is widely preferred in functional approaches because of its simplicity. Indeed calculations in all other gauges lead to one or more additional obstacles that have to be overcome. But although the Landau has been intensively investigated for many years by now, there is no consensus yet about the solution for the Green functions in the IR. Two possible candidates have emerged, and even the variant that they correspond only to different non-perturbative extensions of the Landau gauge is discussed. I will describe the features of both solutions in this section.
Furthermore, I will also explain the notion of positivity violation, which is a mathematical criterion for confinement.

\section{The action of Yang-Mills theory}
\label{sec:YMAction}

In this section the Lagrangian of Yang-Mills theory is introduced as the gluonic part of QCD. How to fix the gauge, the BRST symmetry \cite{Tyutin:1975qk,Becchi:1975nq,Becchi:1974md} and properties of the Gribov region are described.

\subsection{The gauge invariant action}

The Lagrangian density of QCD is \cite{Fritzsch:1973pi,Yang:1954ek}
\begin{align}\label{eq:actionQCD}
 \mathcal{L}_{QCD}&=\bar{q}(-\slashed{D}+m )q + \mathcal{L}_{YM},\\
 \mathcal{L}_{YM}&=\frac{1}{2} \,tr\lbrace F_{\mu \nu }F_{\mu \nu} \rbrace \label{eq:actionYM}.
\end{align}
The first term in $\mathcal{L}_{QCD}$ describes the propagation of the massive quarks and their interaction with gluons via the covariant derivative $D_\mu$:
\begin{align}
 D_\mu=\partial_\mu+i\,g\,A_\mu.
\end{align}
The second term, $\mathcal{L}_{YM}$, constitutes the purely gluonic part of QCD and thus contains the propagation of gluons and their self-interactions. While it is not possible to find a gauge invariant Lagrangian for quarks without gluons, this is not true in the opposite direction, i.e., $\mathcal{L}_{YM}$ is gauge invariant and can be considered on its own without quarks. It is this term that is investigated in this thesis.

The quantity $F_{\mu \nu}$ is called the field strength tensor and is given by
\begin{align}
F_{\mu \nu}&=\partial_\mu A_\nu -\partial_\nu A_\mu +i\,g\,[A_\mu,A_\nu].
\end{align}
The gauge field $A_\mu$ lives in an algebra defined by the hermitian generators $T^r$ of a generic gauge group. They obey the relations
\begin{align}
[ T^r,T^s]&=i f^{rst}T^t,\\
tr \lbrace T^r T^s \rbrace &= T_f \delta^{rs},
\end{align}
with $T_f=\mhalfo$ for $SU(N)$.
The decomposition of the gauge field is
\begin{align}
 A_\mu=A_\mu^r T^r
\end{align}
and similarly for the field strength tensor:
\begin{align}
 F_{\mu \nu}&=F_{\mu \nu}^r T^r,\\
 F_{\mu \nu }^r&=\partial_\mu A_\nu^r-\partial_\nu A_\mu^r-g\,f^{rst}A_\mu^s A_\nu^t.
\end{align}
In components the YM Lagrangian reads
\begin{align}\label{eq:Lagrangian-YM}
 \mathcal{L}_{YM}&=\frac{1}{4} F_{\mu \nu }^rF^r_{\mu \nu}.
\end{align}

The transformation of the gluon field under which the QCD Lagrangian is invariant is given by
\begin{align}\label{eq:gauge-trans}
A^U_\mu(x)=U(x)\,A_\mu(x)\,U(x)^{-1}+\frac{i}{g}(\partial_\mu U(x))\,U(x)^{-1},
\end{align}
where $U(x)$ is
\begin{align}
U(x)=e^{i\,g\, \omega(x)}
\end{align}
with $\omega(x)$ the Lie algebra valued gauge parameter:
\begin{align}
 \omega(x)=\omega^r(x)T^r.
\end{align}
Eq. (\ref{eq:gauge-trans}) is called a gauge transformation. In infinitesimal form it reads
\begin{align}\label{eq:infinitesimalGT}
 A_\mu^r \rightarrow A_\mu^r+\delta A_\mu^r=A_\mu^r - \partial_\mu \omega^r -g\,f^{rst} \omega^s A_\mu^t=A_\mu^r-D_\mu^{rs} \omega^s,
\end{align}
where the covariant derivative in the adjoint representation $D_\mu^{rs}$ is defined as
\begin{align}
 D_\mu^{rs}=\delta^{rs}\partial_\mu+g\,f^{rst} A_\mu^t.
\end{align}

As mentioned above the gauge group for the strong interaction is the compact simple Lie group $SU(3)$. In this thesis, however, I will for a large part adopt the gauge group $SU(N)$. The reason is that not only $SU(3)$ per se is of interest, but also general $SU(N)$ and especially $SU(2)$. In the latter case calculations can become easier and especially in the maximally Abelian gauge  $SU(2)$ is often employed as the Lagrangian simplifies considerably, see Chapter \ref{chp:MAG}. General $SU(N)$ or also other groups like $G_2$, see, for example, refs. \cite{Maas:2007af,Liptak:2008gx}, are investigated to determine the dependence of certain aspects of Yang-Mills theories on the gauge group. Also investigations in the so-called large $N$ limit are performed \cite{'tHooft:1973jz}.

\subsection{Fixing the gauge}
\label{ssec:gaugeFixing}

To work with functional methods we cannot use the action \eref{eq:Lagrangian-YM} as problems appear at several points. For example, in the canonical quantization, see, e.g., ref. \cite{Pascual:1984zb}, the commutator relations cannot be fulfilled, as the conjugate momentum of $A_\mu$, given by
\begin{align}
 \Pi_\mu(x)=\frac{\partial \mathcal{L}}{\partial (\partial_0 A_\mu(x))}=F_{0\mu},
\end{align}
has no zero component $\Pi_0$. But this contradicts the equal-time commutation relation
\begin{align}
 [A_\mu(x), \Pi_\nu(y)]=i\,g_{\mu\nu}\de(x-y).
\end{align}
If, however, a certain gauge is chosen, $\Pi_0$ does no longer vanish and one can proceed as usual.

In the path integral the source of the problems is that one integrates over all gauge equivalent gauge field configurations, although the idea is only to take into account gauge non-equivalent, i.e., physically different configurations. In other words the defining symmetry of the theory makes the integration in the path integral overcomplete.
Furthermore, one cannot calculate the propagator of the theory, as the gluon two-point function has zero eigenvalues. Consequently it is singular and cannot be inverted to get the propagator.

It is useful to introduce here the notion of a gauge orbit, which is the set of field configurations connected by a gauge transformation:
\begin{align}
 O[A]:=\lbrace A_\mu' | A_\mu'=A^U_\mu \rbrace.
\end{align}
$A^U$ is given in \eref{eq:gauge-trans}.
Ideally one takes only one representative per gauge orbit. The idea by Faddeev and Popov to achieve this was to restrict the integration in the path integral to a hyperplane \cite{Faddeev:1967fc}. This can be done by inserting unity, given by
\begin{align}\label{eq:Delta}
1=\Delta[A]\int \mathcal{D}U \delta(f[A^U]),
\end{align}
into the path integral. The delta functional defines the hyperplane $f[A]=0$ and $\mathcal{D}U$ is an integration over group space. The factor $\Delta[A]$ is the Jacobian arising from the transition from field variables $A$ to gauge transformation variables $U$. The integration over $U$ can be absorbed into the normalization of the path integral and the factor $\Delta[A]$ can be calculated as
\begin{align}
 \Delta^{rs}[A]&=det \left(\frac{\de f^r[A(x)]}{\de \omega^s(y)}\right)=:det\,M^{rs}(x,y),
\end{align}
where the color indices have been made explicit.
$M(x,y)$ is known as the Faddeev-Popov operator. Its determinant can be localized by the introduction of a pair of Lie algebra valued Grassmann fields, the so-called Faddeev-Popov ghosts $\bar{c}$ and $c$:
\begin{align}
 det\,M^{rs}(x,y)=\int \mathcal{D}[\bar{c}c]e^{\int dx\,dy\,\bar{c}^r(x)\,M^{rs}(x,y)\,c^s(y)}.
\end{align}
Although they have zero spin, they obey Fermi statistics, which underlines their status as auxiliary and not physical fields.

As an alternative to the restriction to a hyperplane one can relax this condition to a Gaussian distribution over the gauge orbit with the mean value defined as the original gauge fixing condition. One example are linear covariant gauges with a gauge fixing parameter $\xi$ that gives the width of the distribution around the Landau gauge fixing condition $\partial_\mu A_\mu=0$. In the limit of $\xi \rightarrow 0$ the Landau gauge is recovered. One should note that such a Gaussian distribution includes all gauge copies. Since it is normalized it does not lead to the same problems as the unnormalized integration before.

The gauge fixing part which is added to the Lagrangian density is then given by
\begin{align}\label{eq:Lagrangian-gaugeFixing}
 \mathcal{L}_{gf}= \frac{1}{2\xi}(\partial_\mu A^r_\mu(x))^2-\int dy \,\bar{c}^r(x)\,M^{rs}(x,y)\,c^s(y).
\end{align}
Also in the Landau gauge the full expression has to be added,
\begin{align}
 \mathcal{L}_{YM} \rightarrow \mathcal{L}_{YM}+\mathcal{L}_{gf},
\end{align}
and only after inverting the gluon two-point function one can set $\xi =0$. In the path integral the factor $exp(-\frac1{2\xi}(\partial_\mu A_\mu^r)^2)$ reduces to a delta functional in the limit $\xi \rightarrow 0$.

\subsection{The Faddeev-Popov operator and the Gribov region}

In the remainder of this section we will confine ourselves mostly to the Landau gauge in order to discuss some specific properties of the Faddeev-Popov operator and the Gribov region. In Landau gauge it is given by
\begin{align}
 M^{rs}(x,y)=M^{rs}(x)\delta(x-y)=-\partial_\mu D_\mu^{rs}\delta(x-y).
\end{align}
It is easy to demonstrate that the Landau gauge fixing is not complete as there are still gauge copies left. However, the fact that the gauge is not fixed completely by a local gauge fixing condition is not a specific property of the Landau gauge, but valid for all local gauge fixing conditions \cite{Singer:1978dk}. Starting with a gauge configuration fulfilling the Landau gauge condition, $\partial_\mu A_\mu=0$, we can perform a gauge transformation according to \eref{eq:gauge-trans} demanding that the result again fulfills the Landau gauge condition:
\begin{align}
 \partial_\mu A_\mu =0 \rightarrow \partial_\mu A_\mu - \partial_\mu D_\mu \omega\overset{!}{=}0 \quad \Rightarrow \quad M \omega\overset{!}{=}0.
\end{align}
Hence if the Faddeev-Popov operator has zero modes, there are still gauge equivalent configurations left.

The first to treat this issue was Gribov \cite{Gribov:1977wm}. He proposed to restrict the configuration in the path integral to the region $\Omega$, where the Faddeev-Popov operator $M(A)$ is strictly positive:
\begin{align}
 \Omega:=\{A | \partial_\mu A_\mu=0, M(A)>0\}.
\end{align}
Positivity entails that the operator is invertible, which is important as its inverse corresponds to the propagator of the Faddeev-Popov ghosts.
Today the region $\Omega$ is known as the Gribov region, which has the following properties in Landau gauge:
\begin{itemize}
 \item The vacuum configuration, $A_\mu=0$, lies in the Gribov region \cite{Zwanziger:2003cf}. This can easily be verified by observing that it fulfills $\partial_\mu A_\mu=0$ and the Faddeev-Popov operator reduces to the Laplacian, $M(0)=-\Box$, which is a positive operator. Since perturbation theory is an expansion around $A=0$, one can understand that it yields good results, as long as the quantum fluctuations do not become large enough to feel the presence of the Gribov horizon.
 \item The Gribov region is bounded in all directions \cite{Zwanziger:2003cf}. In this it differs decisively from the Gribov region of the maximally Abelian gauge \cite{Capri:2008vk,Capri:2010an}, which will be discussed in Section \ref{ssec:MAGSolution}.
 \item It is convex, i.e., two arbitrary configurations $A_1$ and $A_2$ within the Gribov region can be combined to a new configuration $A_3$ that lies again within the Gribov region as follows \cite{Zwanziger:2003cf}:
 \begin{align}
  A_3=\alpha A_1 + (1-\alpha) A_2, \qquad 0\leq\alpha \leq1.
 \end{align}
 \item Every gauge orbit passes at least once through the Gribov region \cite{Dell'Antonio:1991xt}. This property is important so that one can restrict the integration to the Gribov region without missing any gauge orbits.
 \item Unfortunately there are still gauge copies within the Gribov region so that the gauge fixing is not complete \cite{vanBaal:1991zw,vanBaal:1997gu}. However, expectation values are not influenced by these additional copies \cite{Zwanziger:2003cf}.
\end{itemize}

The statement that every gauge orbit passes through the Gribov region is related to the fact that the Landau gauge condition can be derived by minimizing the functional
\begin{align}\label{eq:R}
 R[A]:=\frac1{2}\int dx A_\mu(x)A_\mu(x)
\end{align}
with respect to infinitesimal gauge transformations, \eref{eq:infinitesimalGT}:
\begin{align}
 \delta R[A]&=\int dx A^r_\mu(x) \delta A^r_\mu(x)=\int dx A^r_\mu(x)(-D^{rs}_\mu(x)\omega^s(x))=\nnnl
&=\int dx A^r_\mu(x)(-\partial_\mu \omega^r(x)-g\,f^{rst}A^t_\mu(x) \omega^s(x))=\nnnl
&=\int dx (\omega^r(x)\partial_\mu A^r_\mu(x)).
\end{align}
In order to have an extremum $\partial_\mu A_\mu^r$ must vanish. The type of the extremum can be determined by the second derivative:
\begin{align}
 \delta^2 R[A]&=\int dx (\omega^r(x)(-\partial_\mu D_\mu^{rs}(x)\omega^s_\mu(x))=\int dx\, \omega^r(x)M^{rs}(x) \omega^s(x).
\end{align}
Here we recognize the Faddeev-Popov operator $M^{rs}$, i.e., if it is positive, we have a minimum of the functional $R[A]$. Thus the restriction to the Gribov region corresponds, as asserted above, to the minimization of the functional $R[A]$.

As the Gribov region is still plagued by gauge copies \cite{vanBaal:1991zw,vanBaal:1997gu}, one can think about better ways of fixing the gauge. A natural choice for a unique gauge fixing is the set of gauge field configurations that corresponds to the absolute minimum of $R[A]$. This region is known as the fundamental modular region. It possesses a topologically non-trivial boundary where degenerate global minima exist that have to be identified. Taking the global minimum is also known as absolute Landau gauge and taking one arbitrary minimum as minimal Landau gauge. In lattice calculations one can employ algorithms that correspond to the latter or approximate the former and indeed finds an influence on the correlation functions \cite{Maas:2008ri,Bornyakov:2008yx}. More on different Landau gauges on the lattice can be found in Section \ref{ssec:GribovRegionOnLattice}.

\subsection{Replacing gauge symmetry by the BRST symmetry}

In the action fixed to a hyperplane in field configuration space gauge invariance is explicitly broken. However, there is another symmetry that takes its place. It is named BRST symmetry after Becchi, Rouet, Stora \cite{Becchi:1975nq,Becchi:1974md} and Tyutin \cite{Tyutin:1975qk}. It is very useful in proving renormalizability and unitarity of a theory, see, for example, refs. \cite{Kugo:1979gm,Piguet:1995er}. The corresponding transformations can be derived from the standard gauge transformation by replacing the gauge parameter $\omega$ by a ghost field $c$. Due to this choice the pure Yang-Mills part is trivially invariant. The invariance of the gauge fixing part is explained below.

It is convenient to introduce the gauge fixing condition via an auxiliary Lie-algebra valued field $b$ that takes the role of a Lagrangian multiplier:
\begin{align}
\int Db^r\,e^{-\int dx \left(i b^r f^r[A] \right )}=N \delta(f^r[A]).
\end{align}
It is not dynamical and called Nakanishi-Lautrup field \cite{Nakanishi:1972pt,Lautrup:1967}.
Relaxing the gauge fixing condition $f[A]=0$ into $f[A]=i\,\xi\,b/2$ leads to
\begin{align}
 \int Db^r\,e^{-\int dx \left(i b^r f^r[A]+ \frac{\xi}{2} b^r  b^r \right )}=N e^{-\frac1{2\xi}(f[A])^2}
\end{align}
and corresponds to the Gaussian averaging over the gauge orbit in linear covariant gauges as described above. $N$ is some normalization factor and $\xi$ is called a gauge fixing parameter. It determines the width of the Gaussian distribution. Specifically $\xi=0$ is the original gauge fixing condition. However, one should keep in mind that this is only well defined if the solution to $f[A]=0$ is unique, what is not the case non-perturbatively.

In Landau gauge the off-shell BRST transformation reads:
\begin{align}\label{eq:BRST}
s\,A_\mu^r&=-D_\mu^{rs} c^s, \\
s\,c^r&=-\mhalfo g\,f^{rst} c^s c^t, \\
s\,\bar{c}^r&=i \, b^r,\\
s\,b^r&=0.
\end{align}
Integrating out the Nakanishi-Lautrup field yields the on-shell form, where we just have to replace $b^r$ by $-i\,(\partial_\mu A_\mu^r)/\xi$. A substantial property of the BRST symmetry is its nilpotency, $s^2=0$, in the off-shell case. Thus BRST can also be defined as follows: The gauge parameter of the gauge transformation is taken as the anti-commuting field $c$. Requiring nilpotency of this transformation, i.e., $s^2 A_\mu=0$, determines $s\,c$. Finally the fields $\bar{c}$ and $b$ are introduced as a BRST doublet, i.e., they have the trivial BRST transformations $s\,\bar{c}=i\,b$ and $s\,b=0$.

The nilpotency property allows an easy way to fix the gauge without the need for a path integral \cite{Kugo:1981hm}. This is based on the observation that one can add any quantity that is the result of a BRST transformation, a so-called BRST exact quantity, to the Lagrangian without spoiling its BRST invariance. As the gauge fixing condition $f[A]$ has ghost number zero and the BRST transformation itself raises the ghost number by one, we introduce the factor $\bar{c}$ to get ghost number zero and add $s(\bar{c} f[A])$. In the case of the Landau gauge, where $f[A]=\partial_\mu A_\mu$, this prescription to fix the gauge directly leads to the known gauge fixing terms:
\begin{align}\label{eq:BRST-LG}
 \mathcal{L}_{gf}&=s(\bar{c}^r f[A]^r)=s(\bar{c}^r \partial_\mu A^r_\mu)=\nnnl
 &=i\,b^r\,(\partial_\mu A^r_\mu)-\bar{c}^r\, \partial_\mu (-D^{rs}_\mu c^s)= i\,b^r (\partial_\mu A^r_\mu)-\bar{c}^r\,M^{rs}\,c^s.
\end{align}
One minus sign stems from the anti-commutativity property of $s$ and $\bar{c}$. This method can be used also for other gauges and allows the use of gauge fixing conditions depending on ghost fields \cite{Kugo:1981hm}. It will be employed for the maximally Abelian gauge in Chapter \ref{chp:MAG}, where it significantly simplifies the gauge fixing procedure.
As the expectation value of any gauge invariant quantity remains unaffected by adding such a BRST exact form, one can very nicely see in this way of quantization that all physical observables are independent of the chosen gauge.

\section{Aspects of the asymptotic infrared regime}
\label{sec:aspectsIR}

Solving a Dyson-Schwinger equation (DSE) numerically for the complete momentum region is a challenging task and it proves useful to know the asymptotic behavior of the correlation functions \cite{Maas:2005xh,vonSmekal:1997vx}. For the ultraviolet (UV) perturbation theory naturally provides a good guideline. For the IR regime it is more complicated to determine the qualitative behavior and thus it is worthwhile to develop analytic methods for its investigation. But the low momentum region is not only of interest for providing input for numerical calculations. One can also learn about the realization of different scenarios of confinement \cite{Alkofer:2006fu}. In this section I will give a short overview over confinement scenarios directly related to correlation functions. The description of an additional scenario, the dual superconductor picture, is deferred to Section \ref{sec:dualSC} as it does not translate directly into conditions for the correlation functions. A purely mathematical criterion for confinement, violation of positivity, is described later in Section \ref{sec:IR-Landau}.

\subsection{The Gribov-Zwanziger confinement scenario}

As mentioned in Section \ref{ssec:gaugeFixing}, the usual gauge fixing is not sufficient for the non-per\-tur\-bative regime. An improvement can be achieved by restricting the integration in field configuration space to the first Gribov region. This is explicitly realized by the so-called Gribov-Zwanziger action \cite{Gribov:1977wm,Zwanziger:1989mf}. Its derivation and IR analysis are described in Chapter \ref{chp:GZ}. Here I only mention the qualitative consequences of this improved gauge fixing for the theory.

The main statements of the Gribov-Zwanziger confinement scenario in the Landau gauge are that the gluon propagator vanishes at zero momentum \cite{Zwanziger:1991gz} and that the ghost propagator is IR enhanced, i.e., it diverges stronger than a simple pole \cite{Zwanziger:1992qr,Zwanziger:1993dh}. The interesting point is that these statements are realized in the Gribov-Zwanziger action already at the perturbative level. The reason is that the existence of the Gribov horizon, which is a non-perturbative object, is taken into account. The tree-level gluon propagator is
\begin{align}
 D_{A,\mu\nu}^{rs}(p^2)=\delta^{rs}\left(g_{\mu\nu}-\frac{p_\mu p_\nu}{p^2} \right)\frac{p^2}{p^4+2\,N\,g^2\gamma^4},
\end{align}
where $N$ is the number of colors and $\gamma$ a mass parameter. $\gamma$ is not free but determined by the horizon condition, see Section \ref{ssec:GZAction}. This condition has to be enforced in order to make the theory well-defined in the first place \cite{Gribov:1977wm}.

Naively the bare ghost propagator goes like $1/p^2$, but it turns out that the horizon condition leads to a cancelation of this term at one-loop level and the IR leading part of the ghost propagator goes like $1/p^4$. One can show this cancelation diagrammatically for the exact ghost propagator \cite{Zwanziger:2009je} and perturbatively it was checked up to two loops in ref. \cite{Gracey:2005cx}.

\subsection{The Kugo-Ojima confinement scenario}
\label{ssec:KugoOjima}

Although being based on completely different considerations as the Gribov-Zwanziger scenario, the mechanism proposed by Kugo and Ojima leads qualitatively to the same predictions for the IR behavior of the gluon and ghost propagators \cite{Kugo:1979gm,Kugo:1995km}. The basis of the Kugo-Ojima construction is the existence of a global, non-perturbatively defined BRST symmetry. It was long not clear if the standard BRST, as given in \eref{eq:BRST}, could be extended in this sense to the non-perturbative regime. Recent progress in this respect, however, leads exactly to this conclusion \cite{vonSmekal:2007ns,vonSmekal:2008ws,vonSmekal:2008es,Smekal:2009pd}.

The importance of the BRST symmetry rests on the fact that the corresponding charge defines the physical state space of the underlying quantum field theory by its cohomology. In contrast to the total state space, which possesses an indefinite metric, the physical state space must have a positive metric in order to allow a probabilistic interpretation of expectation values. According to Kugo and Ojima physical states $\phi \in \mathcal{V}_{phys}$ obey the condition
\begin{align}
 Q_B|\phi \rangle=0,
\end{align}
where $Q_B$ is the Noether charge corresponding to the BRST symmetry \cite{Kugo:1979gm,Nakanishi:1990qm}.
Furthermore, one can distinguish between states with positive and zero norm, $\Psi$ and $\chi \in \mathcal{V}_0$, respectively:
\begin{align}
 \langle \Psi | \Psi \rangle >0, \qquad \qquad  \langle \chi | \chi \rangle =0.
\end{align}
The zero norm states $\chi$ are also called daughter states, as they are the BRST variation of unphysical states, the parent states $|\alpha \rangle$:
\begin{align}
 |\chi \rangle=Q_B| \alpha \rangle, \quad | \alpha \rangle \notin \mathcal{V}_{phys}.
\end{align}
Consequently the subspace $\mathcal{V}_0 \in \mathcal{V}_{phys}$ is orthogonal to $\mathcal{V}_{phys}$ and does not influence any amplitude in $\mathcal{V}_{phys}$.
The physical Hilbert space equipped with a positive metric is then given by the completed quotient space
\begin{align}
 H_{phys}=\overline{\mathcal{V}_{phys}/\mathcal{V}_0}.
\end{align}

In summary states fall in one of the following three classes:
\begin{itemize}
 \item BRST non-invariant states $| \alpha \rangle$: These are unphysical states, as the BRST charge does not annihilate them, $Q_B |\alpha\rangle \neq 0$.
 \item BRST exact states $|\chi \rangle$: They are the BRST variation of the states $|\alpha \rangle$. 
 \item BRST invariant but not BRST exact states $|\Psi \rangle$: They define real physical states, i.e., they are BRST invariant, $Q_B|\Psi\rangle=0$, and have a positive norm, $\langle \Psi | \Psi \rangle >0$.
\end{itemize}

Kugo and Ojima could show that the states $|\alpha \rangle$ and $|\chi\rangle$ always appear in so-called quartets, which is given by a "Faddeev-Popov conjugated" pair of BRST doublets \cite{Kugo:1979gm}. Using the notation $|k,N\rangle$, where $k$ represents all quantum numbers except for the ghost number $N$, the four particles of a quartet can be denoted as follows:
\begin{align}
 &|\phi_1\rangle=|k, \, N\rangle,&\quad &|\phi_2\rangle=Q_B|\phi_1\rangle=|k,\,N-1\rangle,\nnnl
 &|\phi_3\rangle=|k, \, -N\rangle, &\quad &|\phi_4\rangle=Q_B|\phi_3\rangle=|k,\, -(N+1)\rangle.
\end{align}
$|\phi_1\rangle$ and $|\phi_2\rangle$ are one BRST doublet, $|\phi_3\rangle$ and $|\phi_4\rangle$ the other. Furthermore, the states $|\phi_1\rangle$ and $|\phi_3\rangle$ belong to the class of BRST non-invariant states, and $|\phi_2\rangle$ and $|\phi_4\rangle$ to the BRST exact states. The contributions of these four states always cancel and they will never appear as asymptotic observable states.

The only other possibility of states are states $\Psi$ that are BRST invariant but not BRST exact, i.e., they obey $Q_B|\Psi \rangle =0$ and cannot be written as a BRST variation of another state. These states are called singlet states and should represent the physical states of the theory. For QCD this means that the fundamental, confined fields of quarks, gluons and ghosts should belong to some quartet, whereas physical states such as mesons, baryons and glueballs should be BRST singlets.

Based on the global color charge Kugo and Ojima derived a simple criterion for color confinement in Yang-Mills theory. Using the equation of motion of the gluon field the conserved global color current can be written as
\begin{align}
 J^a_\nu:=\partial_\nu F_{\mu\nu}^a+\{Q_B,D_\mu^{ab}\bar{c}^b\}, \qquad \partial_\nu J^a_\nu=0.
\end{align}
The related charge is the global color charge
\begin{align}
 Q^a:=G^a+N^a:=\int d^3 x \left(\partial_i F_{0i}^a+\{Q_B,D_0^{ab}\bar{c}^b\}\right),
\end{align}
where $G^a$ and $N^a$ correspond to the first and second terms in the integral. The first criterion of Kugo and Ojima is that $\partial_i F_{0i}^a$  contains no discrete massless pole. The charge $G^a$ then vanishes, as the integral is over a total derivative. If there were massless contributions, $G^a$ would be ill-defined. The second criterion requires that $N^a$ is zero and consequently the total color charge $Q^a$ is also zero.

In the charge $N^a$ the anti-ghost field appears. It belongs to the so-called elementary quartet and has a massless contribution from the asymptotic field $\bar{\gamma}^a$. The asymptotic contributions due to the composite operator $g f^{abc}A_\mu^c\bar{c}^b$ are given by $u^{ab} \partial_\mu \bar{\gamma}^b$ characterized by the dynamical parameter $u^{ab}$, which can be determined from the correlation function
\begin{align}
 \int dx \,e^{i\,p(x-y)}\langle \left(D_\mu^{ae}c^e\right)(x) gf^{bcd}A_\nu^d(y) \bar{c}^c(y) \rangle =\left(g_{\mu\nu}-\frac{p_\mu p_\nu}{p^2}\right) u^{ab}(p^2).
\end{align}
The condition for the charge $N^a$ to be well-defined can be inferred from the asymptotic behavior of $(D_\mu^{ab} \bar{c}^b)(x)$, given by
\begin{align}
 (D_\mu^{ab} \bar{c}^b)(x) \overset{x_0 \rightarrow \pm \infty}{=} (\de^{ab}+u^{ab})\partial_\mu \bar{\gamma}^b(x),
\end{align}
as
\begin{align}
 u^{ab}\equiv u^{ab}(0)=-\de^{ab}.
\end{align}

In summary the two confinement criteria due to Kugo and Ojima are:
\begin{enumerate}
 \item There is no discrete massless pole in $\partial_i F_{0i}^a$.
 \item The parameter $u^{ab}$ has to be $-\de^{ab}$.
\end{enumerate}

The last point was taken up again by Kugo sixteen years after his article with Ojima, ref. \cite{Kugo:1979gm}. In ref. \cite{Kugo:1995km} he derived how the Landau gauge ghost propagator, parametrized by
\begin{align}
 D^{ab}_c(p^2)=-\delta^{ab}\frac{c_c(p^2)}{p^2}
\end{align}
is related to the parameter $u$:
\begin{align}
 c_c(0)=\frac{1}{1+u}.
\end{align}
Hence the ghost propagator is more IR singular than a simple pole, if the criterion $u=-1$ by Kugo and Ojima is met.

\section{Solutions of Landau gauge Yang-Mills theory in the infrared}
\label{sec:IR-Landau}

The non-perturbative calculation of propagators and vertices is involved and in the course of history the emerging picture underwent some changes and improvements. The best investigated gauge in this context is Landau gauge for which I will give a short review in this section.

\subsection{From infrared slavery to ghost dominance in the Landau gauge}

The Landau gauge is a preferred gauge in functional approaches for several reasons. First of all, it is the gauge with the simplest form of the action. There are only two fields which interact via three vertices. This simplicity is supported by additional information such as, for example, the fact that the system of the transverse components of Green functions is closed, see, e.g., ref. \cite{Fischer:2008uz}, and the longitudinal terms can be discarded. Furthermore, there have been arguments that the ghost-gluon vertex stays bare in the IR \cite{Marciano:1977su,Taylor:1971ff}. This was used as input at the beginning, but having understood the gauge better and better it could be proven directly from functional equations \cite{Lerche:2002ep,Fischer:2009tn,Huber:2009wh}.

A well-known example of early non-perturbative calculations of the gluon propagator in the Landau gauge is due to Mandelstam \cite{Mandelstam:1979xd}. Motivated by perturbation theory he neglected the ghost loop, which only contributes a small amount at high momenta, and the four-gluon vertex, which only occurs at two-loop level. With an approximated three-gluon vertex he obtained an IR divergent gluon propagator and also an IR singular running coupling. As such a gluon propagator apparently is perfectly suited to yield a linear rising potential by single gluon exchange, this picture was widely accepted and became known under the name IR slavery. The results of the Mandelstam approximation were subsequently confirmed in refs. \cite{Brown:1988bn,Hauck:1996sm,Worrall:1985pt}. However, this is a perfect example of how assumptions based on perturbation theory can lead in the wrong direction in the non-perturbative regime. In the late nineties it was shown in refs. \cite{vonSmekal:1997is,vonSmekal:1997vx} that the ghost propagator yields the most important contribution to the IR sector of Landau gauge Yang-Mills theory. In fact, it was found that the ghost contributions were dominating all DSEs \cite{Alkofer:2004it,Alkofer:2008jy,Fischer:2009tn} supporting the scenarios of ghost dominance as proposed by Gribov and Zwanziger \cite{Gribov:1977wm,Zwanziger:1992qr} and Kugo and Ojima \cite{Kugo:1979gm,Kugo:1995km}.

This solution is characterized by power laws for the dressing functions $c_A(p^2)$ and $c_c(p^2)$ of the gluon and ghost propagators, respectively:
\begin{align}
 D_A(p^2)= d_A \cdot (p^2)^{\de_A}, \quad D_c(p^2)= d_c \cdot (p^2)^{\de_c}.
\end{align}
$d_A$ and $d_c$ are momentum independent coefficients and the qualitative behavior is determined only by the exponents, called infrared exponents (IREs).
An important feature of these power laws is that the exponents depend only on one parameter, usually denoted by $\ka$, as they are related by a so-called scaling relation:
\begin{align}\label{eq:scal-rel-LG}
 \ka:=-\de_c=\de_A/2.
\end{align}
The value of $\ka$ can be calculated analytically. The most reliable value is $0.5953\ldots$ \cite{Lerche:2002ep,Zwanziger:2001kw}. It is obtained for a bare ghost-gluon vertex and changes only slightly when an IR constant, but momentum dependent dressing is employed \cite{Lerche:2002ep}.

The fact that the ghost-gluon vertex is not IR enhanced was originally used as an input to solve the DSEs. It relied on an argument by Taylor \cite{Marciano:1977su,Taylor:1971ff} and was confirmed by lattice \cite{Cucchieri:2006tf,Ilgenfritz:2006he,Maas:2007uv,Cucchieri:2008qm} and DSEs studies \cite{Schleifenbaum:2004id,Alkofer:2008dt}. The self-consistency of this assumption was confirmed later on \cite{Alkofer:2004it} and finally it was shown that this is indeed the only possible IR solution \cite{Fischer:2009tn,Huber:2009wh}. This proof was possible by combining the two distinct systems of functional renormalization group equations and DSEs as suggested in \cite{Fischer:2006vf}. A different approach with the same result was taken in ref. \cite{Alkofer:2008jy}, where as an assumption the existence of a stable skeleton expansion in the IR was used. It turned out that such an assumption necessarily holds as can be derived from functional renormalization group equations \cite{Huber:2009wh}. Details on the proof can be found in Section \ref{ssec:skeletonExpansion}, where this emerges merely as a side result from the general analysis of the systems of functional equations.

\subsection{The decoupling solutions and the scaling solution}
\label{ssec:decoupling-scaling}

\begin{figure}[t]
 \begin{center}
  \includegraphics[width=0.43\textwidth]{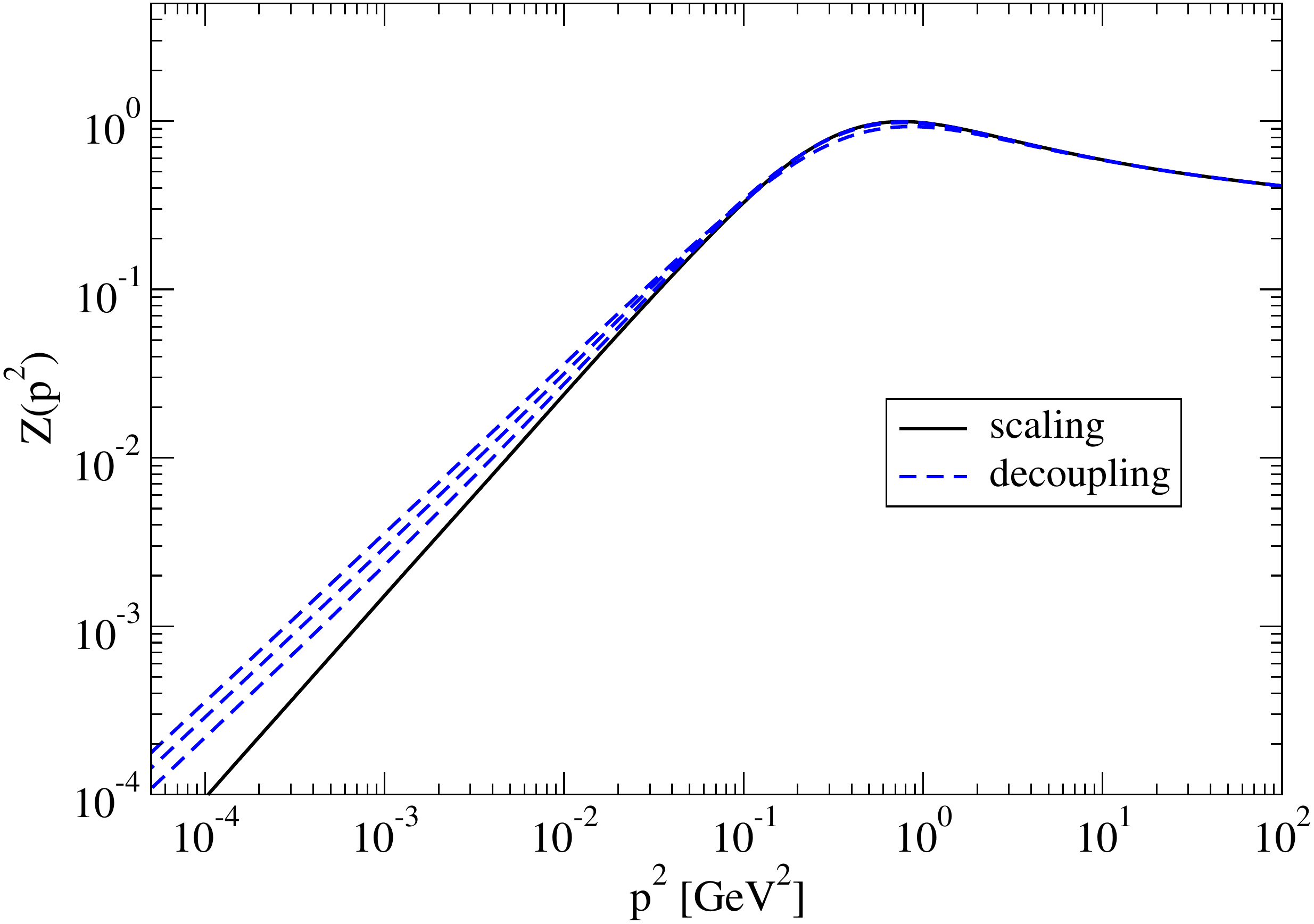}
  \hskip1cm
  \includegraphics[width=0.43\textwidth]{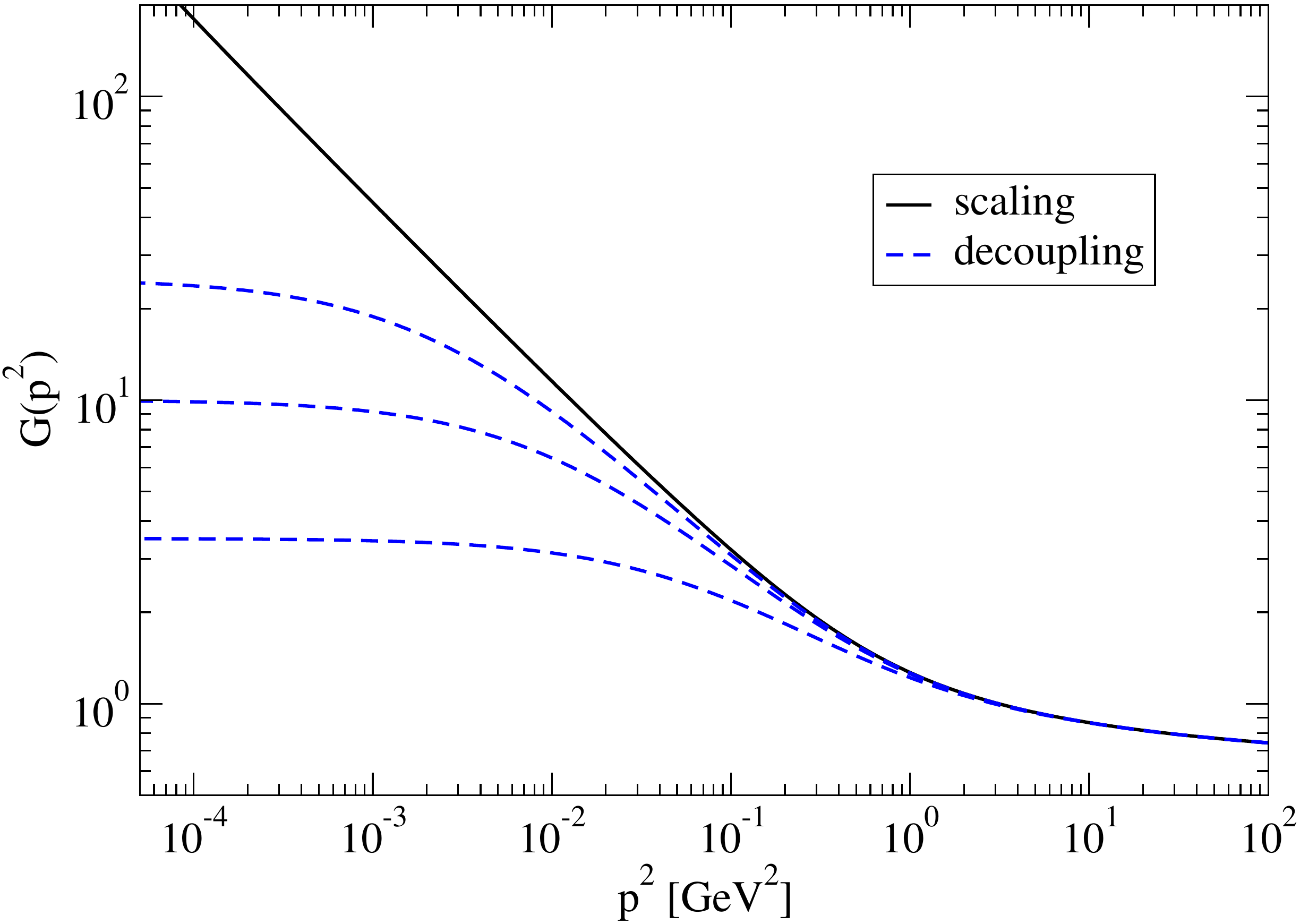}
  \caption{\label{fig:scaling-decoupling}Numerical solutions of the gluon and ghost two-point DSEs. Plotted are the gluon (left) and ghost (right) dressing functions for the decoupling and scaling solutions. Printed with permission from Elsevier from ref. \cite{Fischer:2008uz}.}
 \end{center}
\end{figure}

Although the uniqueness of this solution in terms of scaling laws was established, the results did not agree with calculations on the lattice, see, for example, refs. \cite{Bloch:2003sk,Bogolubsky:2005wf,Sternbeck:2005tk,Ilgenfritz:2006he,Cucchieri:2006xi,Sternbeck:2006cg,Bogolubsky:2007bw,Cucchieri:2007rg,Cucchieri:2007md,Oliveira:2007dy,Bogolubsky:2007ud,Cucchieri:2008fc}. For exceptions see refs. \cite{Oliveira:2007dy,Oliveira:2008uf}, the two-dimensional case \cite{Maas:2007uv,Cucchieri:2007rg} and the strong coupling limit \cite{Sternbeck:2008na,Sternbeck:2008mv,Sternbeck:2007ug}. Indeed another solution was found with continuum methods that possesses quite different characteristics: The ghost propagator is not IR enhanced but stays bare in the IR, and the gluon propagator becomes finite instead of going to zero \cite{Dudal:2007cw,Dudal:2008sp,Aguilar:2008xm,Boucaud:2008ji,Fischer:2008uz}. This behavior is responsible for the name \textit{decoupling solution}, as the gluons decouple below the scale given by their mass. One should note that a massive behavior for the gluon propagator is also possible in the formerly found solution, now called \textit{scaling solution}, since the value of the parameter $\ka$ could be $1/2$. But then the ghost propagator would still be IR enhanced via the scaling relation \eref{eq:scal-rel-LG}. In the decoupling solution, however, the propagators do not obey an IR scaling relation and the ghost stays bare in the IR. Nevertheless there is some realization of ghost dominance in the decoupling solution, as the gluon is IR suppressed. Another qualitative difference between the two solutions is that the scaling solution is unique, while there exists a family of decoupling solutions \cite{Fischer:2008uz}. In \fref{fig:scaling-decoupling} both solutions are plotted.

The connection between the two solutions is the renormalization of the ghost propagator DSE \cite{Fischer:2008uz}. Solving the DSEs one requires renormalization conditions and for the ghost propagator it is convenient to specify the renormalization condition at zero momentum. Choosing a finite value for the ghost dressing function yields a decoupling solution, where the value of the ghost dressing function at zero momentum is connected to the mass of the gluon \cite{Fischer:2008uz}. For an infinite ghost dressing function at zero momentum the scaling solution is recovered \cite{vonSmekal:1997is,vonSmekal:1997vx,Fischer:2008uz}. One should note that it is not necessary to fix the value at zero momentum but the normalization prescription can be defined at any arbitrary momentum, since the values for different momenta are uniquely connected. Mathematically the necessary choice of the renormalization condition for the ghost dressing function corresponds to a boundary condition for the DSEs.

This is also related to an argument by Zwanziger that one can in principle cut the integration in field configuration space at the first Gribov horizon without modifying the form of the equations \cite{Zwanziger:2001kw}. What is changed, however, are the boundary conditions. In case one cuts at the first Gribov horizon one uses as boundary condition that the ghost propagator is more enhanced than a simple pole, as derived from the horizon condition \cite{Zwanziger:2009je,Zwanziger:1993dh,Zwanziger:1992qr}. One can also argue for the same boundary condition with the Kugo-Ojima confinement scenario \cite{Kugo:1995km,Kugo:1979gm} without referring to the Gribov region. Unfortunately it is not understood how these two arguments are related, since the implementation of the Kugo-Ojima condition $u=-1$ into the action leads to the Gribov-Zwanziger action \cite{Dudal:2009xh}. However, the Gribov-Zwanziger action breaks the BRST symmetry, while the Kugo-Ojima formalism relies on an unbroken BRST charge.

What speaks in favor that both solutions are valid is the fact that physical quantities do not seem to depend on the deep IR, the only region where the two solutions differ from each other. In ref. \cite{Fischer:2009gk} the transition temperatures of the confinement/deconfinement and the chiral transitions at vanishing chemical potential were calculated with identical results for both solutions.

In the present work I concentrate only on the scaling solution, which is more accessible to analytic approaches and offers explanations for many aspects of Yang-Mills theory. The most important one is certainly the confinement of gluons which can directly be inferred from the IR vanishing gluon propagator, as one can show analytically that the Schwinger function $\Delta(t)$ is not positive \cite{Zwanziger:1991gz}. This, however, is necessary to interpret a particle as physical
\cite{Osterwalder:1973dx,Osterwalder:1974tc}. It boils down to show that
\begin{align}
 \Delta(t):=\int d^3x \int \frac{d^4p}{(2\pi)^4}e^{i\,x\,p}\sigma(p^2)\geq0,
\end{align}
where $\sigma(p^2)$ is a scalar function characterizing the propagator. The Fourier transformation of the gluon propagator at zero momentum is related to the Schwinger function by
\begin{align}
 D(p^2=0)=\int d^4x D(x-y)=\int dt \Delta(t).
\end{align}
Thus if $D(p^2=0)=0$, $\Delta(t)$ is either zero or it has equally distributed positive and negative contributions.
In this case one speaks of maximal positivity violation. But also for the decoupling solution a violation of positivity is observed both by lattice simulations \cite{Cucchieri:2004mf,Sternbeck:2006cg,Bowman:2007du} and functional equations \cite{Fischer:2008uz}. 

A delicate point is the global BRST symmetry of the solutions. If a global symmetry exists that has the same form as the perturbative definition, \eref{eq:BRST}, we know from the seminal paper of Kugo and Ojima how the propagators of the ghost and the gluon should behave in the deep IR \cite{Kugo:1979gm,Kugo:1995km}: The ghost propagator should be IR enhanced and the gluon propagator should vanish at zero momentum. Only the scaling solution fulfills these criteria. However, one cannot exclude a different realization of a global BRST that leads to a different qualitative behavior in agreement with the decoupling solution.

Finally I would like to note that both the decoupling and the scaling solutions fulfill the criterion for a confining Polyakov loop potential \cite{Braun:2007bx} and thus quarks are confined. This is a sufficient criterion depending only on the asymptotic IR part, but in actual calculations it is found that the region responsible for confinement is the mid-momentum regime, where both solutions agree \cite{Braun:2007bx}.
\chapter{Functional equations}
\label{chp:FunctionalEquations}

This chapter serves as a short introduction to Dyson-Schwinger (DSEs) and functional renormalization group equations (FRGEs). Both systems of equations form complete sets describing the theory exactly. As such they are suited for the investigation of non-perturbative phenomena.

The derivation of DSEs is described in Section \ref{sec:DSEs}. First some general relations among the generating functionals are provided in order to allow a self-contained description. Then the process of deriving the equations is explained and the computational tool \textit{DoDSE} based on the presented algorithm is introduced.

FRGEs are treated in Section \ref{sec:FRGEs}. They require the introduction of the so-called effective average action which depends on an artificial momentum scale and can be interpreted as an action interpolating between the classical and the quantum action. After introducing the effective average action, the derivation of FRGEs is sketched.

\section{Dyson-Schwinger equations}
\label{sec:DSEs}

Dyson-Schwinger equations (DSEs) are named after F.~J.~Dyson \cite{Dyson:1949ha} and J.~S.~Schwinger \cite{Schwinger:1951ex,Schwinger:1951hq}. Put into one sentence, DSEs are the equations of motion of Green functions and describe the propagation and interaction of the fields of the theory. The full system of DSEs provides a complete description of the theory. Therefore DSEs can be used to generate a perturbative expansion in the weak coupling regime, but they show their true strength when applied for strong coupling. This makes them a perfect tool for the investigation of Yang-Mills theory and also QCD, where non-perturbative methods are needed to explain such phenomena as confinement and dynamical chiral symmetry breaking. Both are important properties of these theories and cannot be accounted for by standard perturbation theory.

\subsection{The generating functionals}
\label{ssec:generatingFunctionals}

The easiest way to derive DSEs is via the path integral. A derivation with the canonical formalism is possible, but less instructive and definitely more tedious. The interested reader is referred to ref. \cite{Rivers:1988pi}. The path integral also allows for an easier algorithmic implementation.

In the following we will consider a theory with fields $\phi_i$. The multi-index $i$ contains the field type, all indices of the fields, as can be, for example, Lorentz or color indices, and also the space-time or momentum dependence. Equal indices in a product are summed and integrated over. This avoids cumbersome notation. The action is then\footnote{Here we restrict ourselves to three- and four-point functions, but the generalization is straightforward.}
\begin{align}\label{eq:S-phi}
 S[\phi]=&\frac{1}{2!}S_{rs}\phi_r \phi_s - \frac{1}{3!} S_{rst} \phi_r \phi_s \phi_t -
	\frac{1}{4!}S_{rstu} \phi_r \phi_s \phi_t \phi_u.
\end{align}
The statistical factors are chosen such that the coefficients $S_{rs}$, $S_{rst}$ and $S_{rstu}$ denote the bare two-, three and four-point functions, and the choice of signs is a consequence of the definition of the vertices, see \eref{eq:def-vertices} below. From this action the path integral is constructed as
\begin{align}
Z[J]=\int D[\phi] e^{-S + \phi_i J_i}=e^{W[J]},
\end{align}
where $J_i$ is the source of the field $\phi_i$. The path integral $Z[J]$ is also called the generating functional for full Green functions and $W[J]$ that for connected Green functions. A Legendre transform of $W[J]$ with respect to the sources yields the so-called effective action that generates the one-particle irreducible (1PI) Green functions, which are those Green functions that are still connected after one internal line is cut:
\begin{align}
 \Gamma[\Phi]&=-W[J]+\Phi_i J_i.
\end{align}
It depends on the averaged fields $\Phi$ in the presence of external currents $J$,
\begin{equation}
\Phi_{i}:=\left\langle \phi_{i}\right\rangle _{J}=\frac{\delta W[J]}{\delta J_{i}}=Z[J]^{-1}\int D[\phi] \phi_i e^{-S + \phi_j J_j}.
\end{equation}
The current, on the other hand, can be expressed as the derivative of the effective action:
\begin{align}
 J_i&=\frac{\delta \Gamma[\Phi]}{\delta \Phi_i}.
\end{align}

Derivatives of the effective action with respect to fields are abbreviated as $\Gamma_{i_{1}\cdots i_{n}}^{J}$:
\begin{align}\label{eq:def-vertices}
\Gamma_{i_{1}\cdots i_{n}}^{J}:=-\frac{\delta\Gamma[\Phi]}{\delta\Phi_{i_{1}}\cdots\delta\Phi_{i_{n}}}.
\end{align}
The minus sign is chosen for later convenience such that no additional minus signs in the DSEs of vertices appear.
The $\Gamma_{i_{1}\cdots i_{n}}^{J}$ are not yet the physical n-point functions of the theory as the external sources $J$ are still non-vanishing. Setting them to zero physical propagators $D_{ij}$ and vertices $\Gamma_{i_{1}\cdots i_{n}}$ are obtained:
\begin{align}
D_{ij}&:=D_{ij}^{J=0},\\
\label{eq:def-vertices-J0}
\Gamma_{i_{1}\cdots i_{n}}&:=\Gamma_{i_{1}\cdots i_{n}}^{J=0}.
\end{align}
The function $D_{ij}^J$ is given by
\begin{align}
D_{ij}^{J}:=\frac{\delta^{2}W[J]}{\delta J_{i}\delta J_{j}}=\left(\frac{\delta^{2}\Gamma[\Phi]}{\delta\Phi_{i}\delta\Phi_{j}}\right)^{-1}\label{eq:super-prop}.
\end{align}
It is important in the derivation of DSEs that the sources $J$ are set to zero only at the end. Otherwise one would miss contributions.

The propagators $D_{ij}$ are the inverse of the two-point functions $\Gamma_{ij}$. It should be stressed that this is a matrix relation, i.e., if the two-point matrix is not diagonal there is a non-trivial relationship between propagators and two-point functions. This complicates analyses of actions with mixed propagators like the Gribov-Zwanziger action. For a diagonal matrix in case of bosons or an off-diagonal matrix for fermions the situation is simpler, since the propagator can be directly calculated as the inverse of the corresponding two-point function.

\subsection{Derivation of Dyson-Schwinger equations}

Using our multi-index field the formal derivation can be done in a few lines. The real work is then to expand in the physical fields of the theory. DSEs for full, connected and 1PI Green functions can be worked out, but since the full theory can be reconstructed from 1PI correlators we focus on those.

To start we note that the integral of a total derivative vanishes so that
\begin{align}\label{eq:DSE-Z}
\int D[\phi] \left( -\frac{\delta S}{\delta \phi_i} + J_i \right) e^{-S + \phi_j J_j}=\left( -\frac{\delta S}{\delta \phi'_i}\Bigg\vert_{\phi'_i=\delta/\delta J_i} +J_i \right) Z[J]=0.
\end{align}
Employing further derivatives with respect to sources yields the DSEs of the full Green functions.
Substituting $Z[J]$ by $e^{W[J]}$ and using
\begin{align}
e^{-W[J]}\left(\frac{\delta}{\delta J_i}\right)e^{W[J]}= \frac{\delta W[J]}{\delta J_i}+\frac{\delta}{\delta J_i} 
\end{align}
after multiplication of \eref{eq:DSE-Z} from the left with $e^{-W[J]}$ we find
\begin{align}\label{eq:W-DSE}
-\frac{\delta S}{\delta \phi_i}\Bigg\vert_{\phi_i=\frac{\delta W[J]}{\delta J_i}+\frac{\delta}{\delta J_i}} +J_i=0.
\end{align}
This is the generating DSE for connected correlation functions. The DSEs of connected Green functions are obtained by acting with further source derivatives on \eref{eq:W-DSE}. To get the corresponding version for 1PI functions we perform the Legendre transformation of $W[J]$ with respect to all sources. Thereby $\delta W[J]/\delta J_i$ changes to $\Phi_i$ and $\delta/\delta J_i$ becomes
\begin{align}
\frac{\delta}{\delta J_i}=\frac{\delta \Phi_j}{\delta J_i} \frac{\delta}{\delta \Phi_j}=\frac{\delta}{\delta J_i} \frac{\delta W[J]}{\delta J_j} \frac{\delta}{\delta \Phi_j}=\frac{\delta^2 W[J]}{\delta J_i \delta J_j} \frac{\delta}{\delta \Phi_j}=
D_{ij}^J \frac{\delta}{\delta \Phi_j}.
\end{align}
This yields
\begin{align}\label{eq:DSE-master}
-\frac{\delta S}{\delta \phi_i}\Bigg\vert_{\phi_i=\Phi_i+D_{ij}^J  \, \delta/\delta \Phi_j} +\frac{\delta \Gamma[\Phi]}{\delta \Phi_i}=0,
\end{align}
which is the basic equation. All DSEs for 1PI Green functions can be derived from it by further differentiations with respect to the fields.

Let us write down this expression in the case of an action with three- and four-point interactions as given in \eref{eq:S-phi}. The derivative is simply
\begin{align}
\label{eq:1st-deriv-DSE}
\frac{\delta{S}}{\delta{\phi_i}}=S_{is}\phi_s  - \frac{1}{2!} S_{ist} \phi_s \phi_t -
	\frac{1}{3!}S_{irst} \phi_s \phi_t \phi_u.
\end{align}
Before replacing the field $\phi_i$ by $\Phi_i + D_{ij}^J \frac{\delta}{\delta \Phi_j}$ we need to know how the differentiation operator $\frac{\delta}{\delta \Phi_j}$ acts on fields and propagators. For completeness we also include the derivative of an n-point function:
\begin{subequations}\label{eq:derivatives}
\begin{align}
\frac{\delta}{\delta\Phi_{i}}\Phi_{j} & =\delta_{ij},\\
\frac{\delta}{\delta\Phi_{i}}D_{jk}^{J} & =\frac{\delta}{\delta\Phi_{i}}\left(\frac{\delta^{2}\Gamma[\Phi]}{\delta\Phi_{j}\delta\Phi_{k}}\right)^{-1}=\nnnl
 &=-\left(\frac{\delta^{2}\Gamma[\Phi]}{\delta\Phi_{j}\delta\Phi_{m}}\right)^{-1}\left(\frac{\delta^{3}\Gamma[\Phi]}{\delta\Phi_{m}\delta\Phi_{i}\delta\Phi_{n}}\right)\left(\frac{\delta^{2}\Gamma[\Phi]}{\delta\Phi_{n}\delta\Phi_{k}}\right)^{-1}=D_{jm}^{J}\Gamma_{min}^{J}D_{nk}^{J}, \label{eq:der-prop} \\
\frac{\delta}{\delta\Phi_{i}}\Gamma_{j_{1}\cdots j_{n}}^{J} & =-\frac{\delta\Gamma[\Phi]}{\delta\Phi_{i}\delta\Phi_{j_{1}}\cdots\delta\Phi_{j_{n}}}=\Gamma_{ij_{1}\cdots j_{n}}^{J}.\end{align}
\end{subequations}
Eq. (\ref{eq:der-prop}) can be derived from the equation $\partial ( M\,M^{-1})=0$, where $M_{ij}=D^J_{ij}$ and $M^{-1}_{ij}=\Gamma^J_{ij}$.
Using these relations in \eref{eq:1st-deriv-DSE} yields
\begin{align} 
\label{eq:genDSE}
\frac{\delta \Gamma[\Phi]}{\delta \Phi_i}=&S_{is} \Phi_s -  \mhalfo S_{ist}(\Phi_s \Phi_t + D_{st}^J) + \nnnl
	&-\frac{1}{3!}S_{istu} (\Phi_s \Phi_t \Phi_u + 3 \Phi_s D_{tu}^J + D_{sv}^J D_{tw}^J D_{ux}^J \Gamma_{vwx}^J).
\end{align}
In \fref{fig:functDSE} a graphical representation of this equation is shown. Also the differentiation rules from \eref{eq:derivatives} can be depicted graphically as given in \fref{fig:diagRules}. Since all required ingredients are now available in graphical form, it is very convenient to proceed like this.

\fig{t}{fig:functDSE}{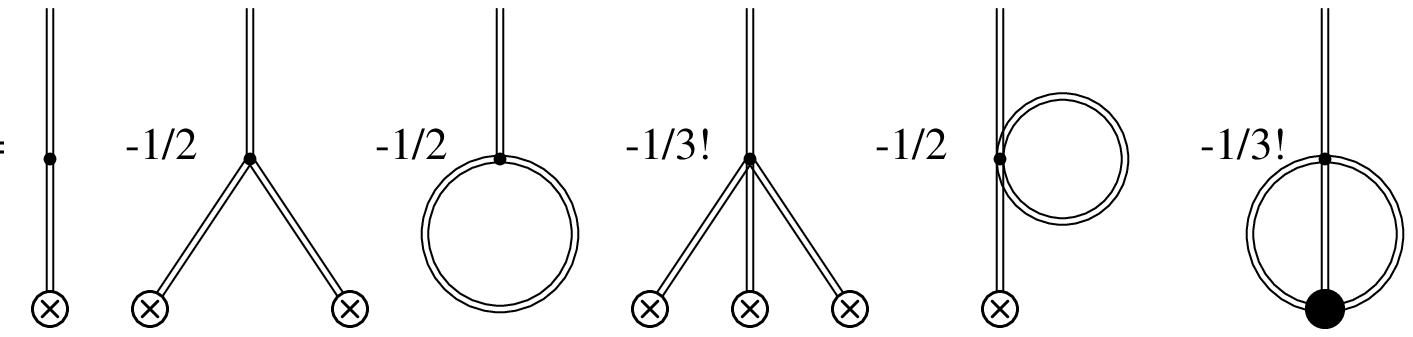}{The generating DSE for 1PI functions. Crosses in circles denote external fields. All internal propagators are dressed and the big blob denotes a dressed 1PI vertex. The double line represents the generic field $\Phi$.}{width=0.7\textwidth}

\fig{b}{fig:diagRules}{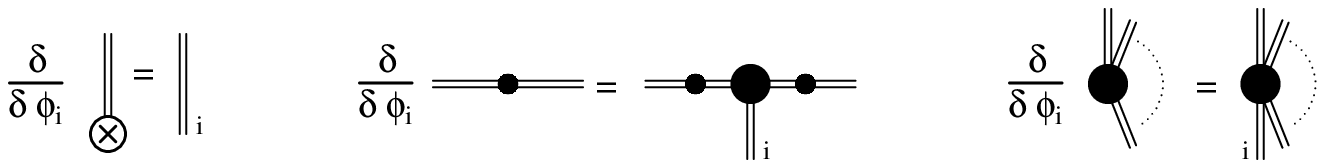}{Diagrammatic rules for differentiating an external field, a propagator or a vertex. The circle with the cross denotes an external field, small blobs denote dressed propagators, and big blobs 1PI vertices. The double line represents the super-field $\Phi$.}{width=0.9\textwidth}

The DSE for a generic two-point function is derived by performing another differentiation of the generating DSE \eref{eq:genDSE} using the diagrammatic replacement rules of \fref{fig:diagRules} in the corresponding diagrammatic representation \fref{fig:functDSE}. The result is shown in \fref{fig:2Point}. Proceeding to higher vertex functions the number of diagrams increases rapidly: For three-point vertices there are 15 generic diagrams and for four-point functions 60. For real applications it is therefore recommendable to exploit possible simplifications.

First, the final number and form of graphs depend on the first differentiation in \eref{eq:genDSE} as the corresponding field determines which bare vertices appear in the diagrams. For example, the DSE of the ghost-gluon vertex in Landau gauge QCD has only four terms, when the first derivative is performed with respect to a ghost field. In this case one can drop all diagrams with bare gluonic vertices. On the other hand, if one starts with the gluon field, all vertices have to be kept and one ends up with twelve graphs. A detailed derivation of these two examples is provided in ref. \cite{Alkofer:2008nt}. Secondly, one can skip some diagrams taking into account where one is going. Simple examples are that for a three-point function we do not have to drag along the bare four-point vertex or diagrams with an external field can be dropped if no further derivatives with respect to this particular field follow.

\fig{t}{fig:2Point}{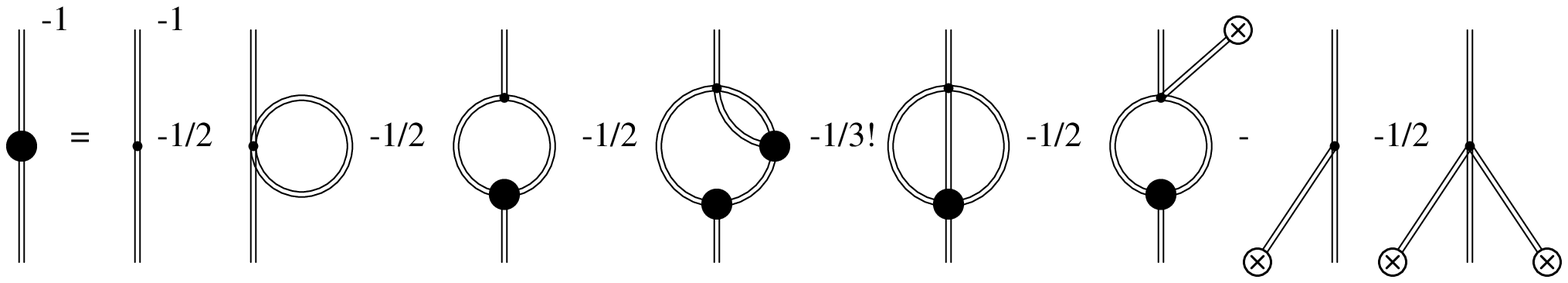}{The DSE for a generic two-point function.}{width=0.95\textwidth}

An important point is keeping the internal indices general and do not set them to specific values too early in the derivation process in order to try to simplify the calculation. This may discard some diagrams.

The algorithm described so far did not take into Grassmann fields. As they represent only a technical complication and the basic structure remains the same, their inclusion is described in Appendix \ref{chp:GrassmannFields}.

\subsection{\textit{DoDSE}: Derivation of Dyson-Schwinger equations}

Deriving DSEs can easily become very tedious: First of all, in the case of actions with a small number of interactions the number of terms blows up so fast that only the lowest n-point functions can be done by hand. Secondly, the interest in using DSEs of actions with many interaction terms is increasing as we get DSEs better under control. Lagrangians with up to eleven vertices are used and necessitate a quick and easy method to derive DSEs. The algorithm presented above allows the implementation into a symbolic programming language and thereby the automated derivation of DSEs.

For such a purpose several programming languages are available, e.g., \textit{FORM} \cite{Vermaseren:2000nd} or \textit{Mathematica} \cite{Wolfram:1999}. The latter was chosen due to its greater accessibility, its more user-friendly interface and the possibility to draw the results directly as Feynman diagrams. The result is a package called \textit{DoDSE}, which is short for "Derivation of Dyson-Schwinger Equations". For the beginner it is a useful tool to derive DSEs and represent them graphically and for the advanced user it provides further possibilities like step-by-step calculations, keeping external fields or using it for actions with mixed two-point functions. For simple applications, like, for example, summing up the IR exponents of a given graph, the primary output of \textit{DoDSE} can be used for further calculations in \textit{Mathematica}. However, if one would like to obtain the integral in full glory with color and Lorentz indices and so on, some more effort is required.

For this purpose \textit{DoDSE} can use Feynman rules provided by the user to write down the complete expressions for a DSE. They can then be further processed with other functions as required by the user. Automating this process proved very useful for the calculation of the numeric values of IR exponents. While in normal Landau gauge this is a manageable task by hand, it can become arbitrarily complicated for other examples like the maximally Abelian gauge or the Gribov-Zwanziger action where there are more diagrams with more dressing functions to calculate, see Chapters \ref{chp:MAG} and \ref{chp:GZ}, respectively, for more details. A short guide to \textit{DoDSE} and the derivation of the DSEs for the maximally Abelian gauge and the Gribov-Zwanziger action can be found in Appendix \ref{chp:DoDSE}.

\section{Functional renormalization group equations}
\label{sec:FRGEs}

FRGEs are used in a variety of fields in physics, for example, in the investigation of ultracold fermion gases, e.g., \cite{Diehl:2009ma,Diehl:2007th}, for supersymmetric models, e.g., \cite{Gies:2009az,Synatschke:2010ub,Synatschke:2009nm}, in gravitation, e.g., \cite{Manrique:2009tj,Reuter:1996cp,Reuter:2001ag,Eichhorn:2010tb,Eichhorn:2009ah} or for the phase diagram of QCD, e.g., \cite{Schaefer:2006sr,Schaefer:2006ds,Braun:2007bx,Schaefer:2008hk,Braun:2009gm}. This list is by far not exhausted. It should only give a glimpse at the vast applicability of FRGEs, for reviews see refs. \cite{Berges:2000ew,Pawlowski:2005xe,Gies:2006wv,Schaefer:2006sr}.

This section is devoted to the description of some properties of FRGEs and of their derivation. Details on how to use FRGEs in the case of interest for this thesis, the IR analysis of Yang-Mills theory, are provided in Section \ref{ssec:FRGEs-IR}.

\subsection{The effective average action}
\label{ssec:effAvAction}

The functional renormalization group is formulated with the help of the effective average action $\Gamma[\Phi]^k$, which is closely related to the effective action or generating functional of 1PI Green functions we encountered in Section \ref{ssec:generatingFunctionals}. The defining property of the effective average action $\Gamma[\Phi]^k$ is the dependence on an artificial momentum scale $k$, denoted by the superscript. The flow equations, derived below in Section \ref{ssec:flowEquations}, describe the dependence on this scale by an integro-differential equation. The quantum fluctuations above the scale $k$ are all integrated out. Thus two special limits of the effective average action are $k\rightarrow 0 $ and $k \rightarrow \infty$, where it has to correspond to the standard effective action $\Gamma[\Phi]$ and the classical action $S$, respectively: In the former case all quantum fluctuations are included and in the latter none. In statistical physics a natural UV cutoff $\Lambda$ can be present. In this case $\Gamma[\Phi]^k$ is equivalent to $S$ for $k \rightarrow \Lambda$. In the following, however, we will focus on the use of the functional renormalization group in quantum field theory.

The dependence on $k$ is introduced as an IR cutoff in the definition of the path integral:
\begin{align}
W^k[J]=ln \,Z^k[J]=ln \int D\phi \, e^{-S[\phi] +j_i \phi_i-\frac{1}{2}\phi_i R^k_{ij} \phi_j}.
\end{align}
We used again the notation from the last section, i.e., writing out the integration in the last term of the exponent gives
\begin{align}
 \frac1{2}\phi_i R^k_{ij} \phi_j= \frac1{2} \int \ddotp{q} R^k_{ab}(q) \phi_a(-q)\phi_b(q),
\end{align}
where $R^k_{ab}$ is an IR cutoff function with the following properties:
\begin{itemize}
 \item It has to vanish for $k \rightarrow 0$ to obtain the standard effective action.
 \item It has to diverge for $k \rightarrow \infty$ so that the classical action is recovered in this limit.
 \item For small momenta $q^2<k^2$ it is proportional to $k^2$, which can be interpreted as an effective mass that constitutes an IR cutoff for fluctuations with small momenta.
 \item Finally, it has to vanish for large momenta $q^2>k^2$ so that it does not interfere with the high momentum behavior of Green functions.
\end{itemize}
The effective average action is then defined via a modified Legendre transformation:
\begin{align}
 \Gamma^k[\Phi]=-W^k[J]+J_i \Phi_i - \frac1{2}\Phi_i R^k_{ij} \Phi_j
\end{align}
with
\begin{align}\label{eq:def-Phi-k}
 \Phi_i=\frac{\delta W^k[J]}{\delta J_i}=\langle \phi_i \rangle.
\end{align}

\subsection{Flow equations}
\label{ssec:flowEquations}

The dependence on the scale $k$ is used to derive the so-called flow equations or functional renormalization group equations. They describe how one gets from the classical action $S$ to the effective action $\Gamma[\Phi]$ by lowering $k$, which is equivalent to integrating out more and more fluctuations. The trajectory from $S$ to $\Gamma[\Phi]$ is called a flow. It can be interpreted as moving through theory space, starting at the microscopic theory and heading towards a macroscopic description. The trajectory depends on the chosen IR cutoff function $R^k$, but the endpoint is always the same, because it is determined by physics. Thus approximations and truncations can only be checked by comparing the endpoints, not the flows.

\begin{figure}[t]
 \begin{center}
\includegraphics[width=0.68\textwidth]{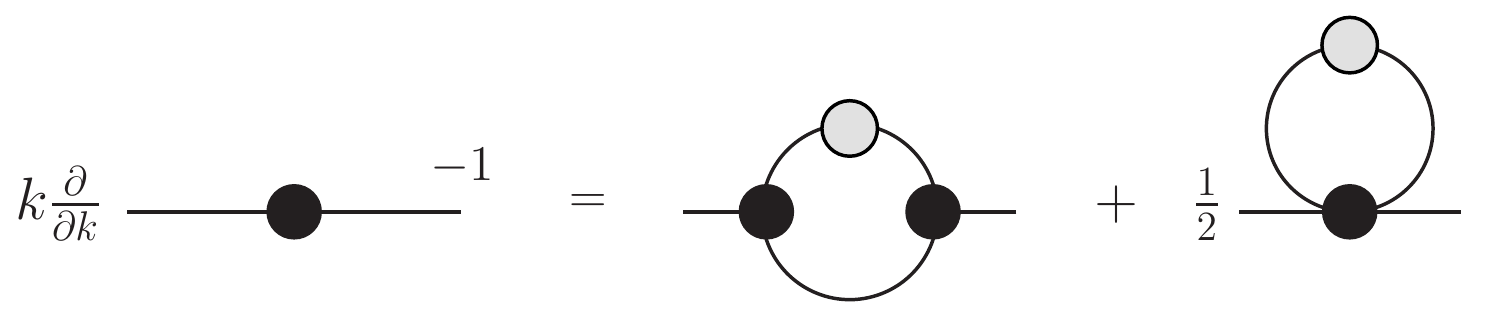}
 \caption{\label{fig:RG-example} The FRGE obtained from \eref{eq:FRGE-2p} for a theory with one field. A grey blob denotes the regulator insertion $\partial_k R^k$, a black blob a dressed n-point function. Internal propagators are all dressed.}
\end{center}
\end{figure}

Differentiating $\Gamma[\Phi]^k$ with respect to $k$ yields
\begin{align}\label{eq:deriv-k}
 \partial_k \Gamma^k[\Phi]=& -\partial_k W^k[J]-\frac{\delta W^k[J]}{\delta J_i} \partial_k J_i+ \frac{\partial J_i}{\partial k}  \Phi_i- \frac1{2} \Phi_i \partial_k R^k_{ij}\Phi_j=\nnnl
= & \left\langle \frac1{2} \phi_i \partial_k R^k_{ij} \phi_j \right \rangle -\frac1{2} \Phi_i \partial_k R^k_{ij}\Phi_j=\nnnl
= & \frac1{2}  \partial_k R^k_{ij} \left( \left \langle \phi_i  \phi_j \right\rangle - \Phi_i \Phi_j \right)=\nnnl
= & \frac1{2}  \partial_k R^k_{ij} G_{ij}^J.
\end{align}
where $\partial_k:=\partial/\partial k$ and \eref{eq:def-Phi-k} was used to cancel the second and third terms in the first line.
Furthermore, $\langle \phi_i \phi_j \rangle$ was decomposed as $G^J_{ij}+\langle \phi_i \rangle \langle \phi_j \rangle= G^J_{ij}+\Phi_i \Phi_j$, where
\begin{align}
 G^J_{ij}:=\frac{\delta^2 W^k[J]}{\delta J_i \delta J_j}
\end{align}
is the connected propagator in presence of the sources $J$. Its inverse is the two-point function but with an additional contribution from $R^k$:
\begin{align}
 \de_{ij}=\frac{\de \Phi_i}{\de \Phi_j}=\frac{\de J_l}{\de \Phi_j} \frac{\de}{\de J_l} \frac{\de W^k[J]}{\de J_i}=
 \frac{\de \left(\Gamma^k[\Phi]+\frac{1}{2} \Phi_m R^k_{mn} \Phi_n \right)}{\de \Phi_j \de \Phi_l}  \frac{\de^2 W^k[J]}{\de J_l  \de J_i}=
 \left(\Gamma^{k,J}_{jl}[\Phi] + R^k_{jl} \right) G^J_{li}
\end{align}
with
\begin{align}
 \Gamma^{k,J}_{ij}:=\frac{\de^2 \Gamma^k[\Phi]}{\de \Phi_i \de \Phi_j}
\end{align}
and
\begin{align}
 J_l=\frac{\de(\Gamma^k[\Phi]+\frac{1}{2}\Phi_m R^k_{mn} \Phi_n )}{\de \Phi_l}.
\end{align}
Thus \eref{eq:deriv-k} can also be written as
\begin{align}
 \partial_k \Gamma^k[\phi]=& \frac1{2} \left(\Gamma^{k,J}_{ij}[\Phi] + R^k_{ij} \right)^{-1} \partial_k R^k_{ij}.
\end{align}
From this equation functional identities for all Green function can be obtained by further differentiation. I will illustrate the procedure for the two-point function of a theory with three- and four-point functions:
\begin{align}\label{eq:FRGE-2p}
 \partial_k& \Gamma^{k,J}_{ij}=\frac{\de^2}{\de \Phi_i \de\Phi_j} \partial_k \Gamma[\Phi]^k=\nnnl
 &=\frac{\de^2}{\de \Phi_i \de\Phi_j}\frac1{2} \left(\Gamma^{k,J}_{mn} + R^k_{mn} \right)^{-1} \partial_k R^k_{mn}= \frac{1}{2} \frac{\de}{\de \Phi_i}  G^J_{mr} \Gamma^{k,J}_{rjs} G^J_{sn} \partial_k R^k_{mn} =\nnnl
 &= \frac{1}{2} G^J_{mt} \Gamma^{k,J}_{tiu} G^J_{ur} \Gamma^{k,J}_{rjs} G^J_{sn} \partial_k R^k_{mn} +
  \frac{1}{2} G^J_{mr}  \Gamma^{k,J}_{rjs} G^J_{st} \Gamma^{k,J}_{tiu} G^J_{un} \partial_k R^k_{mn} +
  \frac{1}{2}  G^J_{mr} \Gamma^{k,J}_{irjs} G^J_{sn} \partial_k R^k_{mn}.
\end{align}
The following derivatives have been used:
\begin{subequations}\label{eq:derivatives-RG}
\begin{align}
\frac{\de G_{ij}^J}{\de \Phi_l}&=\frac{\de\left(\Gamma^{k,J}_{ij} + R^k_{ij} \right)^{-1}}{\de \Phi_l}= G^J_{im} \Gamma^{k,J}_{mln} G^J_{nj},\\
\frac{\delta}{\delta\Phi_{i}}\Gamma^{k,J}_{j_{1}\cdots j_{n}} & =-\frac{\delta\Gamma^k[\Phi]}{\delta\Phi_{i}\delta\Phi_{j_{1}}\cdots\delta\Phi_{j_{n}}}=\Gamma^{k,J}_{ij_{1}\cdots j_{n}}.\end{align}
\end{subequations}
Again one has to keep in mind that propagators and vertices only correspond to physical quantities once the external sources are set to zero at the end of the derivation. Doing so in \eref{eq:FRGE-2p}, the FRGE depicted in \fref{fig:RG-example} is obtained for a theory with one field. The inclusion of Grassmann fields goes along the same lines as for DSEs, see Appendix \ref{chp:GrassmannFields}.

Performing more differentiations with respect to fields one can derive flow equations for all higher n-point functions, which can also be presented graphically like DSEs. The appearing integrals are all one-loop and only dressed quantities appear. This is a decisive difference to DSEs, which contain also bare n-point functions. Furthermore, every integral has an insertion of the regulator $R^k$. Suitable choices of it for the IR analysis are discussed in Section \ref{ssec:FRGEs-IR}.
\chapter{Scaling solutions}
\label{chp:scalingSolutions}

In this chapter a method to assess the possible existence and form of a scaling solution for a given theory is derived. The original goal of the underlying work was to investigate the IR behavior of Yang-Mills theory in the maximally Abelian gauge (MAG). However, it soon became clear that the system of interactions given by its Lagrangian requires a more refined treatment than provided by any previous work, the reason being the large number of terms in the DSEs. Subsequently a generic method was developed that can handle a large set of interactions. It allows to determine for an arbitrary system of interactions if a scaling solution is possible and what its qualitative features are. The main proof, albeit relying only on simple mathematics like combinatorics and inequalities, is rather technical. However, it provides a simple recipe for finding possible scaling solutions.

The underlying idea is to combine information from DSEs and FRGEs as suggested in ref. \cite{Fischer:2006vf}. Both systems of equations describe the complete content of a theory and their combination allows to extract enough additional information to derive all possible scaling relations.
With the method of a combined analysis it was even possible to prove that there is only one scaling solution in the Landau gauge \cite{Fischer:2009tn}. This is often called uniqueness of the scaling solution and should not be confused with uniqueness of the solution, see the discussion on scaling and decoupling solutions in Section \ref{ssec:decoupling-scaling}. The generic method for obtaining the scaling relations of more general cases, originally developed for the study of the MAG, has been presented in ref.~\cite{Huber:2009wh}.

The first section of this chapter provides an introduction to the method of IR power counting, which is employed in Section \ref{sec:basicRelations} to derive several general statements about the IR behavior of dressing functions. Section \ref{sec:solution-IR} explains the method to extract possible scaling solutions from a given Lagrangian, which is exemplified in Section \ref{sec:otherGauges} by employing it for linear covariant and ghost anti-ghost symmetric gauges. Up to this point all considerations are for the special case of a Lagrangian diagonal in the fields at the two-point level. Section \ref{sec:mixed2Point} provides the generalization to the case of mixed two-point functions.

\section{Infrared power counting}
\label{sec:IRPC}

In order to determine the qualitative IR behavior of correlation functions I rely in this thesis on a technique called IR power counting. It is based on the assumption of a power law behavior of all dressing functions in the IR. The qualitative information is then contained in the corresponding exponents and the analysis can be shifted from the integral equations to the exponents. In the following I will explain how they are extracted from DSEs and FRGEs. The role of possible or required cancelations in the functional equations is discussed at the end of this section.

\subsection{Integrals of DSEs at low external momenta}

The goal of this subsection is to explain how the IR behavior of an integral can be assessed. The starting point are the fully dressed propagators of the theory, parameterized by
\begin{align}
D_{ab}^{(i)}(p)=P_{ab}^{(i)}\frac{\,c^{(i)}(p^2)}{p^2},
\end{align}
where $P^{(i)}_{ab}$ is the part containing color and Lorentz structure and $c^{(i)}$ is some dressing function. The propagators are labeled by the superscript $(i)$.
A basic assumption supported by renormalization group arguments is that the dressing function $c^{(i)}$ obeys a simple power law in the IR:
\begin{align}
c^{(i),IR}(p^2)=d^{(i)}\cdot(p^2)^{\de^{(i)}}.
\end{align}
Here $d^{(i)}$ is a constant coefficient and the exponent $\de^{(i)}$ is called infrared exponent (IRE). It describes the qualitative IR behavior of the propagator. The assumption of power law behavior does not take into account the possibility of a logarithmic momentum dependence if $\de^{(i)}$ is zero. One speaks of IR enhancement/suppression if $\de^{(i)}<0$ or $\de^{(i)}>0$, respectively, and clearly the propagator vanishes if $\de^{(i)}>1/2$. The generalization to several dressing functions corresponding to different tensors is straightforward.

In the following we investigate DSEs at low external momenta. Such momenta occur within the integrals in factors like $1/(p-k)^2$, where one momentum is a loop momentum and the other an external one. Hence the integrand itself is dominated by low internal momenta and all dressing functions take their IR form in the limit of vanishing external momenta. Here we also assume a power like behavior for vertices in the IR with their own IREs. We will only treat the case that all momenta of a vertex vanish simultaneously. This is called uniform \cite{Alkofer:2008jy} or global scaling \cite{Fischer:2009tn}.

If all dressing functions are given by power laws the integral can in principle be solved by methods like Negative Dimensional Integration (NDIM), see, for instance, \cite{Dunne:1987qb,Dunne:1987am,Halliday:1987an,Ricotta:1990nd,Anastasiou:1999ui,Suzuki:2002vg,Suzuki:2002ak,Suzuki:2000us,Suzuki:1999ap,Suzuki:1997yz}, or employing the Mellin-Barnes representation, see, e.g., \cite{Boos:1987bg}. It is vital that these methods can handle non-integer exponents which is not the case for most standard methods like, for example, integration by parts \cite{Chetyrkin1981a,Tkachov1981}.

The simplest case is the one-loop two-point function, where the result can be written down as a simple closed expression:
\begin{align}\label{eq:2-point}
\int\frac{d^{d}k}{\left(2\pi\right)^{d}}\left(k^{2}\right)^{\nu_{1}}\left(\left(k-p\right)^{2}\right)^{\nu_{2}}=\left(4\pi\right)^{-\frac{d}{2}}\frac{\Gamma\left(\frac{d}{2}+\nu_{1}\right)\Gamma\left(\frac{d}{2}+\nu_{2}\right)\Gamma\left(-\frac{d}{2}-\nu_{1}-\nu_{2}\right)}{\Gamma\left(-\nu_{1}\right)\Gamma\left(-\nu_{2}\right)\Gamma\left(d+\nu_{1}+\nu_{2}\right)}\left(p^{2}\right)^{\frac{d}{2}+\nu_{1}+\nu_{2}} .
\end{align}
The convergence of the integral gives constraints on the exponents $\nu_1$ and $\nu_2$: UV convergence dictates $d/2+\nu_1+\nu_2 \leq0$ and IR convergence $d/2+\nu_1\geq0$ and $d/2+\nu_2\geq0$. Also for the one-loop three-point integral a solution in closed form is known \cite{Boos:1987bg,Anastasiou:1999ui}. The qualitative behavior if all external momenta scale equally can easily be assessed. A detailed numerical analysis, however, is complicated by the appearing Appell's function $F_4$ that have several singularities in the space of the two variables given by ratios of the external momenta. Series representations that converge for Euclidean momenta can be found in refs. \cite{Exton:1994de,Alkofer:2008dt,Huber:2007da}.

Detailed solutions for higher n-point functions can be derived and allow the assessment of uniform scaling. Numerically the closed forms are, however, of no use due to their complex form. Fortunately we only need the IREs of given diagrams. To obtain these it suffices in the case of uniform scaling to count the powers of all momenta in the integrals as upon integration the internal momenta turn into external momenta.

In the case of higher n-point functions with $n\geq3$ one could wonder what happens if only a subset of external momenta vanishes. Is it possible that additional IREs occur that have to be taken into account? In principle the answer is yes. In ref. \cite{Alkofer:2008jy} it was shown that taking into account these additional kinematic IREs a self-consistent solution can be derived that is an extension of the uniform case. Numerical evidence for the existence of additional IREs of three-point functions can be found in ref. \cite{Alkofer:2008dt,Huber:2008mq}, where also the in general non-trivial dependence on the kinematics is confirmed. Interestingly additional divergences only appear in the longitudinal part of the three-point functions \cite{Alkofer:2008dt,Fischer:2009tn}, which is irrelevant for Landau gauge as the set of transverse dressing functions is closed \cite{Fischer:2008uz}. Also higher n-point functions do not show kinematic IR divergences in their transverse parts \cite{Fischer:2009tn}. Hence kinematic divergences do not influence the IR fixed point of Landau gauge qualitatively, but the non-trivial kinematic dependence on external momenta can have a quantitative impact on quantities like $\ka$ \cite{Alkofer:2008dt}.

\subsection{Dyson-Schwinger equations in the infrared}
\label{ssec:DSEsIR}

Knowing how to obtain the IRE of a given diagram we can continue to the analysis of the DSEs themselves. The key observation is rather simple: Every diagram on the right-hand side of a DSE can only be as divergent as the one on the left-hand side and it cannot be more divergent as this would render the equation inconsistent. On the other hand one diagram has to scale exactly as the one on the left-hand side. This allows us to write down simple inequalities for the IREs.

Let me illustrate this with a simple example. The DSE of the gluon propagator in Landau gauge is depicted in \fref{fig:LG-g-DSE}. With the IREs $\delta_A$, $\delta_c$, $\ka_{AAA}$, $\ka_{Acc}$ and $\ka_{AAAA}$ of the gluon and ghost propagators, the three-gluon, the ghost-gluon and the four-gluon vertices, respectively, the integrals can be evaluated for low external momentum $p$ in four dimensions as
\begin{align}
d_A^{-1} (p^2)^{1-\delta_A} =& p^2-(p^2)^{1+\ka_{AAA}+2\delta_A} L^{g-loop}+(p^2)^{1+\ka_{Acc}+2\delta_c} L^{gh-loop} -\nnnl
&- (p^2)^{1+\ka_{AAAA}+3\delta_A} L^{sunset} - (p^2)^{1+2\ka_{AAA}+4\delta_A} L^{squint}.
\end{align}
The tadpole was not taken into account as it does not depend on the external momentum and can thus be absorbed in the renormalization when employing dimensional regularization. $d_A$ denotes the coefficient in the power law of the gluon propagator dressing function and the $L$s constant terms for the gluon-loop, the ghost-loop, the sunset and the squint diagrams. They depend on the IREs and the $L$s contain also the constant coefficients from the power laws.
\fig{t}{fig:LG-g-DSE}{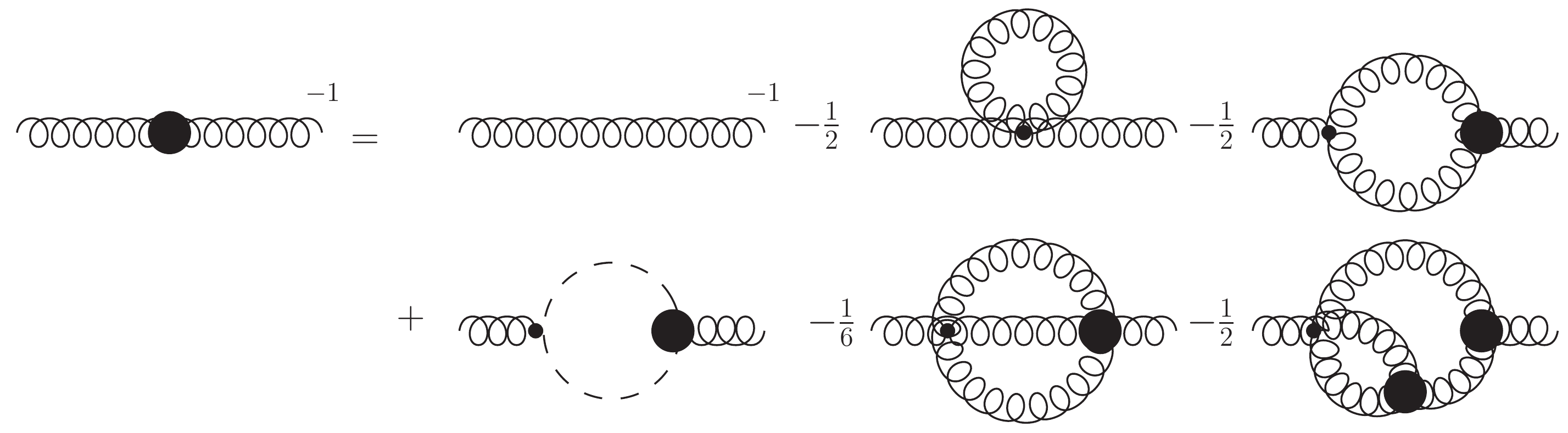}{The gluon two-point function DSE in the Landau gauge: On its left-hand side the dressed gluon two-point function and on its right-hand side the bare gluon two-point function and the tadpole, the gluon-loop, the ghost-loop, the sunset and the squint diagrams. Small blobs represent bare vertices and large blobs dressed 1PI vertices. All internal propagators are dressed.}{width=\linewidth}

Every diagram on the right-hand side could be leading. Thus we can extract the following inequalities:
\begin{align}
 -\de_A\leq&\ka_{AAA}+2\delta_A \label{fig:g-DSE-ineq-gLoop}\\
 -\de_A\leq&\ka_{Acc}+2\delta_c\\
 -\de_A\leq&\ka_{AAAA}+3\delta_A\\
 -\de_A\leq&2\ka_{AAA}+4\delta_A.
\end{align}
It is worth pointing out that at this point we can get a boundary for the gluon propagator IRE by combining \eref{fig:g-DSE-ineq-gLoop} with the inequality $\ka_{AAA}\leq0$, which just reflects that the three-gluon vertex has a contribution from the bare vertex in its DSE. We rewrite \eref{fig:g-DSE-ineq-gLoop} and plug in the boundary on $\ka_{AAA}$:
\begin{align}
0 \leq \ka_{AAA}+3\delta_A \leq 3\delta_A \quad \Rightarrow \quad \de_A\geq 0.
\end{align}
Thus the gluon propagator cannot be IR enhanced and confinement scenarios based on a $1/p^4$ behavior of the gluon propagator can be ruled out in the Landau gauge. A stricter argument for the non-negativity of the gluon propagator IRE can be found in ref. \cite{Alkofer:2008jy}, where a similar argument based on the three-gluon vertex DSE has been derived and possible cancelations have been investigated.

One could now proceed by writing down all inequalities obtained from DSEs up to a certain order. For DSEs with many terms this will get quite tedious and it is not sure how much useful information can be derived from this program. In Section \ref{subsec:ineqs} I will describe a better way of proceeding by determining the most restrictive inequalities. It will be found that all inequalities required for the analysis can be written down in closed form.

\subsection{Functional renormalization groups equations in the infrared}
\label{ssec:FRGEs-IR}

As the IR analysis relies also on FRGEs, I explain here shortly how their IR behavior may be assessed. Although at the end it turns out that the IREs can be counted similarly as in DSEs, some special care is required as the regulator function $R^k$ appears.

The choice of the regulator $R^k$ is free, as long as the constraints listed in Section \ref{ssec:effAvAction} are obeyed. For an IR analysis there are some especially attractive choices which have been discussed in ref. \cite{Fischer:2006vf}. One possibility is of the form
\begin{align}
 R^k_{ij}(p^2)=\Gamma^0_{ij}(p^2)\de_\epsilon(p^2-k^2),
\end{align}
where $\de_\epsilon$ is a smeared $\delta$-functional around $k^2$. Such a regulator projects out modes with $p^2\approx k^2$ and the momentum dependence of physical, i.e., in the limit $k\rightarrow 0$, n-point functions in the IR is not altered.

A second possibility is
\begin{align}
 R^k_{ij}(p^2)=\Gamma^0_{ij}(p^2)r(p^2/k^2).
\end{align}
The physical two-point functions then become
\begin{align}
G^{-1}_{ij}(p^2)=\Gamma^0_{ij}(p^2)(1+r(p^2/k^2))
\end{align}
and again the asymptotic behavior for small $p$ is not altered qualitatively.

\begin{figure}[t]
 \begin{center}
\includegraphics[width=0.68\textwidth]{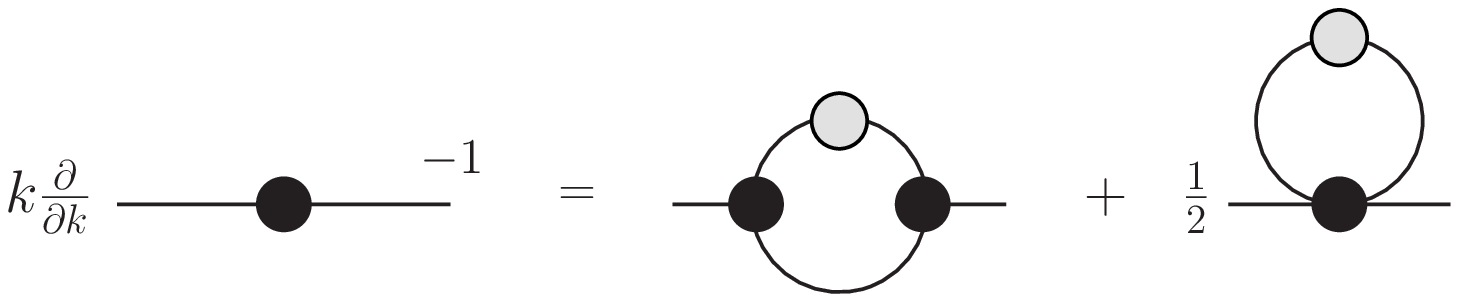}
 \caption{\label{fig:RG-2p} A typical two-point FRGE. The grey blob denotes the regulator insertion $\partial_k R^k$, the black blob dressed n-point functions. Internal propagators are all dressed.}
\end{center}
\end{figure}

The IRE of a diagram in an FRGE is determined in the usual way by counting the IREs of all quantities. Let us consider as an example a diagram from a two-point FRGE as depicted in \fref{fig:RG-2p}:
\begin{align}
\partial_k \Gamma_{ij}(p^2)= G_{mt} \Gamma^k_{tiu} G_{ur} \Gamma^k_{rjs} G_{sn} \partial_k R^k_{mn}+ \frac{1}{2}  G_{mr} \Gamma^k_{irjs} G_{sn} \partial_k R^k_{mn}.
\end{align}
As the regulator takes the momentum dependence of a two-point function, its IRE cancels in the power counting with that of a propagator. Hence we can just ignore regulators in the power counting and count one propagator instead of two propagators and a regulator.

An alternative way of determining the IR behavior of an n-point function is given in ref. \cite{Fischer:2009tn} using integrated flow equations. The flow equations take then a similar form as DSEs but with only dressed vertices in the diagrams. As in these equations the integral over the scale $k$ is performed, also the classical term of an n-point function, defined by the limit $k\rightarrow \infty$, appears similar to DSEs. For details I refer to ref. \cite{Fischer:2009tn}.

\subsection{The role of cancelations}

Shifting the analysis from the integrals to the level of the IREs may be a delicate step. The reason are possible cancelations in the IR leading terms. A detailed analysis is necessary to assess if the IR leading terms cancel and the next leading terms take over. If cancelations occur they are likely linked to some symmetry and do not appear at random. An example is the longitudinal part of the gluon propagator in the linear covariant gauge. A naive power counting analysis yields for its IRE the same result as for the transverse part, but it is known that it does not acquire a dressing at any momentum. Details can be found in Section \ref{ssec:linCovGauge}.

The occurrence of cancelations can often not be completely disproved, but one might exclude some types of them. They can happen between different diagrams or between different contributions of one and the same diagram. These two types of cancelations can be ruled out for the two-point DSEs in Landau gauge, since there is only one IR leading diagram in each equation and the ghost-gluon vertex, the only appearing vertex in these equations, has only one relevant dressing function. For vertices there are always several IR leading diagrams and cancelations could happen between these. Also other vertices except the ghost-gluon vertex with several relevant dressing functions appear. Another possibility is a cancelation between parts stemming from different momentum regions of an integral. This could happen, for instance, via a change of the sign of a dressing function. However, this again can be excluded in the Landau gauge: The dressing functions of both propagators are necessarily positive and also the ghost-gluon vertex dressing function does not change sign as can be inferred from lattice data \cite{Cucchieri:2006tf,Ilgenfritz:2006he,Maas:2007uv,Cucchieri:2008qm} and other calculations \cite{Schleifenbaum:2004id,Alkofer:2008dt}.

On the other hand there are cases when cancelations are required. The most prominent example is probably the ghost propagator of Landau gauge. In its DSE the bare two-point function appears that leads to the inequality $\de_c\geq0$ and thus an IR suppressed propagator. As long as the bare ghost two-point function is in the DSE, one cannot get an IR enhanced propagator. However, it is possible to cancel the bare term in the renormalization by adopting the renormalization condition
\begin{align}
 \left[D_c(p^2)p^2\right]^{-1} \xrightarrow{p^2\rightarrow0} 0
\end{align}
for the ghost propagator $D_c(p^2)$. This condition is not chosen without reason but due to the Kugo-Ojima and Gribov-Zwanziger confinement scenarios as discussed in Section \ref{sec:aspectsIR}.
Nevertheless it has to be mentioned that also a different renormalization prescription may be adopted which leads then to the decoupling solution, see Section \ref{sec:IR-Landau}.

The cancelation of the bare two-point function in the DSE of an IR suppressed two-point function is indeed crucial, as otherwise the bare term is leading. We will neglect in the following the bare two-point functions in the DSEs and determine a posteriori if a cancelation is required. In both systems investigated we will find that such cancelations are possible: In case of the Gribov-Zwanziger action the same reasoning as in the standard Landau gauge can be applied and in the MAG the renormalization of the diagonal gluon propagator apparently plays the same role as that of the ghost in the Landau gauge connecting the obtained scaling solution with the massive solution found on the lattice \cite{Mendes:2006kc} and in the refined Gribov-Zwanziger scenario \cite{Capri:2010an,Capri:2008ak}.

Having clarified the role of possible cancelations it remains to be stressed that the analysis is exact from the point on where we go to the level of the IREs. This is quite remarkable as we will be able to make statements about the complete, infinitely large tower of integral equations without any truncations or approximations.

\section{Basic relations derived from functional equations}
\label{sec:basicRelations}

There is quite much information about the relations between the IREs of Green functions that can be extracted from functional equations. This section is devoted to the study of these relations. First I will show how to derive two classes of constraints for the IREs. Then a formula for the IRE of an arbitrary diagram will be derived. Finally I point out the connection between the inequalities provided by FRGEs and the skeleton expansion, which was employed in earlier investigations of DSEs.

In order to keep this part as general as possible I will not refer to any special Lagrangian except in some examples for illustration. Therefore, generic, not necessarily different fields denoted by $A$, $B$, $C$, \ldots\, are used. The only restriction on them is that they are of mass dimension $(d-2)/2$. Otherwise the canonical dimensions of correlation functions will change. Note that in general no real simplifications occur if the equations are written down for a definite Lagrangian. Even worse, in many cases writing out all the equations would lead to extremely large expressions the treatment of which is not feasible. The MAG, which is the topic of Section \ref{chp:MAG}, is a perfect example for such a case. In this and the following sections all expressions are only valid for four dimensions except when denoted otherwise. The corresponding expressions valid in $d$ dimensions are given in App. \ref{chp:details-scaling}, to which all detailed calculations of this section are deferred.

\subsection{Inequalities from FRGEs and DSEs}
\label{subsec:ineqs}

At first sight it seems that DSEs and FRGEs allow to derive a plethora of inequalities as described in Section \ref{ssec:DSEsIR}. However, on closer inspection one can see that a part of them is contained within others. Therefore we should look for the minimal set of inequalities required. It turns out that there are two groups which play a distinguished role. The first connects the IRE of a vertex and the IREs corresponding to its legs, and the second relates the IREs of propagators among themselves. The first group consists of infinitely many inequalities and the second of only a few. In fact we will see that there are at most only as many members of this group as there are primitively divergent vertices, i.e., vertices appearing in the Lagrangian. This is the decisive aspect that makes the IR analysis a manageable task.

Before we start a few explanations on the notation are required.
Throughout this thesis I use the letter $\ka$ for IREs of n-point functions and the letter $\de$ for IREs of propagators. The reason for this is the discrimination of two-point functions and propagators as it will be required in Section \ref{chp:GZ}. The subscripts of the Greek letters denote the fields of the external legs. In the case of propagators and two-point functions the abbreviations $\de_i:=\de_{ii}$ and $\ka_i:=\ka_{ii}$ are used, respectively. Furthermore, vertices are abbreviated by monomials of their fields, e.g., $ABCC$ corresponds to a four-point function with one $A$, one $B$ and two $C$ legs. Such a monomial can also denote a DSE. Then the order of fields refers to the order in which the derivatives are applied as this can lead to different realizations of the equation. The field $\phi_i$ corresponds to that field indicated by its index (see also Section \ref{ssec:generatingFunctionals}).

We start with three-point functions. In their FRGE there appears a triangle diagram as given in \fref{fig:RG-3-point}. The corresponding inequality is
\begin{align}\label{eq:RG-3-point}
\ka_{ABC}\leq 3\ka_{ABC}+\delta_A+\delta_B+\delta_C \qquad \Rightarrow \qquad \ka_{ABC}+\mhalfo(\delta_A+\delta_B+\delta_C)\geq0.
\end{align}
This equation already gives a hint at the general form of these inequalities: Given the IRE of a vertex, we can add half the sum of the propagator IREs corresponding to the legs and we get a non-negative quantity:
\begin{align}\label{eq:verts-props-inequal1}
\setlength{\fboxsep}{3mm}
\fbox{$ \displaystyle \ka_{{i_1}\ldots {i_r}}+\frac{1}{2}\sum_{i}k_{{i}}^{{i_1}\ldots {i_r}}\delta_{{i}} \geq 0.$}
\end{align}
The symbol $k_{{i}}^{{i}\ldots {j}}$ denotes the number of times the field $\phi_i$ appears in the vertex $\phi_{i}\cdots \phi_{j}$.
It remains to be shown that this inequality is true for all n-point functions.
\fig{t}{fig:RG-3-point}{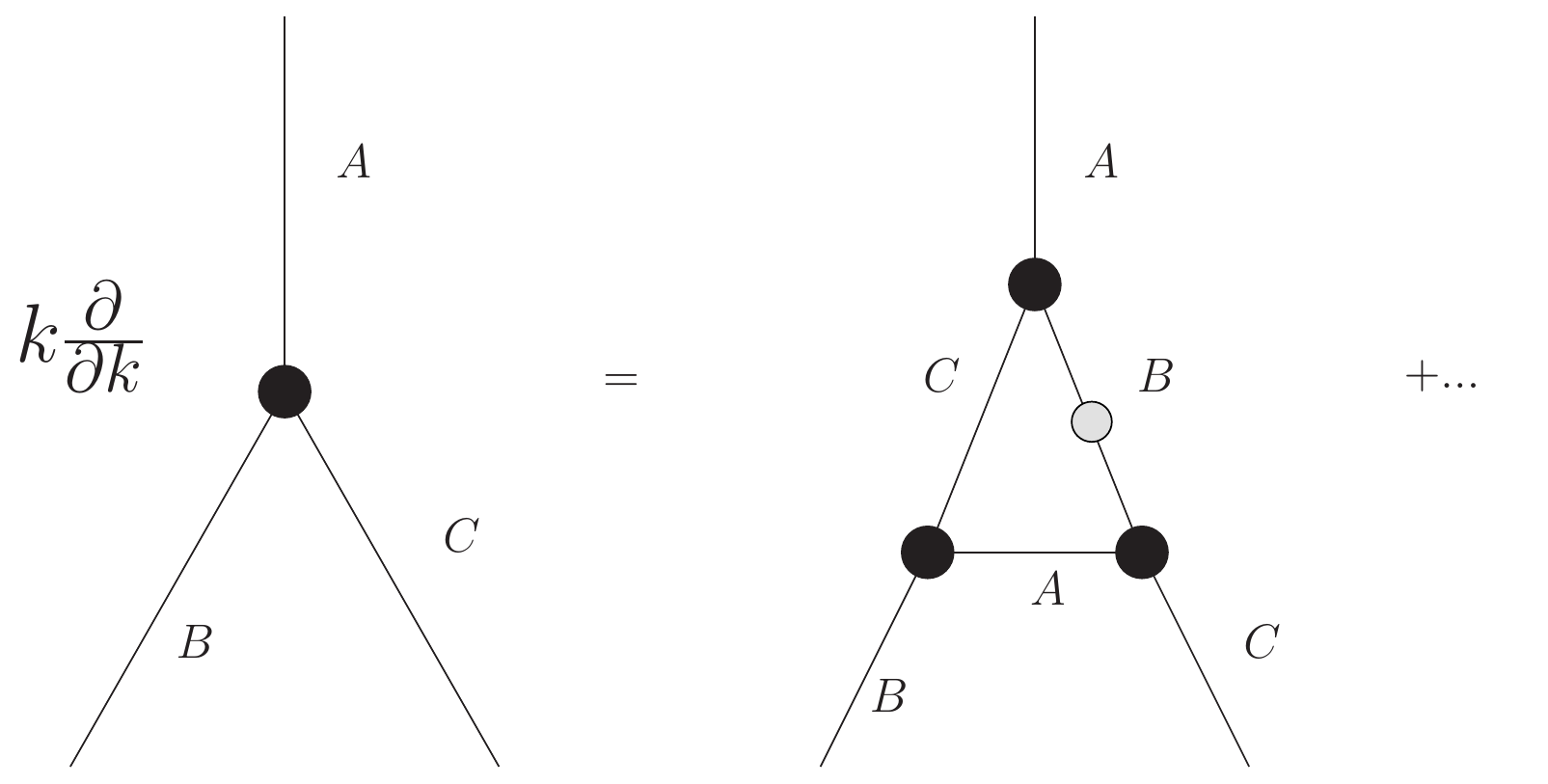}{One specific diagram in the FRGE of a generic three-point function. Internal lines represent dressed propagators, black blobs dressed vertices. The grey blob is a regulator insertion.}{width=0.48\textwidth}

\begin{figure}[t]
\begin{center}
 a) \includegraphics[width=0.43\textwidth]{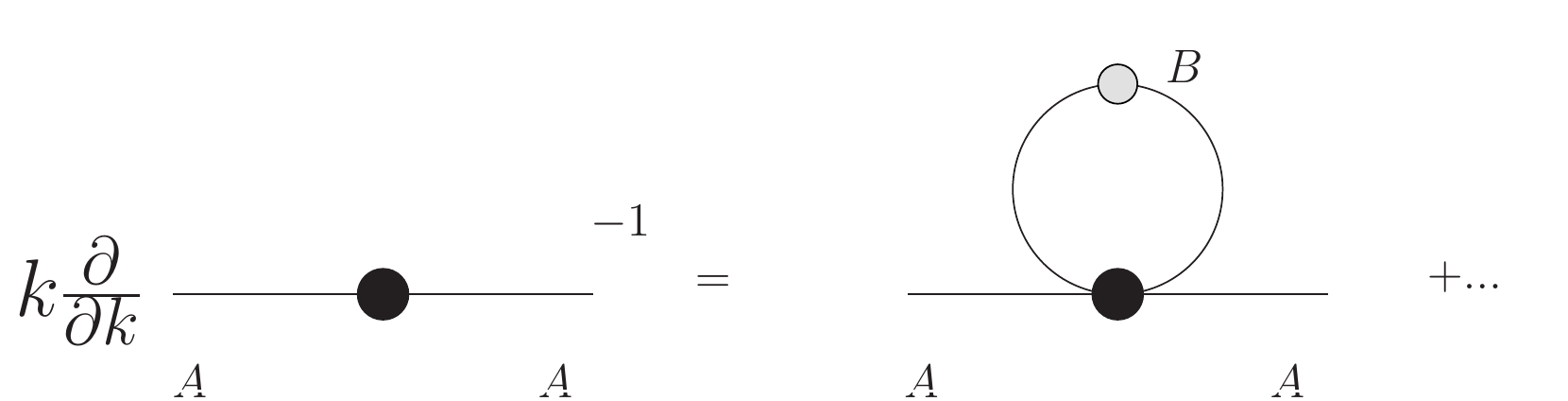}\hskip1cm
 b) \includegraphics[width=0.43\textwidth]{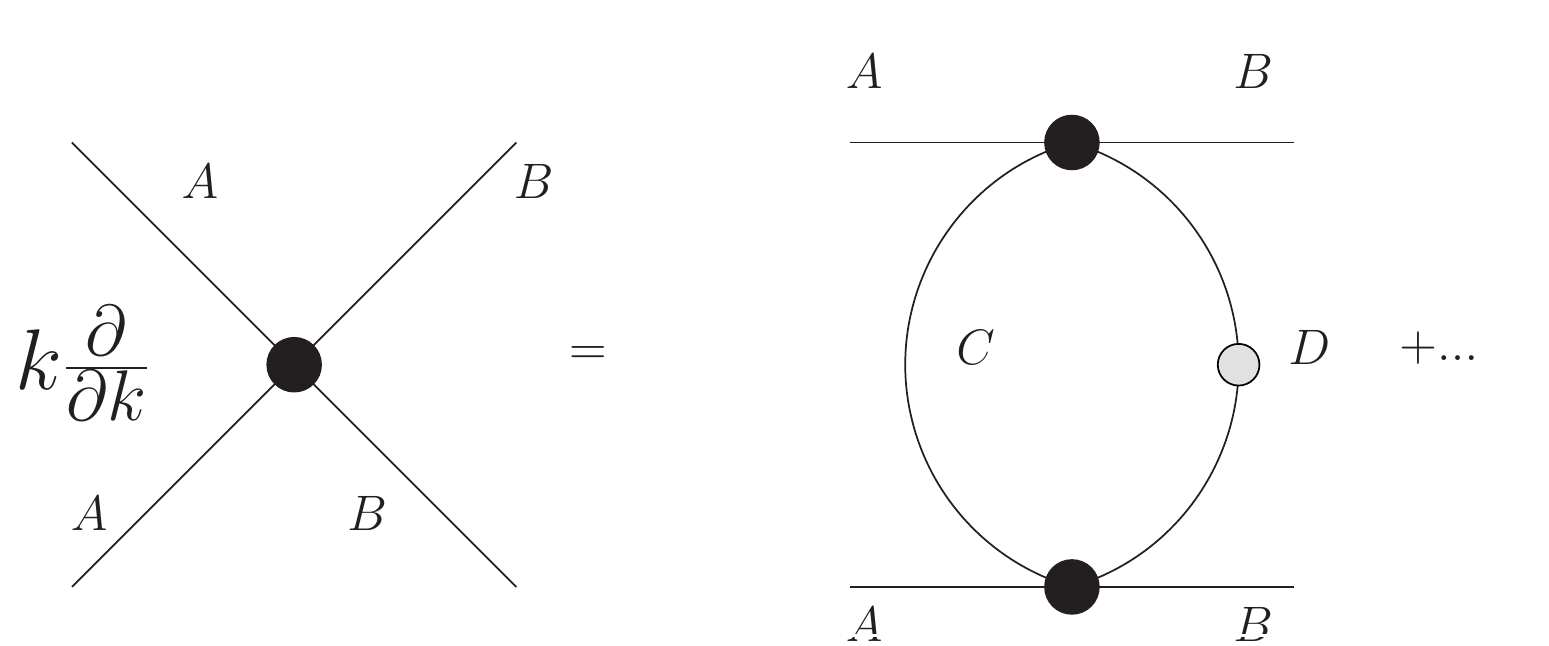}
\caption{\label{fig:RG-4-point} Parts of the FRGEs of generic four-point functions.}
\end{center}
\end{figure}

For the general four-point inequality corresponding to \eref{eq:RG-3-point} we have to make a detour via a special case. The FRGEs from which we infer these inequalities are given in \fref{fig:RG-4-point}. They read
\begin{align}
 -\delta_A-\delta_B  &\leq \ka_{AABB},\\
 \ka_{AABB}&\leq2\ka_{ABCD}+\delta_{C}+\delta_{D}.
\end{align}
Combining them yields
\begin{align}
 \ka_{ABCD}+\frac{1}{2} \left(\delta_{A}+\delta_{B}+\delta_{C}+\delta_{D}\right) \geq0,
\end{align}
what is the expected inequality for a four-point function. It should be noted that the inequalities of this group can only be derived from FRGEs, since the appearance of the bare vertex in DSEs leads to a different, less restrictive numerical factor in front of the propagator IREs in \eref{eq:verts-props-inequal1}.

\begin{figure}[t]
\begin{center}
 a) \includegraphics[width=0.43\textwidth]{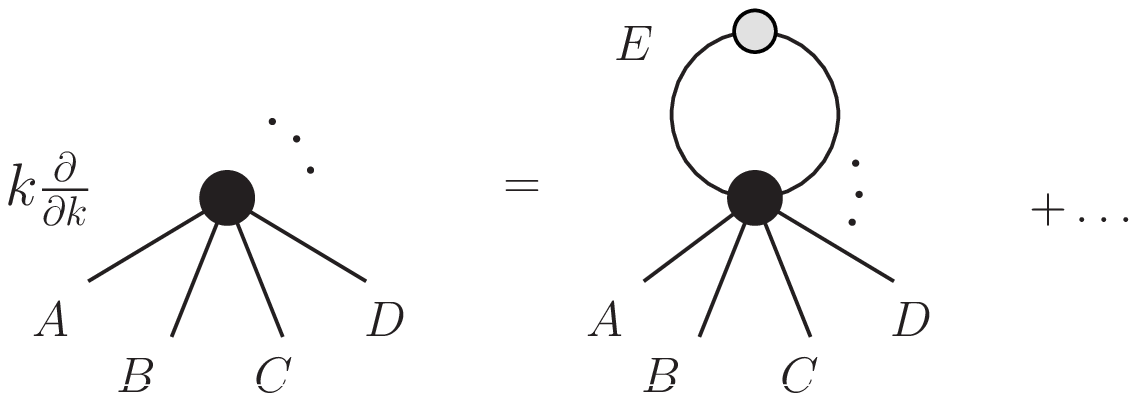}\hskip1cm
 b) \includegraphics[width=0.43\textwidth]{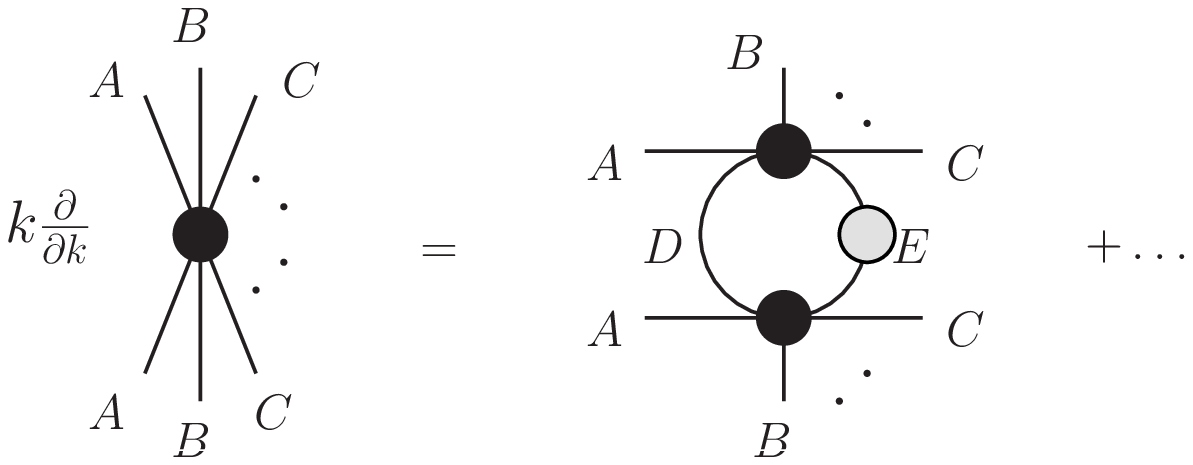}
\caption{\label{fig:RG-3+6+} Parts of the FRGEs of generic n-point functions.}
\end{center}
\end{figure}

In a similar way one can prove by induction the validity of \eref{eq:verts-props-inequal1} for higher n-point functions. I give here only the general structure of the proof and refer for details to Appendix \ref{chp:details-scaling}. For the proof one needs two relations which can be inferred from \fref{fig:RG-3+6+}. The figure on the right-hand side relates two n-point functions with an m-point function where every field appears twice and $m=2n-4$. The connection of this m-point function to an (m-2)-point function is given in the left figure. Plugging the one inequality into the other establishes an inequality between an n-point and a (2n-6)-point function. For the latter one can repeatedly use relations obtained from the type of FRGE depicted on the left-hand side of \fref{fig:RG-3+6+} until one gets to the equation as given in \fref{fig:RG-4-point} on the left. Thus one can establish \eref{eq:verts-props-inequal1} for all n-point functions.

Eq. (\ref{eq:verts-props-inequal1}) allows an interesting observation concerning the so-called skeleton expansion. The corresponding discussion is deferred to Section \ref{ssec:skeletonExpansion}.

Now we turn to the second group of inequalities. They can be derived directly from \eref{eq:verts-props-inequal1} by noting that for all interactions appearing in the Lagrangian, the primitively divergent vertices, the corresponding bare vertices appear in their DSEs. Thus their IRE is non-positive,
\begin{align}\label{eq:primDivVerts}
 \ka^{prim.\,div.}_{ABC\cdots}\leq0,
\end{align}
and we infer from \eref{eq:verts-props-inequal1} that
\begin{align}\label{eq:verts-props-inequal2}
\setlength{\fboxsep}{3mm}
\fbox{$ \displaystyle \frac{1}{2}\sum_{i}k_{{i}}^{{i_1}\ldots {i_r}}\delta_{{i}} \geq 0 \qquad \forall \text{ primitively divergent vertices}.$}
\end{align}
This group of inequalities can be derived purely from DSEs, but not from the FRGEs without help from the DSEs.

In the preceding analysis one should always keep in mind that for a realistic Lagrangian it is not sufficient to replace the generic fields $A$, $B$ etc. by the fields of the Lagrangian, but one has to check if the corresponding diagrams really exist. The reason for the disappearance of a diagram can be as simple as a trivial color factor or as complex as a non-manifest symmetry. The former case will be encountered in the analysis of the MAG but only for diagrams not relevant in the IR. Thus they do not invalidate the result.

\subsection{Infrared exponent of an arbitrary diagram}

We established in Section \ref{sec:IRPC} how the IRE of any given diagram can be evaluated by power counting. This is now formulated in very general terms by considering an $l$-loop diagram $v$ with $n_i$ internal propagators of the field $\phi_i$, $m_i$ external legs of the field $\phi_i$, $n^b_{i\ldots k}$ bare vertices of the type $\phi_i \cdots \phi_k$ and $n^d_{i\ldots k}$ dressed vertices of the type $\phi_i \cdots \phi_k$:
\begin{align}\label{eq:arb-diag-1}
\ka_{v}= & l\frac{d}{2}+\sum_{i}n_i(\delta_{{i}}-1)+\sum_{vertices,r\geq3}n^{d}_{{i_1}\ldots {i_r}}(\ka_{{i_1}\ldots {i_r}}+c_{{i_1}\ldots {i_r}})+\nnnl
 & +\sum_{vertices,r\geq3}n^{b}_{{i_1}\ldots {i_r}}c_{{i_1}\ldots {i_r}}-c_{v}.
\end{align}
Here $c_{{i}\ldots {j}}/\ka_{{i}\ldots {j}}$ denote the canonical dimension/IRE of the vertex $\phi_{i}\cdots \phi_{j}$.
Eq. (\ref{eq:arb-diag-1}) is valid for $d$ dimensions, but in the following I only give the results for $d=4$, since this makes the expressions more transparent. The arguments, however, are the same and the details for $d$ dimensions can be found in Appendix \ref{sec:details-scaling-App}.
The double subscripts of the $n^d$ and $n^b$ indicate all possible combinations of $r$ fields.
E.g., for the Landau gauge the term corresponding to $r=3$ is
\begin{align}
n^d_{AAA}\left(\ka_{AAA}+\frac1{2}\right)+n^d_{A\bar{c}c}\left(\ka_{A\bar{c}c}+\frac1{2}\right)+\frac1{2}n^b_{AAA}+\frac1{2}n^b_{A\bar{c}c}.
\end{align}

Using topological relations that connect the numbers of propagators, vertices, loops and legs \eref{eq:arb-diag-1} can be rewritten:
\begin{align}\label{eq:IREArbitraryDiagram}
\ka_{v} = & -\frac{1}{2}\sum_{i}m_{i}\delta_{{i}}+\sum_{vertices,r\geq3}n^{d}_{{i_1}\ldots {i_r}}\left(\ka_{{i_1}\ldots {i_r}}+\frac{1}{2}\sum_{i}k_{{i}}^{{i_1}\ldots {i_r}}\delta_{{i}}\right)+\nnnl
 & +\sum_{vertices,r\geq3}n^{b}_{{i_1}\ldots {i_r}}\left(\frac{1}{2}\sum_{i}k_{{i}}^{{i_1}\ldots {i_r}}\delta_{{i}}\right).
\end{align}
Details on the derivation of this formula are given in Appendix \ref{sec:details-scaling-App}. The reason for rewriting it in this way is that now it depends only on the external legs and the numbers of vertices and we got rid of the dependence on internal propagators.

Eq. (\ref{eq:IREArbitraryDiagram}) is not only useful in general, but it allows also to determine a lower bound for the IRE of a diagram. To this end we resort to the results of Section \ref{subsec:ineqs} which are given by eqs. (\ref{eq:verts-props-inequal1}) and (\ref{eq:verts-props-inequal2}). Comparing them to \eref{eq:IREArbitraryDiagram} we see that they correspond exactly to the terms in parentheses. Even the fact that \eref{eq:verts-props-inequal2} is only valid for primitively divergent diagrams is mirrored as the corresponding coefficient in \eref{eq:IREArbitraryDiagram}, $n^{b}_{{i_1}\ldots {i_r}}$, can only be non-zero for primitively divergent vertices. Hence the first term in \eref{eq:IREArbitraryDiagram} is a lower bound for the IRE of the diagram $v$ due to the non-negativity of the other two expressions. Even more, since the first term only depends on the external legs it is the same for all diagrams appearing in a DSE. This establishes a lower bound for the IRE of an arbitrary vertex function. It is called the maximally IR divergent solution:
\begin{align}\label{eq:Maximal-IR-Divergent}
\setlength{\fboxsep}{3mm}
\fbox{$ \displaystyle \ka_{v,max}= -\frac{1}{2}\sum_{i}m_{i}\delta_{{i}}.$}
\end{align}

\subsection{Skeleton expansion of higher vertices}
\label{ssec:skeletonExpansion}

Before continuing with the IR analysis I want to make a short detour to a tool that was used repeatedly in IR analyses of Green functions: the skeleton expansion. It can be seen as a loop expansion of a vertex using only dressed propagators and vertices; for more details on the original idea see, e.g., ref. \cite{Bjorken:1967qf}. In functional equations one usually restricts the set of vertices in the expansion to those appearing in the Lagrangian. One can then derive the IREs of all vertices in Landau gauge \cite{Alkofer:2004it,Huber:2007kc}. The assumption that the skeleton expansion does not explicitly diverge, meaning that higher orders do not feature more IR divergent diagrams, provides sufficient additional information to prove the uniqueness of the IR scaling solution of Landau gauge \cite{Alkofer:2008jy}. The key observation thereby is that higher orders of the skeleton expansion are constructed by adding further loops using certain combinations of correlation functions. These insertions into diagrams must not make the diagrams more IR divergent and hence they must have non-negative IREs.

Originally the validity of the skeleton expansion was an assumption, but the functional renormalization group clarifies how and why the skeleton expansion indeed works for power counting. The explanation is given by \eref{eq:verts-props-inequal1}, which was derived from FRGEs. The inequalities provided by it correspond exactly to the inequalities derived from the insertions used in the skeleton expansion, see ref. \cite{Alkofer:2008jy} for details. Thus the use of the skeleton expansion is well justified. However, \eref{eq:verts-props-inequal1} is even more powerful, because it covers all possible vertices in contrast to the skeleton expansion, which could - at least in the form as it was used - provide inequalities only for the vertices appearing in the Lagrangian.

\section{Solution for the system of IREs}
\label{sec:solution-IR}

The information collected up to now will be used to determine the scaling solution(s) of a given Lagrangian. For this purpose we analyze the DSEs of the two-point functions. However, instead of considering single cases we again keep things general.
The analysis proceeds by \textit{assuming} that one diagram in a two-point function DSE is leading without specifying which one. This will lead to a new inequality that has only to be fulfilled in this one case. Taking into account the inequalities obtained earlier, we can discard some solutions and come up with a set of possible IR scaling solutions for the given system. This set can be empty, consist of a unique solution or of several different ones.

\subsection{Analysis of two-point equations}
\label{ssec:analysis-2Point}

We consider a generic two-point function DSE. Fig. \ref{fig:DSE-2-point} illustrates possible diagrams if only three- and four-point interactions appear in the Lagrangian. In the following we do not refer to specific diagrams and the analysis covers \textit{all} possible diagrams. For the leading diagram the following equation can be derived from \eref{eq:IREArbitraryDiagram}
\begin{align}\label{eq:DSE-2-point}
 \ka_{i}&=-\frac1{2}\sum_i m_i \de_i+\sum_{\substack{vertices}} n_{{i_1} \ldots {i_r}} \left(\frac1{2}\sum_i k_i^{i_1 \ldots i_r} \de_i \right)
+\sum_{\substack{dressed\\vertices}}  n^d_{{i_1} \ldots {i_r}} \ka_{{i_1} \ldots {i_r}},
\end{align}
where $\ka_i$ is the IRE of the two-point function under investigation. The only values depending on the specific leading diagram are $n_{{i_1} \ldots {i_r}}$ and $n^d_{{i_1} \ldots {i_r}}$, but we can leave them arbitrary for now. Using again some topological relations and the lower bound for the IRE of a vertex, \eref{eq:Maximal-IR-Divergent}, we arrive at the inequality
\begin{align}\label{eq:DSE-2-point-leading-ineq}
- n^b_{{i_1} \ldots {i_r}} \left(\frac1{2}\sum_i k_i^{i_1 \ldots i_r} \de_i \right)\geq0.
\end{align}
The details of the calculation can be found in Appendix \ref{sec:Prop-Eqs}. The only information from the leading diagram that enters here is the type of its bare vertex as given in $n^b_{{i_1} \ldots {i_r}}$, i.e., we get as many different possible inequalities as there are bare vertices.
\fig{t}{fig:DSE-2-point}{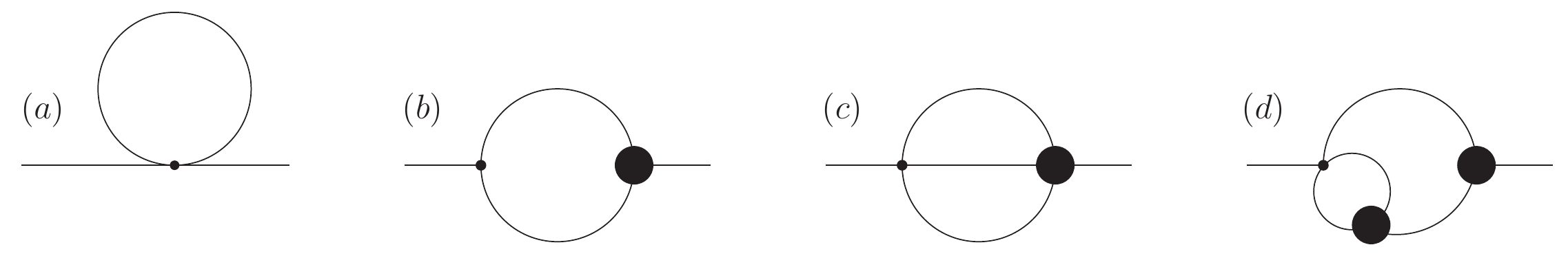}{Topologies of diagrams appearing in the DSE of a two-point function for three- and four-point interactions only. Conventional names for the diagrams are (a) tadpole, (c) sunset, (d) squint.}{width=\textwidth}

Interestingly \eref{eq:DSE-2-point-leading-ineq} looks exactly the same as the inequalities given in \eref{eq:verts-props-inequal2}, but with the opposite sign. Hence, both inequalities can only be fulfilled if they are saturated, i.e.,
\begin{align}\label{eq:DSE-2-point-leading-eq}
\setlength{\fboxsep}{3mm}
\fbox{$ \displaystyle \frac1{2} \hat{n}^b_{{i_1} \ldots {i_r}} \sum_{j}\de_{j} k_{j}^{{i_1} \ldots {i_r}}=0.$}
\end{align}
The hat was added in order to indicate that this equality was only derived for one specific interaction, namely the one appearing with a bare vertex in the IR leading diagram. From eqs. (\ref{eq:verts-props-inequal1}), (\ref{eq:primDivVerts}) and (\ref{eq:DSE-2-point-leading-eq}) it follows directly that the IRE of the corresponding vertex is zero:
\begin{align}\label{eq:leading-eq}
\setlength{\fboxsep}{3mm}
\fbox{$ \displaystyle \hat{\ka}_{{i_1} \ldots {i_r}}=0.$}
\end{align}
Thus for a scaling solution at least one vertex does not get IR enhanced. This vertex is called the IR leading vertex for reasons that will become clear soon.

In order to obtain valid scaling relations one has to determine all possible realizations of \eref{eq:leading-eq} and determine if they are compatible with the residual system of inequalities. In some cases the scaling relation does not relate all propagator IREs to each other and a second scaling relation has to be obtained from the remaining DSEs. Very often this leads to the trivial solution where all IREs are zero. Known examples where one sector of the theory possesses a scaling relation independent of a second sector are massless QCD \cite{Schwenzer:2008vt} and a fundamentally charged scalar coupled to Yang-Mills theory \cite{Fister:2010yw,Fister:2010ah}: The Yang-Mills sector is not influenced by the quark/scalar so that the usual scaling relation is valid. As the quark/scalar does not couple directly to the ghost, which is the IR dominant field, the quark/scalar propagator is not IR suppressed and its IRE is zero.

It has to be stressed that the notion of a not IR enhanced vertex is not tantamount to a bare vertex in the IR. In general a vertex has several tensors with corresponding dressing functions and the IR analysis just takes into account the most IR divergent one(s). There can always be other dressing functions with a smaller IRE. For the numerical determination of an IRE it suffices to use only the IR leading tensors, but for a complete numerical solution of the equation also the subdominant tensors have to be taken into account. If a vertex has a bare counterpart and is not IR enhanced, then the bare vertex will be part of the IR leading terms unless it gets canceled.

For Landau gauge the issue of choosing the IR leading tensor is easy, as its IR leading vertex, the ghost-gluon vertex, only possesses two tensors of which only one is transverse. The longitudinal tensor does not contribute in Landau gauge and hence is irrelevant \cite{Fischer:2008uz}. The impact of different dressing functions on the value of the parameter $\ka$ was investigated in detail in ref. \cite{Lerche:2002ep}. Thereby several constraints, like the unchanged uniform scaling, were taken into account with the result that $\ka$ only changes slightly. Additionally the IR dressing function was calculated in ref. \cite{Alkofer:2008dt} as a function of the external momenta, where the solutions for the propagators and a bare ghost-gluon vertex were used as input. Even in this simple case the dressing function depends on the kinematics. Thus a more detailed calculation of $\ka$ should take into account the momentum dependence of the ghost-gluon vertex. However, the qualitative behavior of the propagators would not be affected by this.

Having determined the set of possible scaling solutions we have to identify the leading diagrams in the two-point DSEs. To each solution corresponds via \eref{eq:DSE-2-point-leading-eq} a bare vertex. Since all dressed vertices and propagators dropped out in its derivation, all diagrams with the same bare vertex lead to the same scaling relation. However, a word of caution is in order here: This does not necessarily mean that all diagrams with the same bare vertex scale equally. In the case of the MAG such an instance is encountered and will be discussed in detail in Section \ref{ssec:MAGVertices}.

We can summarize the procedure to obtain the possible scaling solutions of a theory as follows:
\begin{enumerate}
 \item Determine the inequalities that are derived from the interactions appearing in the Lagrangian from \eref{eq:verts-props-inequal2}.
 \item Reduce the number of inequalities, if some of them are contained within others.
 \item Try to saturate one inequality after the other according to \eref{eq:DSE-2-point-leading-eq} and see, if you can find any contradictions with the remaining inequalities.
\end{enumerate}

Let me illustrate the simpleness of the method just described by showing how it works for Landau gauge:
\begin{enumerate}
 \item There are three interactions in the Lagrangian: the ghost-gluon, the three-gluon and the four-gluon vertices. According to \eref{eq:verts-props-inequal2} they lead to the three inequalities
 \begin{align}
  \frac{1}{2}\de_A+\de_c\geq0,\qquad \frac{3}{2}\de_A\geq0,\qquad 2\de_A\geq0,
 \end{align}
 where $\de_A$ and $\de_c$ are the IREs of the gluon and ghost propagators, respectively.
 They are all valid at all times.
 \item We reduce this to the two inequalities
 \begin{align}
  \frac{1}{2}\de_A+\de_c\geq0,\qquad \de_A\geq0.
 \end{align}
 \item
  \begin{enumerate}
   \item Saturating the second inequality yields $\de_A=0$, which in turn gives a non-negative IRE of the ghost propagator: $\de_c\geq0$. The equation $\de_A=0$ is obtained from diagrams with bare three-gluon or four-gluon vertices. As such diagrams do not appear in the ghost two-point function DSE, we still have to determine its IR leading diagram. Based on the general argument given above we use that the only option for a scaling relation involving the ghost propagator IRE is $\de_A+2\de_c=0$. Consequently we find $\de_c=0$, i.e., this case corresponds to the trivial solution.

   Since the ghost two-point function DSE in the Landau gauge as depicted in \fref{fig:LG-gh-DSE} has only two diagrams, one could also investigate it directly: For the bare two-point function we find $\de_c=0$ without any effort. For the loop diagram we get with the analysis described above $\de_A+2\de_c=0$, which also leads to $\de_c=0$ for $\de_A=0$.
   \item The first inequality corresponds to the case when the ghost-gluon vertex is the IR leading vertex. Hence the scaling relation is $\de_A+2\de_c=0$ and the ghost-loop and the mixed loop in the gluon and ghost two-point DSEs, respectively, are IR leading. Conventionally the parameter $\ka:=-\de_c$ is used.
  \end{enumerate}
\end{enumerate}

\fig{t}{fig:LG-gh-DSE}{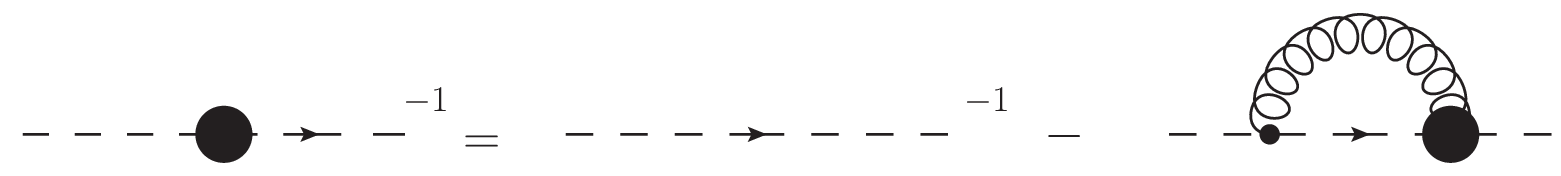}{The ghost two-point function DSE in the Landau gauge.}{width=0.75\linewidth}

The generalization to $d$ dimensions is straightforward. The only difference is that the canonical dimensions do not factor out so that additional terms proportional to $d/2-2$ appear. Appendix \ref{chp:details-scaling} contains all required formulae in $d$ dimensions. For the scaling relation we use \eref{eq:leadingDiagramSimpleEq}:
\begin{align}\label{eq:LGScalingRelationd}
 \de_A+2\de_c+\frac{d}{2}-2=0.
\end{align}

Following a simple list of instructions we could derive the known scaling relation of Landau gauge \cite{vonSmekal:1997is,vonSmekal:1997vx}. What remains to be determined is the IR behavior of the vertices.

\subsection{Infrared exponents of vertices}

From the IR leading diagrams in the two-point DSEs one can construct the IR leading diagrams of higher vertices by adding more legs via insertions of the IR leading vertex. For the higher vertices the corresponding IRE is given by the maximally IR divergent solution, \eref{eq:Maximal-IR-Divergent}, since we only add vertices for which the other two terms, as appearing in \eref{eq:IREArbitraryDiagram}, are zero due to eqs. (\ref{eq:DSE-2-point-leading-eq}) and (\ref{eq:leading-eq}).

As an explicit example we consider again the Landau gauge. From the arguments above the solution for a vertex with $m$ gluon and $2n$ ghost legs follows as
\begin{align}\label{eq:LGIREVertices}
 \ka_{m,2n}=-\frac{m}{2}\de_A-n\,\de_c=(n-m)\ka,
\end{align}
which was found by a different route in ref. \cite{Alkofer:2004it}.

For the generalization to $d$ dimensions one plugs the scaling relation given in \eref{eq:LGScalingRelationd} into the maximally IR divergent solution in $d$ dimensions as provided by \eref{eq:maxIRDivSol-App}:
\begin{align}\label{eq:LGIREVerticesd}
 \ka_{m,2n}=-\frac{m}{2}\de_A-n\,\de_c+\left(\frac{d}{2}-2 \right) \left(1-\frac{m}{2}-n \right)=(n-m)\ka+(1-n)\left( \frac{d}{2}-2 \right).
\end{align}
This is in agreement with an earlier investigation \cite{Huber:2007kc,Alkofer:2007hc}.

The Landau gauge represents the ideal case, where all vertices can be constructed from an IR leading diagram of a two-point DSE. However, it can also happen that this is not possible, e.g., if the IR leading vertex has four legs. How would one construct then a three-point function from a two-point function by inserting four-point functions? The answer is one cannot do so, as can be inferred by simple combinatorial considerations: Out of vertices with an even number of legs one cannot construct one with an odd number of legs.

\section{Examples of infrared scaling relations}
\label{sec:otherGauges}

The method described in the previous section can directly be used for the investigation of different gauges. Two gauges investigated in the past besides the Landau gauge are linear covariant gauges and ghost anti-ghost symmetric gauges \cite{Alkofer:2003jr}. In their analysis a truncation scheme with bare vertices was used \cite{Alkofer:2003jr}. For linear covariant gauges a non-trivial scaling solution was only found, when the longitudinal part of the gluon propagator scales equally as the transverse part and for ghost anti-ghost symmetric gauges only the trivial solution was found. I generalize these results here by including dressed vertices, but the main outcome of ref. \cite{Alkofer:2003jr} is corroborated.

\subsection{Infrared analysis of the linear covariant gauge}
\label{ssec:linCovGauge}

As described in Section \ref{ssec:gaugeFixing} linear covariant gauges can be considered as a generalization of the Landau gauge by relaxing the restriction to the hyperplane $\partial_\mu A_\mu$ into a Gaussian distribution on the gauge orbit. The corresponding gauge fixing part of the Lagrangian is given in \eref{eq:Lagrangian-gaugeFixing}. An immediate consequence of this gauge choice is that the gluon propagator is no longer purely transverse.

A peculiar feature of the linear covariant gauge is that the longitudinal part of the gluon propagator does not acquire a dressing but stays proportional to the gauge fixing parameter $\xi$:
\begin{align}\label{eq:gluonPropLinCov}
 D_{A,\mu\nu}^{ab}(p^2)=\de^{ab}\left(g_{\mu\nu}-\frac{p_\mu p_\nu}{p^2} \right) \frac{c_A(p^2)}{p^2}+\xi \frac{p_\mu p_\nu}{p^4}.
\end{align}
This can be inferred either from a Slavnov-Taylor identity \cite{Alkofer:2000wg,Pascual:1984zb} or from the fact that the longitudinal part of the gluon propagator, $\langle (\partial_\mu A_\mu) (\partial_\nu A_\nu) \rangle$, corresponds to the second moment of the Gaussian distribution
\begin{align}
 e^{-\frac{1}{2\xi}(\partial_\mu A_\mu)^2}
\end{align}
and is hence proportional to $\xi$ \cite{Cucchieri:2008zx}.

To include the longitudinal part of the gluon into the formalism developed so far we interpret it as an additional field. Its propagator has the IRE $\de_{A,long}$. This allows to distinguish also for the vertices between transverse and longitudinal parts. Naively the same inequalities are obtained for the longitudinal gluon field as for the transverse one. This could change if some vertices vanished when they are contracted with a longitudinal gluon propagator. However, as the Landau gauge result showed the ghost-gluon vertex to be the IR leading vertex, we obtain for the longitudinal gluon propagator IRE the same result as for the transverse one, as the longitudinal part of the ghost-gluon vertex does not vanish. The result $\de_A=\de_{A,long}=-2\de_c$ leads together with the triviality of the longitudinal part, $\de_{A,long}=0$, to $\de_A=-2\de_c=0$. This allows two possible conclusions: Either there is no non-trivial scaling relation for linear covariant gauges, or the naive application of the IR analysis is insufficient here. Most probably there are some cancelations related to the triviality of the longitudinal part of the gluon propagator. How these can be made manifest within DSEs is currently unknown.

\subsection{Infrared analysis of the ghost anti-ghost symmetric gauge}

The most general Lagrangian of dimension four constructed from gluons and Faddeev-Popov ghosts which is Lorentz invariant, globally gauge invariant, Hermitian\footnote{For the hermiticity of the Lagrangian the following transformation properties of the ghosts are chosen \cite{Kugo:1979gm}:
\begin{align}
 c^\dagger=c,\qquad \bar{c}^\dagger=-\bar{c}.
\end{align}
}, BRST invariant and anti-BRST invariant without topological terms is \cite{Baulieu:1981sb,ThierryMieg:1985yv,Alkofer:2003jr}
\begin{align}
 \mathcal{L}_{ga}=&\frac{1}{4}F^2_{\mu\nu}+\frac{1}{2\xi}(\partial_\mu A_\mu)^2+\frac{\rho}{2}\left(1-\frac{\rho}{2} \right) \frac{\xi}{2} f^{abc}f^{ade}\bar{c}^b c^c \bar{c}^d c^e +\nnnl
 &-\frac{\rho}{2}(D_\mu \bar{c})\partial_\mu c-\left(1-\frac{\rho}{2} \right) (\partial_\mu \bar{c}) D_\mu c.
\end{align}
The gauge fixing parameter $\rho$ allows to interpolate between the Landau gauge ($\rho=0,2$) and the ghost anti-ghost symmetric case ($\rho=1$).

The IR analysis of this Lagrangian can be done along the same lines as for the Landau gauge. The vital difference between the two gauges is clearly the occurrence of a quartic ghost interaction, which leads to a new inequality due to \eref{eq:verts-props-inequal2}:
\begin{align}
 \de_c\geq 0.
\end{align}
It is now straightforward to conclude that only the trivial solution can be obtained. The inequalities from the vertices in the Lagrangian are
\begin{align}
 \de_A\geq0,\qquad \de_c \geq0,\qquad \de_A+2\de_c\geq0.
\end{align}
If the first inequality is saturated, i.e., $\de_A=0$, the IR leading vertex is purely gluonic and we still need to determine the IR leading vertex of the ghost two-point function DSE. In both possible cases, quartic ghost vertex and ghost-gluon vertex, we obtain $\de_c=0$ and thus the trivial solution. If we start with the inequality $\de_c=0$, the arguments are the same. Finally, for $\de_A+2\de_c=0$ we immediately get $\de_A=\de_c=0$, as both $\de_A$ and $\de_c$ are non-positive. Hence all possibilities lead to the trivial solution.

This confirms the result from ref. \cite{Alkofer:2003jr}, where bare vertices were employed, also in the case of dressed vertices.
Again we can only conclude that the existence of a scaling solution is not possible or that more effort is required to expose some not manifest symmetries which lead to cancelations.

The purpose of this section was to demonstrate how the method for obtaining scaling relations facilitates the analysis compared to earlier investigations. Before we turn to systems which have not been investigated yet in Chapters \ref{chp:MAG} and \ref{chp:GZ}, I will generalize the method to the case of Lagrangians where the fields mix at the two-point level.

\section{Extension to actions with mixed two-point functions}
\label{sec:mixed2Point}

The method developed in the preceding sections works very nicely for many actions. There are, however, also Lagrangians where the fields mix on the level of two-point functions. In this case additional complications arise, because a propagator is no longer the inverse of the corresponding two-point function. In fact the connection between propagators and two-point functions becomes a matrix relation. This requires a more detailed analysis to determine the relation between their IREs. The identity $\de_i=-\ka_i$, valid before, does only hold as a special case and in general the IREs of a propagator and the corresponding two-point are related by an equation involving also the IREs of other propagators/two-point functions. Thus some formulae of Sections \ref{sec:basicRelations} and \ref{sec:solution-IR} have to be generalized.

The first important observation is that the inequalities (\ref{eq:verts-props-inequal1}) and (\ref{eq:verts-props-inequal2}) obtained from FRGEs and DSEs do not change, the reason being simply that in the case of mixing two-point functions the number of diagrams in the equations grows, but all diagrams from before will still be there in general. Furthermore, the new diagrams do not lead to new relevant inequalities.

The formula for the IRE of an arbitrary diagram, however, is altered by the appearance of the new propagators. A detailed derivation is given in Appendix \ref{sec:Prop-Eqs}. For the sake of simplicity I only give the result if there are two mixing fields, denoted by $A$ and $V$. The final expression is
\begin{align}\label{eq:IREArbitraryDiagram-mixed}
\ka_{v} = & -\frac{1}{2}\sum_{i=A,V}m_{i}\delta_{{i}}+\frac1{2}n_{AV}(2\de_{AV}-\de_A-\de_V)+\nnnl
 & +\sum_{vertices,r\geq3}n^{d}_{{i_1}\ldots {i_r}}\left(\ka_{{i_1}\ldots {i_r}}+\frac{1}{2}\sum_{i=A,V}k_{{i}}^{{i_1}\ldots {i_r}}\delta_{{i}}\right)+\sum_{vertices,r\geq3}n^{b}_{{i_1}\ldots {i_r}}\left(\frac{1}{2}\sum_{i=A,V}k_{{i}}^{{i_1}\ldots {i_r}}\delta_{{i}}\right).
\end{align}
The new term contains $n_{AV}$, which is the number of mixed propagators, and the combination $\Delta_{AV}:=2\de_{AV}-\de_A-\de_V$ of the IREs of the propagators. The value of the latter has to be considered for each case. It should be non-negative or the IRE is unbounded from below, as one can find diagrams with an arbitrary number of mixed propagators. They are explicitly constructed by appropriately replacing dressed quantities by their respective DSEs.

Finally, we have to redo the analysis of the leading diagram in two-point function DSEs. The resulting inequality corresponding to \eref{eq:DSE-2-point-leading-ineq} also contains a new term proportional to $\Delta_{AV}$:
\begin{align}
\ka_i&+\frac1{2}\sum_j \de_j m_j-n^b_{{i_1}\ldots {i_r}}\left(\frac{1}{2}\sum_{j}k_{{j}}^{{i_1}\ldots {i_r}}\delta_{{i}}\right) -\nnnl
 &-\frac1{2}(2\de_{AV}-\de_A-\de_V)\left( n_{AV}+\sum_{\substack{dressed\\vertices}} n^d_{{i_1} \ldots {i_r}} \bar{k}^{{i_1} \ldots {i_r}}_{AV} \right) \geq0.
\end{align}
The symbol $\bar{k}^{{i_1} \ldots {i_r}}_{AV}$ denotes the number of times a mixed propagator is contained in the IR leading diagram of the vertex $\phi_{i_1}\cdots \phi_{i_r}$.

In the remaining analysis $\Delta_{AV}$ plays a special role. If it can be shown to be zero, all new terms vanish and one can proceed as before. If it is not zero, more care is required. Since an analysis only makes sense for positive values, it is clear from \eref{eq:IREArbitraryDiagram-mixed} that the leading diagrams are those with the lowest possible number of mixed propagators, because for every one of them the IRE is raised. Hence one determines first the IREs of all diagonal two-point functions and subsequently those of the mixed ones. An application of this procedure can be found in Section \ref{sssec:CaseII}.
\chapter{The infrared regime of the maximally Abelian gauge}
\label{chp:MAG}

For quantizing Yang-Mills theory naturally a variety of gauges is available, each with its own advantages and disadvantages. Depending on the aspect one is interested in one chooses one gauge or another: Not only that the complexity of the calculations is influenced by such a choice, but also some properties are more accessible in certain gauges. In this chapter I discuss the maximally Abelian gauge, which is especially amenable for investigating the role of Abelian field configurations. These are of relevance for the dual superconductor picture of confinement, see, for example, \cite{Mandelstam:1974pi,'tHooft:1975pu,DiGiacomo:1999fb,DiGiacomo:1999fa,Carmona:2002ty,DiGiacomo:2007qj}. I will first give a short overview of this confinement scenario and the related hypothesis of Abelian IR dominance \cite{Ezawa:1982bf}. Then I will introduce the maximally Abelian gauge \cite{'tHooft:1981ht,Min:1985bx} and use the method devised in Chapter \ref{chp:scalingSolutions} to obtain its scaling solution, which supports the hypothesis of Abelian IR dominance. These results have been published in refs. \cite{Huber:2008ea,Huber:2009wh}. I also present additional explicit numerical calculations yielding values for the IREs.

\section{The dual superconductor confinement scenario}
\label{sec:dualSC}

The absence of free chromoelectric charges led to the idea that the QCD vacuum may be a sort of superconductor \cite{Mandelstam:1974pi,'tHooft:1975pu}. In a conventional superconductor of type I the magnetic field is expelled due to the Mei\ss ner-Ochsenfeld effect except for a thin layer below the surface. If, however, the magnitude of the external magnetic field is raised, superconductivity breaks down and there is a sharp transition to the non-superconducting phase. For a type-II superconductor there is an additional phase in between, where there is still zero resistance, but the magnetic flux can penetrate the material by means of flux tubes, also called Abrikosov vortices. This mixed phase is characterized by two critical field strengths: The first one gives the field strength where the magnetic flux starts to penetrate the material and the second gives the value when superconductivity breaks down.

One might think of the QCD vacuum in a similar fashion, namely being a \textit{dual superconductor} where not the magnetic but the chromoelectric field lines are confined into vortices. This picture provides a natural explanation for the formation of flux tubes between quarks.
A possible reason for the formation of such vortices is the condensation of chromomagnetic monopoles, which can be identified after reducing the gauge symmetry from $SU(N)$ to $U(1)^{N-1}$ by choice of an Abelian gauge \cite{'tHooft:1981ht}. The most prominent one is the MAG as introduced in ref. \cite{'tHooft:1981ht} which makes the gauge field as diagonal as possible, because the diagonal part is identified as the Abelian part, see Section \ref{ssec:abelianPart}. Hence one speaks of the diagonal and the off-diagonal parts of the algebra. Indeed in early lattice calculations the condensation of magnetic monopoles was observed in this gauge \cite{Kronfeld:1987ri,Kronfeld:1987vd}. Alternative possibilities for the realization of a dual superconductor exist if no magnetic monopoles occur, e.g., in the Landau gauge, where also a dual Mei\ss ner effect was observed \cite{Suzuki:2004dw}.

Directly connected to the dual superconductor scenario is the \textit{hypothesis of Abelian IR dominance} \cite{Ezawa:1982bf}. Based on the assumption that chromomagnetic monopoles cause confinement Ezawa and Iwazaki argued that the Abelian parts of the gauge fields should be dominant at large distances, since classic magnetic monopoles live in the Cartan subalgebra \cite{Ezawa:1982bf}. This is the algebra constructed from the maximal set of commuting generators and is thereby directly related to the Abelian part of a gauge field. The IR dominance of Abelian configurations in turn means that off-diagonal fields do not propagate over large distances. In subsequent years this was usually attributed to the occurrence of a mass for off-diagonal fields due to which these fields decouple according to the Appelquist-Carazzone theorem \cite{Appelquist:1974tg}. In the present work another another possibility is realized, namely an IR suppression of the off-diagonal propagators without a mass. The IR dominance of the Abelian degrees of freedom is realized by an IR divergent diagonal gluon propagator.

Although never explicitly proven indications of Abelian dominance were found in many instances. One is the fact that the string tension between a static pair of quarks has almost the same value if calculated from the diagonal part of gauge fields alone as from the complete fields, see, for example ref. \cite{Suzuki:1989gp}. Related to this is the notion of monopole dominance, which means that the string tension calculated from the monopole part of the diagonal fields is again close to the true value \cite{Stack:1994wm,Shiba:1994ab}.

There are also lattice simulations that calculated the propagators in the MAG. This allows direct comparisons to the results of functional methods. Early calculations can be found in refs. \cite{Bornyakov:2003ee,Amemiya:1998jz}. The most recent data with the lowest momenta available so far is in refs. \cite{Mendes:2006kc,Mendes:2008ux}. The difference to Landau gauge results is immediately obvious as all three propagators (diagonal gluon, off-diagonal gluon, ghost) become finite at vanishing momentum. Furthermore, there are less statistical errors. The results support the hypothesis of Abelian IR dominance as the propagators of the off-diagonal gluon and the ghost fields are suppressed at low momenta compared to that of the diagonal gluon field.

\section{The maximally Abelian gauge}
\label{sec:MAG}

This section describes how the Abelian part of the gluon field is identified and how the gauge is fixed in the MAG. The gauge fixing procedure is done in such a way to allow a continuous interpolation to the Landau gauge. From the action of the MAG the DSEs of the two-point functions are derived.

\subsection{Identifying the Abelian part of the gauge field}
\label{ssec:abelianPart}

One way to determine the importance of the Abelian part of Yang-Mills theory is to use the freedom to choose a gauge. By minimizing the norm of the off-diagonal gauge fields one can emphasize the role of the diagonal part. For this it is necessary to identify the diagonal part of the gauge field what is done via the generators of the corresponding algebra.
The generators of an algebra $T^r$ are defined by
\begin{align}\label{eq:structure-constants}
 [T^r,T^s]=i\,f^{rst} T^t,
\end{align}
where $f^{rst}$ are the structure constants. In the following we employ a widely used convention where the indices of Abelian generators (and corresponding fields later on) are given by $i,\,j,\, \ldots$ and those of the other generators by $a,\,b,\,\ldots$. Furthermore $r,\,s,\,\ldots$ represent both indices.
The Abelian part of an algebra, the so-called Cartan subalgebra, is defined by the maximal set of generators that commute among each other:
\begin{align}\label{eq:sc-Abelian}
 [T^i,T^j]=0.
\end{align}
The matrices for the Abelian generators can be chosen diagonal so that one speaks of the \textit{diagonal part}. The diagonal generators are given in $SU(N)$ by
\begin{align}
T^j=\left(\frac{2}{j(j+1)}\right)^{1/2} \times diag(\underbrace{1,\ldots,1}_{j \text{ times}},-j,0,\ldots,0), \quad j=1,\ldots, N-1.
\end{align}
The other generators are off-diagonal matrices and so one denotes the corresponding fields as \textit{off-diagonal}. This nomenclature avoids confusion about an Abelian field in a non-Abelian field theory, albeit sometimes in the literature the term non-Abelian field is used for the off-diagonal field alone.

For later it is convenient to determine the non-zero structure constants. Clearly those with at least two diagonal indices vanish due to \eref{eq:sc-Abelian}. If only one index is diagonal it is non-zero, but if all three indices are off-diagonal, one has to consider the specific algebra. Only if it possesses three off-diagonal generators it can be non-zero due to the antisymmetry of the structure functions. For $SU(N)$ this is always the case except for $N=2$, as the number of off-diagonal generators is $N^2-N=2$. For $SU(2)$ the only possible combination is then two off-diagonal and one diagonal indices. This leads to a simplification of a Lagrangian based on $SU(2)$ compared to higher $SU(N)$, when diagonal and off-diagonal parts of the fields are treated separately. Tab.~\ref{tab:structureConstants} gives the vanishing structure constants.

\begin{table}[t]
 \begin{center}
  \begin{tabular}{c||c|c}
  & $SU(2)$ & $SU(N>2)$ \\
 \hline
 \hline
 $f^{ijk}$ & 0 & 0\\
\hline
 $f^{ija}$ & 0 & 0\\
\hline
 $f^{abi}$ & $\checkmark$ & $\checkmark$ \\
\hline
 $f^{abc}$ & 0 & $\checkmark$
\end{tabular}
\caption{\label{tab:structureConstants}Overview over structure constants with diagonal ($i$, $j$, \ldots) and off-diagonal indices ($a$, $b$, \ldots).}
\end{center}
\end{table}

The gauge field $A_\mu$ is given by its components in the Lie algebra $A^r_\mu$, so its natural to separate the diagonal and off-diagonal parts as
\begin{align}
 A_\mu=T^i A^i_\mu+T^a B^a_\mu.
\end{align}
In addition to the index convention the off-diagonal gluon field $B$ was given a new symbol in order to alleviate the distinction between the two gluon fields in the following. These two fields may behave quite differently, what can already be inferred from the Yang-Mills Lagrangian, which is split in the same fashion:
\begin{align}\label{eq:L-YM-MAG}
 S_{YM}=\int dx \frac{1}{4}F_{\mu\nu}^r F_{\mu\nu}^r=\int dx \left(\frac{1}{4}F_{\mu\nu}^i F_{\mu\nu}^i+\frac{1}{4}F_{\mu\nu}^a F_{\mu\nu}^a \right)
\end{align}
with
\begin{align}
F_{\mu \nu }^a=&\partial_\mu B_\nu^a-\partial_\nu B_\mu^a-g\,f^{abc}B_\mu^b B_\nu^c-g\,f^{abi}B_\mu^b A_\nu^i-g\,f^{aib}A_\mu^i B_\nu^b
	=\nnnl
	=&D_\mu^{ab} B_\nu^b-D_\nu^{ab} B_\mu^b -g \, f^{abc}B_\mu^a B_\nu^b,\\
F_{\mu \nu }^i=&\partial_\mu A_\nu^i-\partial_\nu A_\mu^i-g\,f^{iab}B_\mu^b B_\nu^c,\\
D_\mu^{ab}:=&\delta^{ab}\partial_\mu+g\,f^{abi} A_\mu^i.\label{eq:MAGCovDeriv}
\end{align}
Explicitly the diagonal and off-diagonal terms in the Yang-Mills Lagrangian read as follows:
\begin{align}
F_{\mu \nu }^aF^a_{\mu \nu}&=2(\partial_\mu B_\nu^a)(\partial_\mu B_\nu^a)-2(\partial_\mu B_\nu^a)(\partial_\nu B_\mu^a)+\nnnl
& - 2 g\, f^{abc} B_\mu^b B_\nu^c ((\partial_\mu B_\nu^a)-(\partial_\nu B_\mu^a))-4g\, f^{abi}B_\mu^b A_\nu^i ((\partial_\mu B_\nu^a)-(\partial_\nu B_\mu^a))+\nnnl
& + 2g^2\, f^{abi}f^{adj}B_\mu^b A_\nu^i( B_\mu^d A_\nu^j-A_\mu^j B_\nu^d)+2g^2\,f^{abc}f^{adi} B_\mu^b B_\nu^c 
(B_\mu^d A_\nu^i -A_\mu^i B_\nu^d) +\nnnl
&+g^2\,f^{abc}f^{ade}  B_\mu^b B_\nu^c B_\mu^d B_\nu^e,\\
F_{\mu \nu }^iF_{\mu \nu }^i&=2(\partial_\mu A_\nu^i)(\partial_\mu A_\nu^i)-2(\partial_\mu A_\nu^i)(\partial_\nu A_\mu^i)+\nnnl
& -2g\,f^{iab}B_\mu^a B_\nu^b((\partial_\mu A_\nu^i)-(\partial_\nu A_\mu^i))+\nnnl
&+g^2\,f^{iab} f^{icd}B_\mu^a B_\nu^b B_\mu^c B_\nu^d.
\end{align}
In summary there are $BBBB$-, $ABBB$-, $AABB$-, $ABB$- and $BBB$-interactions.

It should be stressed that for later convenience the covariant derivative was defined in \eref{eq:MAGCovDeriv} only with the diagonal gluon field and not as its projection from $D_\mu^{rs}=\delta^{rs}\partial_\mu+g\,f^{rst} A_\mu^t$, where also an off-diagonal field would appear.
At this point it does not make a difference, yet, as it acts on an off-diagonal gluon field and the additional term would be zero: $f^{abc} B_\mu^c B_\mu^b=-f^{abc} B_\mu^b B_\mu^c=0$. However, if it acts on a ghost field, this term has to be taken into account. Accordingly we define the "full" covariant derivative as
\begin{align}
 \tilde{D}_\mu^{ab}:=\delta^{ab}\partial_\mu+g\,f^{abc} B_\mu^c+g\,f^{abi} A_\mu^i.
\end{align}

\subsection{Gauge fixing and renormalizability}

Next we will determine the gauge fixing part of the action. For this we need to derive the BRST transformation, where we discriminate again between the diagonal and off-diagonal parts.
Furthermore the Jacobi identity splits into several identities. The modified expressions can be directly derived from the usual ones by ascribing the free indices to either the diagonal or off-diagonal sector. As an example we have a look at the BRST transformation of the gluon field, given by
\begin{align}
s\,A_\mu^r&=-D_\mu^{rs} c^s.
\end{align}
Choosing $r$ to be either a diagonal or off-diagonal index, $i$ and $a$, respectively, we get the BRST transformation for the two gluon fields:
\begin{align}
s\,B_\mu^a&=-\tilde{D}_\mu^{ab} c^b-\tilde{D}_\mu^{ai} c^i
	=-(D_\mu^{ab} c^b-g\,f^{abc}B_\mu^b c^c-g\,f^{abi} B_\mu^b c^i), \\
s\,A_\mu^i&=-\tilde{D}_\mu^{ij} c^j-\tilde{D}_\mu^{ia} c^a
	=-(\partial_\mu c^i-g\,f^{iab}B_\mu^a c^b).
\end{align}

Similarly the transformations for the ghost fields and the Nakanishi-Lautrup fields are derived:
\begin{align}
s\,c^a&=-\mhalfo g\,f^{abc} c^b c^c-g\,f^{abi}c^b c^i,&
s\,c^i&=-\mhalfo g\,f^{iab}c^a c^b,\\
s\,\bar{c}^a&=i \, b^a,&
s\,\bar{c}^i&=i \, b^i,\\
s\,b^a&=0,&
s\,b^i&=0.
\end{align}
The standard Jacobi identity,
\begin{align}
f^{rst}f^{ruv}+f^{rsu}f^{rvt}+f^{rsv}f^{rtu}=0,
\end{align}
splits up into
\begin{align}
f^{abi}f^{acj}+f^{abj}f^{aic}&=0,\label{eq:MAGJacobianId1}\\
f^{abc}f^{adi}+f^{abd}f^{aic}+f^{abi}f^{acd}&=0,\\
f^{rab}f^{rcd}+f^{rac}f^{rdb}+f^{rad}f^{rbc}&=0.
\end{align}

Now we can proceed to fix the gauge. The basic idea is to minimize the norm of the off-diagonal part given by
\begin{align}
 R_{MAG}=\frac1{2} \int dx \, B^a_\mu(x) B^a_\mu(x).
\end{align}
It is found that a (local) minimum corresponds to
\begin{align}\label{eq:MAGCondition}
D^{ab}_\mu B_\mu^b=0.
\end{align}
This is the gauge fixing condition that defines the MAG. In contrast to the Landau gauge condition it is non-linear as the diagonal gluon field appears in the covariant derivative. This will have far-reaching consequences for the Faddeev-Popov operator and the structure of the interactions in the gauge-fixed Lagrangian.

Having the gauge fixing condition, we can proceed in the usual way of BRST quantization, i.e., we calculate the gauge fixing part of the Lagrangian via a BRST variation:
\begin{align}\label{eq:S-MAG}
S_{MAG}&=s\int dx\,\bar{c}^a(\hat{D}_\mu^{ab} B_\mu^b-i \frac{\alpha}{2} b^a)=\nnnl
&=\int dx\, (\frac \alpha{2}{b^a b^a}+i b^a \hat{D}_\mu^{ab}B_\mu^b+\bar{c}^a \hat{D}_\mu^{ab}D_\mu^{bc} c^c-g\,f^{abi} \bar{c}^a(\hat{D}_\mu^{bc} B_\mu^c) c^i-\nnnl
&-g\,f^{bcd}\bar{c}^a \hat{D}_\mu^{ab} B_\mu^c c^d-g^2\,\zeta f^{abi}f^{cdi} B_\mu^b B_\mu^c \bar{c}^a c^d -
(1-\zeta)g\, f^{abi}\bar{c}^a B_\mu^b \partial_\mu c^i),
\end{align}
where the covariant derivative with the hat, $\hat{D}_\mu^{ab}$, is defined as
\begin{align}
 \hat{D}_\mu^{ab}:=\delta^{ab}\partial_\mu+g\,\zeta\,f^{abi}A_\mu^i.
\end{align}
It allows to interpolate between the MAG and the Landau gauge via the parameter $\zeta$: For the MAG it is set to one and for the Landau gauge to zero.
The parameter $\alpha$ is the conventional gauge fixing parameter for the MAG. Alternative formulations for gauges interpolating between the MAG and the Landau gauge can be found in refs. \cite{Dudal:2004rx,Capri:2005zj}.

We do not integrate out the Nakanishi-Lautrup field yet, as we have to face the following problem of the MAG first: The Lagrangian as we have it now is not renormalizable, as there appear divergences due to a quartic ghost interaction \cite{Min:1985bx,Fazio:2001rm}. In order to absorb these divergences into a counterterm, this interaction is introduced into the Lagrangian via a BRST-exact term so that the BRST symmetry is not spoiled:
\begin{align}
S_R&=  s\int dx (-\mhalfo \lambda\, g\,f^{abi} \bar{c}^a \bar{c}^b c^i-\frac 1{4}\lambda'\, g\,f^{abc} \bar{c}^a \bar{c}^b c^c)=\nnnl
&=\int dx (-i\,g\,\lambda\,f^{abi}b^a \bar{c}^b c^i-i\,\mhalfo g\,\lambda'\,f^{abc} b^a \bar{c}^b c^c+\frac1{4}g^2\lambda f^{abi}f^{cdi}\bar{c}^a\bar{c}^b c^c c^d+\nnnl
&+\lambda'\frac1{4}g^2 f^{abc}f^{adi}\bar{c}^b \bar{c}^c c^d c^i+\lambda'\frac1{8}g^2 f^{abc}f^{ade} \bar{c}^b \bar{c}^c c^d c^e)
\end{align}
The new terms have two parameters $\lambda$ and $\lambda'$ that can be set to zero to recover the original MAG Lagrangian. Note that the second term in the first line is zero in $SU(2)$. Now the Nakanishi-Lautrup field is integrated out\footnote{Note that in principle one could also keep the Nakanishi-Lautrup field, but for the coming analysis it is preferable to set this field on-shell as otherwise there are mixing terms at the two-point level.}:
\begin{align}\label{eq:L-MAG-R}
 S'_{MAG,R}&=\int dx\, \Big(\bar{c}^a \hat{D}_\mu^{ab}D_\mu^{bc} c^c-g\,f^{abi} \bar{c}^a(\hat{D}_\mu^{bc} B_\mu^c) c^i-g\,f^{bcd}\bar{c}^a \hat{D}_\mu^{ab} B_\mu^c c^d-\nnnl
 &-g^2\,\zeta f^{abi}f^{cdi} B_\mu^b B_\mu^c \bar{c}^a c^d -(1-\zeta)g\, f^{abi}\bar{c}^a B_\mu^b \partial_\mu c^i-\frac1{2\alpha}(\hat{D}_\mu^{ab} B_\mu^b)^2+\nnnl
 &+\frac{\lambda'^2}{8\alpha} g^2 f^{abc}f^{ade} \bar{c}^b c^c \bar{c}^d c^e-\frac{\lambda}{\alpha}g\,f^{abi} (\hat{D}_\mu^{ac} B_\mu^c) \bar{c}^b c^i -\nnnl
 &-\frac{\lambda'}{2\alpha} g\,f^{abc} (\hat{D}_\mu^{ad} B_\mu^d) \bar{c}^b c^c +\frac{\lambda \lambda'}{2\alpha} g^2 f^{abi}f^{acd} \bar{c}^b c^i \bar{c}^c c^d +\nnnl
 &+\frac1{4}g^2\lambda f^{abi}f^{cdi}\bar{c}^a\bar{c}^b c^c c^d+\lambda'\frac1{4}g^2 f^{abc}f^{adi}\bar{c}^b \bar{c}^c c^d c^i+\lambda'\frac1{8}g^2 f^{abc}f^{ade} \bar{c}^b \bar{c}^c c^d c^e\Big).
\end{align}

The MAG corresponds to the formal limit $\alpha, \,\lambda,\,\lambda' \rightarrow0$ and $\zeta=1$. A final simplification can be obtained by identifying $\alpha$, $\lambda$ and $\lambda'$. This is not only possible because all three parameters are zero for the MAG, but it is also supported by a Ward identity, the so-called diagonal ghost equation \cite{Fazio:2001rm}. It only has the meaning of a Ward identity if $\alpha=\lambda=\lambda'$.
Identifying these three parameters the terms containing diagonal ghosts drop out completely in $S'_{MAG,R}$. Hence the off-diagonal gauge fixing sector does only depend on off-diagonal ghosts. The interactions that appear are $AAcc$, $BBcc$, $ABcc$, $cccc$, $Acc$ and $Bcc$.

The definition of the MAG does only fix the gauge for the off-diagonal gluon fields and the diagonal Yang-Mills part still has a $U(1)^{N-1}$-symmetry. This remnant of the full gauge symmetry has to be fixed as well for functional equations. From several possible gauge fixing prescriptions we choose the Landau gauge, i.e., $\partial_\mu A_\mu^i=0$. The corresponding gauge fixing parameter is $\xi$ and the required Nakanishi-Lautrup field is again $b$ with a diagonal index:
\begin{align}
S_{diag}&=s\int dx\,\bar{c}^i(\partial_\mu A_\mu^i-i\frac \xi{2} b^i)=\nnnl
&=\int dx\, \left( i b^i (\partial_\mu A_\mu^i-i \frac \xi{2} b^i)+\bar{c}^i \partial_\mu (\partial_\mu c^i-g\,f^{abi}B_\mu^a c^b) \right).
\end{align}
Integrating out the auxiliary fields also here yields
\begin{align}
S'_{diag}&=\int dx\, \left(\frac1{2\xi}(\partial_\mu A_\mu^i)^2+\bar{c}^i \partial_\mu (\partial_\mu c^i-g\,f^{abi}B_\mu^a c^b) \right).
\end{align}
In this Lagrangian the diagonal ghosts appear again. However, they can be integrated out after shifting the diagonal ghost field as \cite{Capri:2010an,Capri:2005tj}
\begin{align}
c^i\rightarrow c^i+\frac{\partial_\mu}{\Box}g\,f^{abi}B_\mu^a c^b.
\end{align}
The Jacobian of this transformation is trivial \cite{Capri:2005tj} and so we can get rid of the diagonal ghosts entirely. Their disappearance is not surprising as ghosts also disappear in the pure Abelian gauge theory in the Landau gauge.

Now we can combine the individual parts to the final action:
\begin{align}\label{eq:L-MAG}
 S=S_{YM}+S''_{MAG,R}+S''_{diag},
\end{align}
with
\begin{align}
 S''_{MAG,R}&=S'_{MAG,R}\Big|_{\lambda'=\lambda=\alpha}=\int dx\, \Big(\bar{c}^a \hat{D}_\mu^{ab}D_\mu^{bc} c^c-g\,f^{bcd}\bar{c}^a \hat{D}_\mu^{ab} B_\mu^c c^d - \nnnl
 &-g^2\,\zeta f^{abi}f^{cdi} B_\mu^b B_\mu^c \bar{c}^a c^d-\frac1{2\alpha}(\hat{D}_\mu^{ab} B_\mu^b)^2+\frac{\alpha}{8} g^2 f^{abc}f^{ade} \bar{c}^b c^c \bar{c}^d c^e-\nnnl
 &-\frac{1}{2} g\,f^{abc} (\hat{D}_\mu^{ad} B_\mu^d) \bar{c}^b c^c +\frac1{4}g^2\alpha f^{abi}f^{cdi}\bar{c}^a\bar{c}^b c^c c^d+\alpha\frac1{8}g^2 f^{abc}f^{ade} \bar{c}^b \bar{c}^c c^d c^e\Big),\\
S''_{diag}&=\int dx\, \frac1{2\xi}(\partial_\mu A_\mu^i)^2 .
\end{align}
One should note that this Lagrangian simplifies considerably for the gauge group $SU(2)$ as the structure constants with three off-diagonal indices are zero:
\begin{align}
 S''_{MAG,R}\Big|_{SU(2)}&=\int dx\, \Big(\bar{c}^a \hat{D}_\mu^{ab}D_\mu^{bc} c^c-g^2\,\zeta f^{abi}f^{cdi} B_\mu^b B_\mu^c \bar{c}^a c^d -\nnnl
 &-\frac1{2\alpha}(\hat{D}_\mu^{ab} B_\mu^b)^2 +\frac1{4}g^2\alpha f^{abi}f^{cdi}\bar{c}^a\bar{c}^b c^c c^d\Big),\\
F_{\mu \nu }^aF^a_{\mu \nu}\Big|_{SU(2)}&=2(\partial_\mu B_\nu^a)(\partial_\mu B_\nu^a)-2(\partial_\mu B_\nu^a)(\partial_\nu B_\mu^a) -4g\, f^{abi}B_\mu^b A_\nu^i ((\partial_\mu B_\nu^a)-(\partial_\nu B_\mu^a))+\nnnl
& + 2g^2\, f^{abi}f^{adj}B_\mu^b A_\nu^i( B_\mu^d A_\nu^j-A_\mu^j B_\nu^d).
\end{align}

\subsection{The Dyson-Schwinger equations of the maximally Abelian gauge}
\label{ssec:MAGDSEs}

\begin{figure}[t]
\begin{center}
 \includegraphics[width=0.92\textwidth]{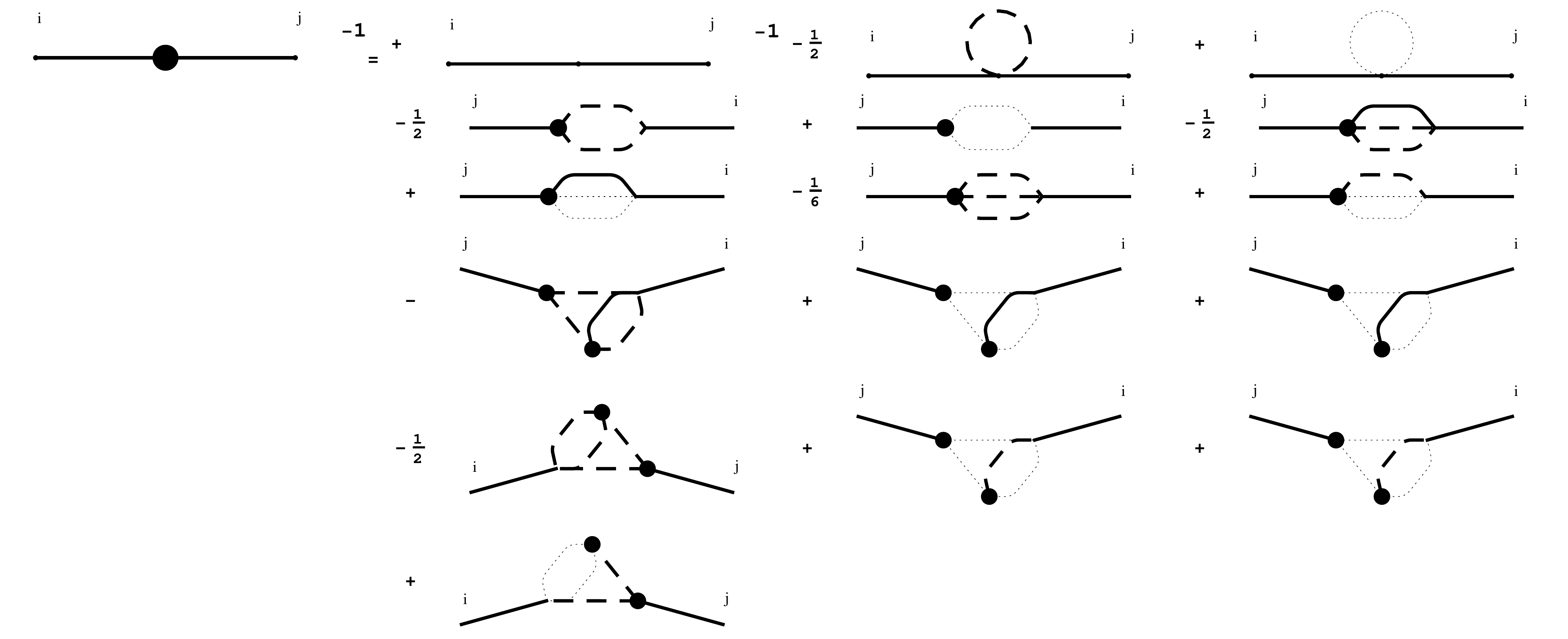}
\caption{\label{fig:MAG-DSEs-AA} DSE of the diagonal gluon two-point function. \textit{DoDSE} cannot draw the usual wiggly lines for gluons, so the convention is that continuous lines are diagonal gluons, dashed ones off-diagonal gluons and dotted ones ghosts.}
\end{center}
\end{figure}
\begin{figure}[t]
\begin{center}
 \includegraphics[width=0.92\textwidth]{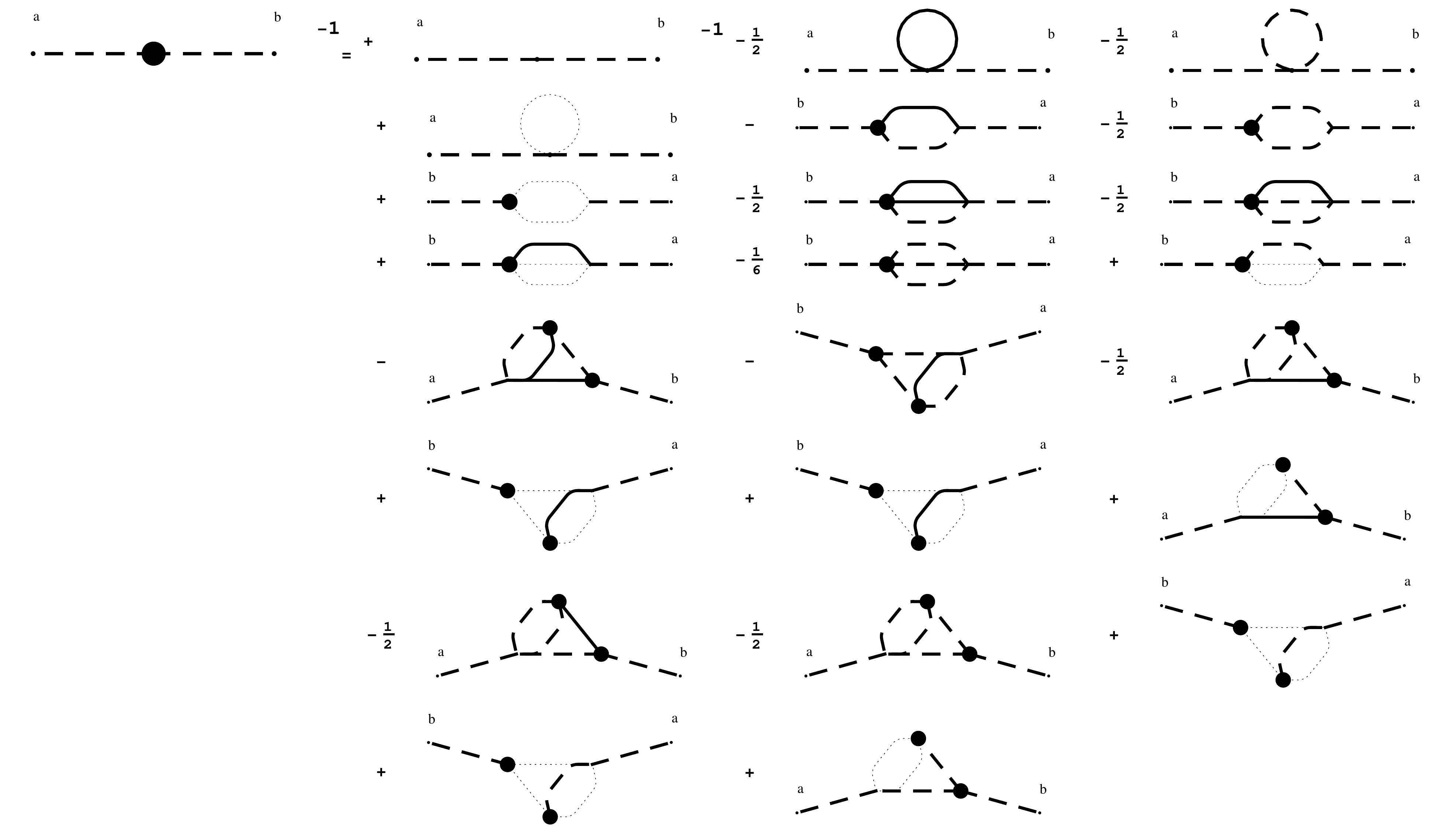}
\caption{\label{fig:MAG-DSEs-BB} DSE of the off-diagonal gluon two-point function.}
\end{center}
\end{figure}
\begin{figure}[t]
\begin{center}
 \includegraphics[width=0.92\textwidth]{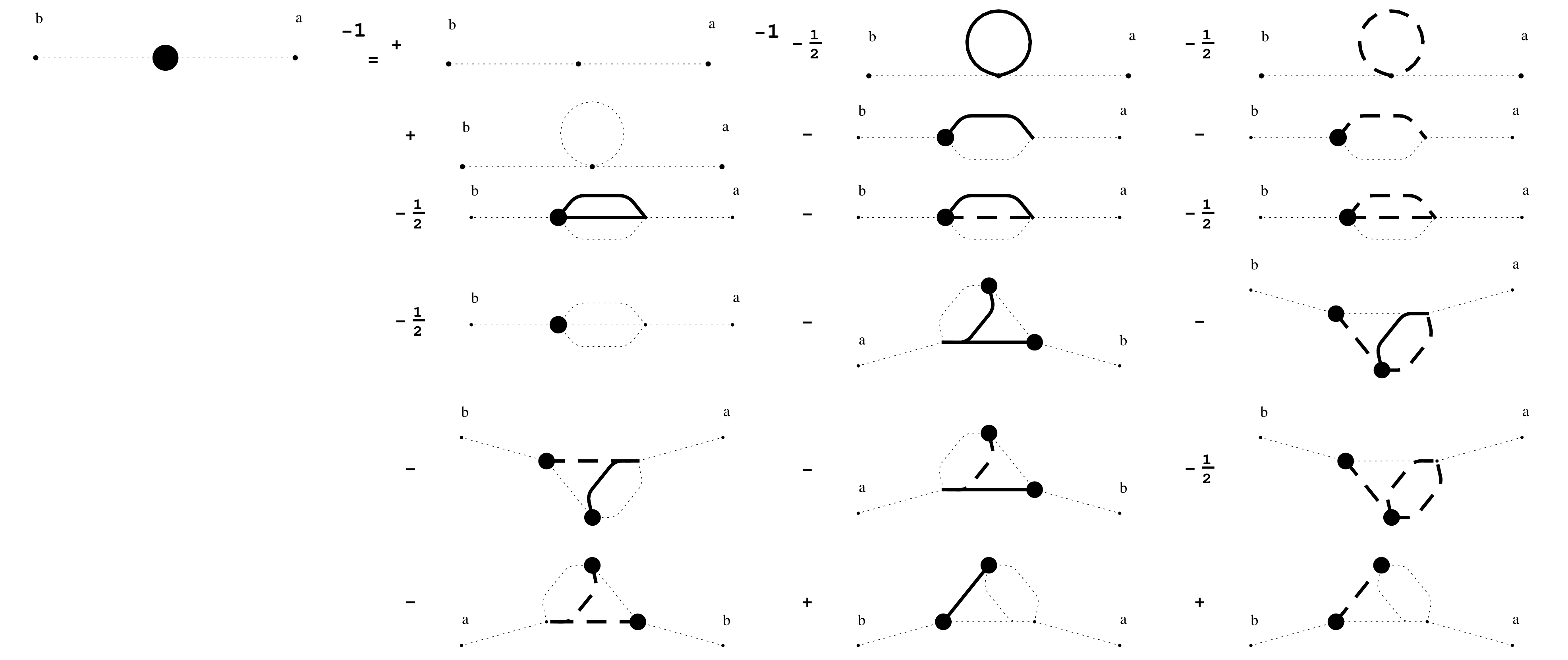}
\caption{\label{fig:MAG-DSEs-cc} DSE of the off-diagonal ghost two-point function.}
\end{center}
\end{figure}

The Lagrangian of the MAG, \eref{eq:L-MAG}, contains many interactions of the three fields. Therefore the derivation of the DSEs becomes quite lengthy when done by hand. Especially the numerous possibilities for different intermediate propagators lead to many terms. For this reason the program \textit{DoDSE} is very useful here. The derivation of the two-point function DSEs is explained in Appendix \ref{sec:DoDSEMAG}. The resulting equations are depicted in figs.~\ref{fig:MAG-DSEs-AA}, \ref{fig:MAG-DSEs-BB} and \ref{fig:MAG-DSEs-cc}. The total number of terms at the two-point level, 16+23+18=57, should be compared with that of the Landau gauge, 6+2=8. This illustrates nicely the increased complexity of the system of equations that has to be treated here.

\begin{figure}[b]
\begin{center}
 a) \includegraphics[width=0.43\textwidth]{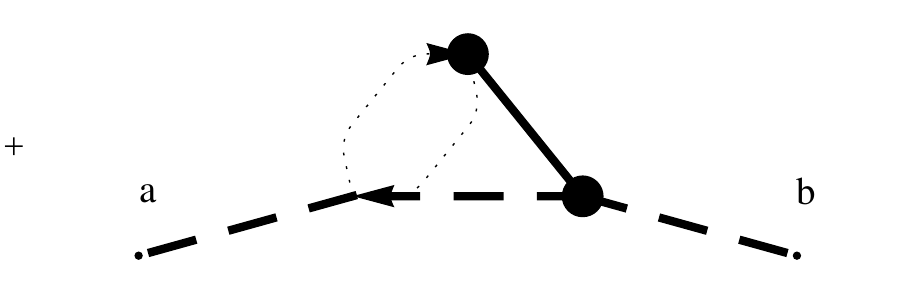}\hskip1cm
 b) \includegraphics[width=0.43\textwidth]{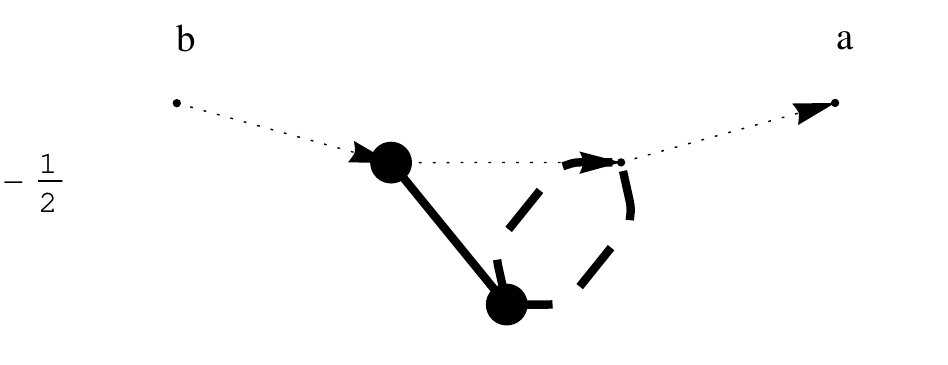}
\caption{\label{fig:MAG-DSEs-2p-van} Diagrams that vanish due to the color algebra.}
\end{center}
\end{figure} 

One issue is that diagrams can vanish due to the color algebra. This has to be checked explicitly for every diagram. At the two-point level this is possible without assumptions for all diagrams except the sunsets, where a dressed four-point function appears. For the three-point function there is, under the assumption that no symmetric tensor $d^{rst}$ appears, only one basis tensor in color space, which is the antisymmetric structure function of $SU(N)$, $f^{rst}$. The dressed three-point functions are then proportional to this color tensor. The two-point functions are all taken as diagonal, i.e., either proportional to $\delta^{ab}$ or $\delta^{ij}$. Contracting the color tensors in all diagrams except for the sunsets reveals that indeed two diagrams vanish: The squint diagrams with a bare $BBcc$ vertex and an internal diagonal gluon line. They are depicted in \fref{fig:MAG-DSEs-2p-van}. Note that other diagrams with a bare $BBcc$ vertex remain.

At this point is has to be stressed that the gauge fixing parameter of the MAG $\al$ is still arbitrary as it appears in the denominator of the bare vertices. Therefore the off-diagonal gluon propagator has a longitudinal part. In order to take into account that the longitudinal part could scale differently as the transverse part one could introduce an extra IRE $\de_{B,long}$ and split the off-diagonal field in transverse and longitudinal parts. However, inserting the bare vertices explicitly and projecting the DSE of the off-diagonal propagator transversely and longitudinally one finds that the equations for both parts is the same. Thus the IR analysis yields $\de_B=\de_{B,long}$. This result can only be invalidated by cancelations between different diagrams or due to contributions from the unknown dressed vertices. Based on this relation we adopt in the following only one common dressing function for both tensors.

For the moment this is as far as we go in excluding cancelations. It cannot be ruled out, though, that nevertheless some diagrams vanish or cancel each other. A clarification of this point would require a more detailed analysis taking into account the detailed Lorentz structure. However, at the moment it seems not manageable to do so even at the IR leading order, as it will be found that the IR dominant diagrams involve dressed four-point functions which possess a plethora of possible Lorentz tensors.

\section{Infrared scaling solution for the maximally Abelian gauge}

Having derived the DSEs of the MAG we can now to proceed to its IR analysis. It will be found that there is a unique scaling relation, but the IREs of vertices with an odd number of legs are ambiguous. In the discussion of this solution in Section \ref{ssec:MAGSolution} differences and connections to the Landau gauge are highlighted. Finally a numerical calculation of the IREs is presented.

\subsection{Obtaining the scaling relation}

As the MAG features three fields and eleven bare interaction vertices, its IR analysis seems much more complicated than that of the Landau gauge. However, with the improved method developed in Chapter \ref{chp:scalingSolutions} only slightly more effort is needed.

The dressed propagators in the MAG are
\begin{align}
 D_A^{ij}(p^2)&=\de^{ij}\frac{c_A(p^2)}{p^2}\left(g_{\mu\nu}-\frac{p_\mu p_\nu}{p^2}\right)+\xi\,\de^{ij}\frac{p_\mu p_\nu}{p^2},\\
 D_B^{ab}(p^2)&=\de^{ab}\frac{c_B(p^2)}{p^2}\left(g_{\mu\nu}-(1-\al)\frac{p_\mu p_\nu}{p^2}\right),\\
 D_c^{ab}(p^2)&=-\de^{ab}\frac{c_c(p^2)}{p^2}.
\end{align}
The reason for choosing only one dressing function for the off-diagonal propagator is given in Section \ref{ssec:MAGDSEs}.
The dressing functions obey power laws in the IR:
\begin{align}
 c_A(p^2)&\overset{p^2\rightarrow 0}{=} d_A \cdot (p^2)^{\de_A},\\
 c_B(p^2)&\overset{p^2\rightarrow 0}{=} d_B \cdot (p^2)^{\de_B},\\
 c_c(p^2)&\overset{p^2\rightarrow 0}{=} d_c \cdot (p^2)^{\de_c}.
\end{align}

Now we follow the procedure outlined in Section \ref{ssec:analysis-2Point}.
The first step is:
\begin{enumerate}
 \item Determine the inequalities that are derived from the interactions appearing in the Lagrangian from \eref{eq:verts-props-inequal2}.
\end{enumerate}
In the MAG this leads to the following inequalities in $SU(2)$:
\begin{align}\label{eq:MAGInequalities}
 \de_A+\de_B\geq0,& \qquad \de_A+\de_c\geq0,& \qquad \de_B+\de_c\geq0, &\qquad 2\de_B\geq0,\nnnl
 2\de_c\geq0,& \qquad \de_A+\frac1{2}\de_B\geq0,&\qquad \de_A+\frac{1}{2}\de_c\geq0.&&
\end{align}
The additional inequalities in $SU(N>2)$ are
\begin{align}\label{eq:inequsSUN}
\frac1{2}\de_A+\frac{1}{2}\de_B+\de_c \geq0,\qquad \frac1{2}\de_A+\frac{3}{2}\de_B\geq0,\qquad \frac{3}{2}\de_B\geq0, \qquad \frac{1}{2}\de_B+\de_c\geq0.
\end{align}

The second step is:
\begin{enumerate}
 \item[2.] Reduce the number of inequalities, if some of them are contained within others.
\end{enumerate}
This yields the following four inequalities for \textit{general} $SU(N)$:
\begin{align}\label{eq:MAGIREIneqs}
 \de_A+\de_B\geq0, \qquad \de_A+\de_c\geq0, \qquad \de_B\geq0,\qquad \de_c\geq0.
\end{align}
The inequalities for $SU(N>2)$, \eref{eq:inequsSUN}, do not provide additional constraints for the system. Hence we will find that $SU(2)$ and $SU(N)$ have the same IR behavior, although their actions are different.

The last step is:
\begin{enumerate}
 \item[3.] Try to saturate one inequality after the other according to \eref{eq:DSE-2-point-leading-eq} and see, if you can find any contradictions with the remaining inequalities.
\end{enumerate}
We first investigate the last two inequalities in \eref{eq:MAGIREIneqs}. They yield $\de_A=\de_B=\de_c=0$. To illustrate this I explain the line of argument for the case $\de_B=0$ in more detail.

The equation $\de_B=0$ is obtained from diagrams with a bare $B$ self coupling. As such bare vertices do not appear in the DSEs for the diagonal gluon or the ghost two-point functions, we still have to determine their IR leading diagrams. In the DSE of the diagonal gluon there are two possibilities for IR leading diagrams, those with bare $AABB$ or $AAcc$ vertices. If the latter are leading we obtain $\de_A+\de_c=0$. However, since both $\de_A$ and $\de_c$ are non-negative if $\de_B=0$, the only possible solution is $\de_A=\de_c=0$. Similarly we get $\de_A=0$, if vertices with a bare $AABB$ vertex are leading. For $\de_c$ the ghost two-point function is analyzed, where we come back to the argument with the $AAcc$ vertex. Hence the emerging solution is that all three propagators scale trivially in the IR, i.e., equally as in the UV. This option is of no interest to us, as we do not expect that perturbative propagators describe the IR regime of Yang-Mills theory. The argument if starting with the equation $\de_c=0$ is completely the same.

The two remaining possibilities are $\de_A+\de_B=0$ and $\de_A+\de_c=0$, which can be fulfilled independently. Thus the scaling relation of the MAG is
\begin{align}\label{eq:MAGScalingRelation}
 \ka_{MAG}:=-\de_A=\de_B=\de_c\geq0.
\end{align}
Before discussing its meaning we determine the IR behavior of vertices.

\subsection{The infrared exponents of vertices}
\label{ssec:MAGVertices}

From the scaling relation found above the IR behavior of vertices can be obtained. A first observation is that the IREs of $AABB$ and $AAcc$ vertices vanish:
\begin{align}\label{eq:leadingsVerticesIREs}
 \ka_{AABB}=\ka_{AAcc}=0.
\end{align}
This can be shown by plugging the scaling relation \eref{eq:MAGScalingRelation} into \eref{eq:verts-props-inequal1} and using that primitively divergent vertices have a non-positive IRE, see \eref{eq:primDivVerts}.

\begin{figure}[t]
 \begin{center}
  \includegraphics[width=0.43\textwidth]{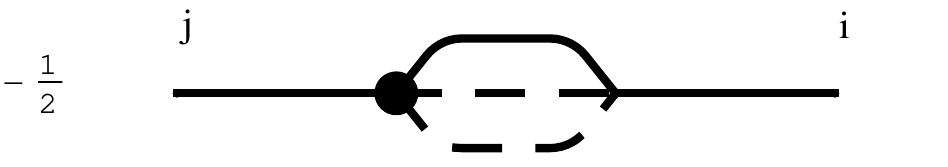}
  \includegraphics[width=0.43\textwidth]{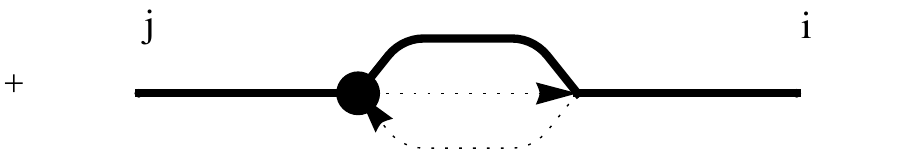}\\
  \includegraphics[width=0.43\textwidth]{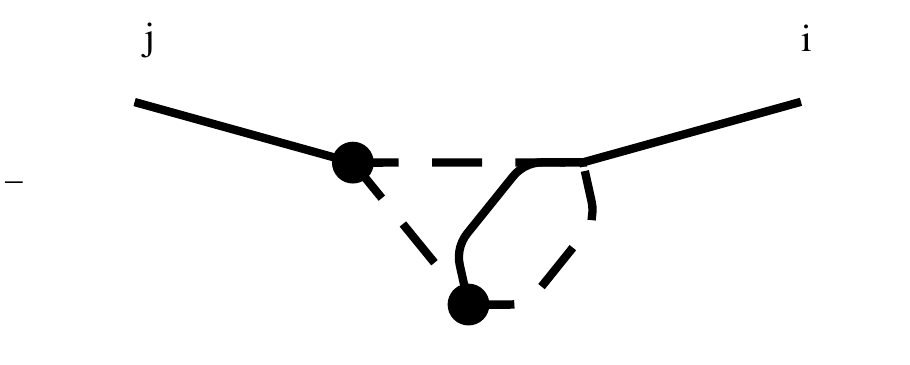}
  \includegraphics[width=0.43\textwidth]{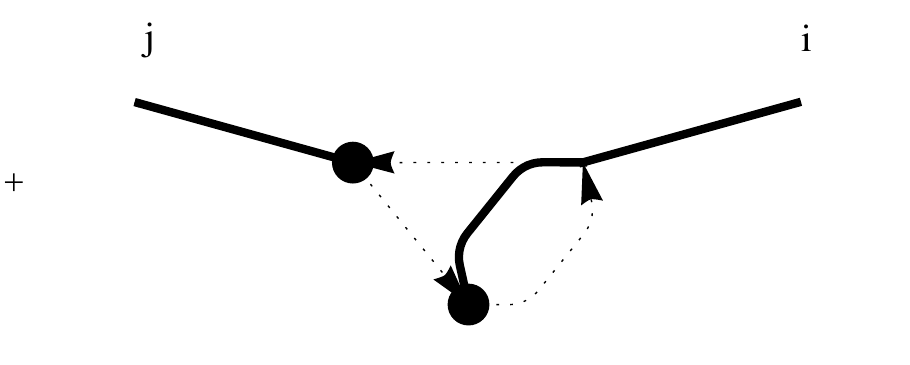}
  \caption{\label{fig:MAG-leadingDiagramsAA} Possibly IR leading diagrams in the DSE of the diagonal gluon two-point function. The sunset diagrams (first row) are definitely at leading order, while it is possible that the squint diagrams (second row) are IR subdominant.}
 \end{center}
\end{figure}

\begin{figure}[t]
 \begin{center}
  \includegraphics[width=0.43\textwidth]{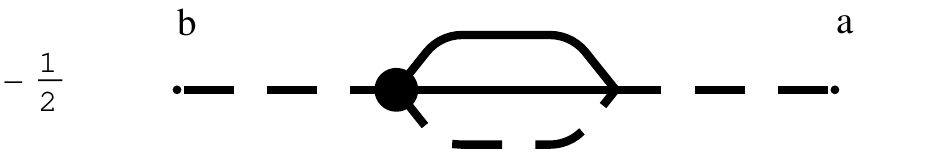}
  \includegraphics[width=0.43\textwidth]{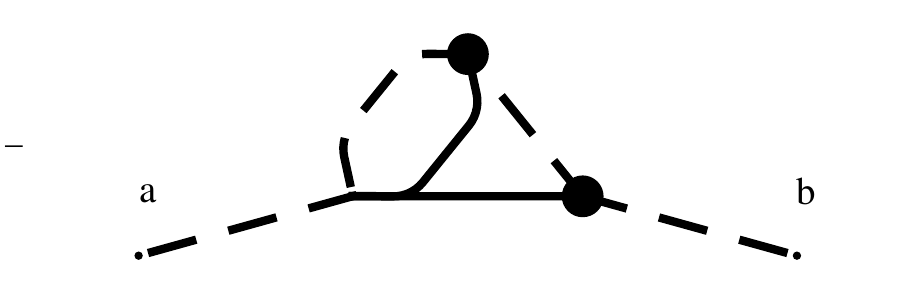}\\
  \includegraphics[width=0.43\textwidth]{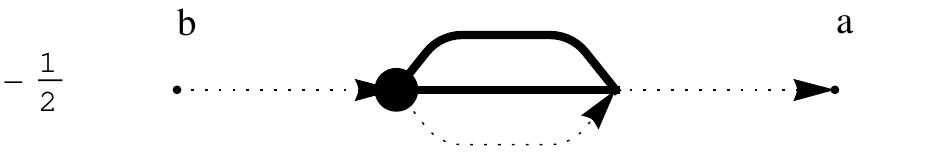}
  \includegraphics[width=0.43\textwidth]{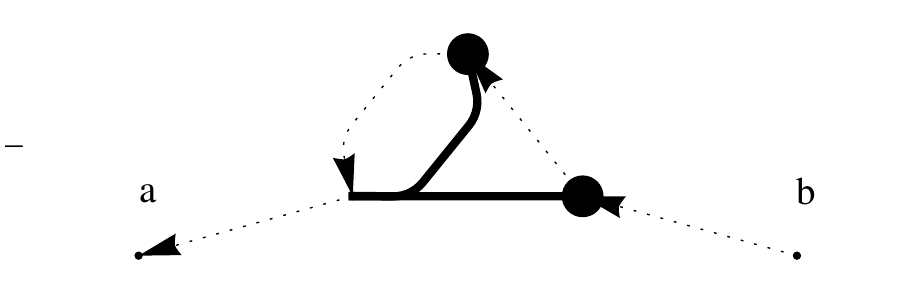}
 \caption{\label{fig:MAG-leadingDiagramsBBcc} Possibly IR leading diagrams in the DSEs of the off-diagonal gluon two-point function (upper row) and of the ghost two-point function (lower row). The sunset diagrams (left) are definitely at leading order, while it is possible that the squint diagrams (right) are IR subdominant.}
 \end{center}
\end{figure}

Next we want to determine the IR leading diagrams so that we can derive the IREs of other vertices from them. However, for the MAG there is an ambiguity due to the fact that there are two classes of possibly leading diagrams. The origin of the scaling relation are diagrams with bare $AABB$ or $AAcc$ vertices, i.e., (neglecting the tadpole) sunset and squint diagrams can be used to derive the scaling relation. The corresponding diagrams are depicted in figs.~\ref{fig:MAG-leadingDiagramsAA} and \ref{fig:MAG-leadingDiagramsBBcc}. As we did not yet specify in our analysis which diagram led to the scaling relation, we look for the consequences if only one class is IR leading.

Let us assume first it is the squint diagram from which we derived the scaling relation and \eref{eq:leadingsVerticesIREs}. This provides enough information to count the IREs of the sunset diagrams which turn out to be at leading order, i.e., if the squints are IR leading also the sunsets are. If, however, we assume that the scaling relation is obtained from the sunset diagrams, we cannot determine the IREs of the three-point vertices necessary to count the IREs of the squint diagrams. So the sunset diagrams are always at leading order, while we cannot say if the squint diagrams scale equally.

\begin{figure}[t]
 \begin{center}
 \includegraphics[width=0.6\textwidth]{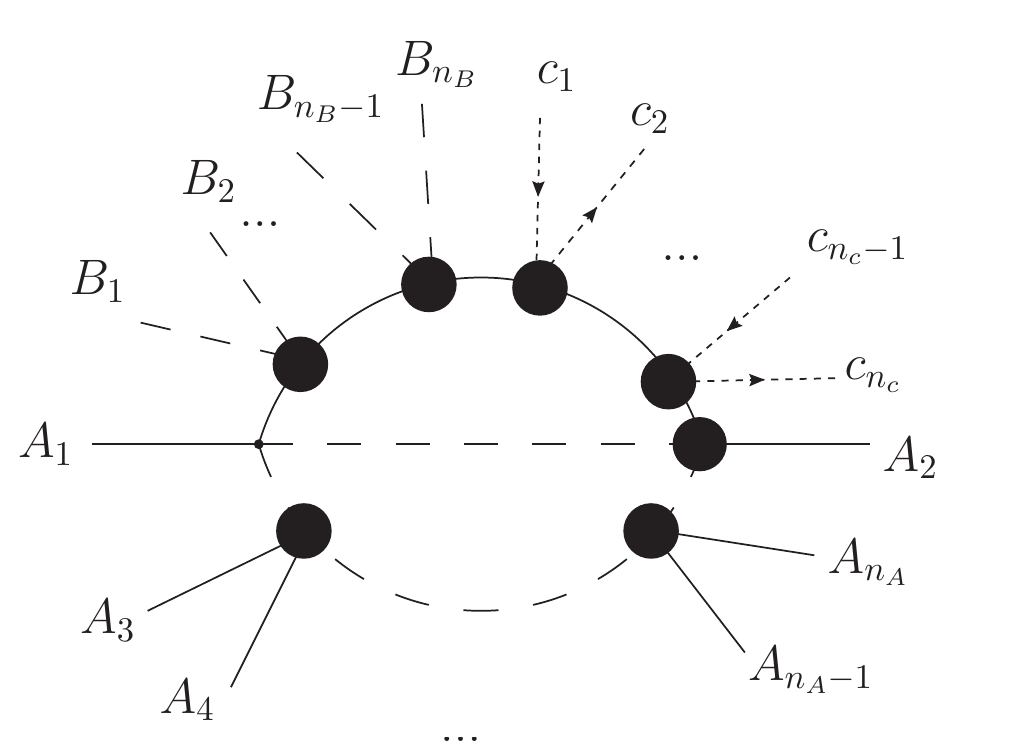}
 \caption{\label{fig:MAGHigherVertices} Consecutively inserting pairs of $A$, $B$ and $c$  fields via $AABB$ and $AAcc$ vertices shows that for graphs with an even number of legs the maximally IR divergent solution is realized.}
 \end{center}
\end{figure}

The reason why we cannot determine the IREs of three-point functions is purely combinatorial. We found that the IR leading vertices have four legs. Consequently it will be easy to construct the IR leading diagrams of an n-point function if n is even: As depicted in \fref{fig:MAGHigherVertices} we plug $AABB$ or $AAcc$ vertices into appropriate IR leading diagrams of two-point functions. The IREs of the resulting diagrams can be inferred from \eref{eq:IREArbitraryDiagram}, but as we only add quantities for which the last two terms are zero, we directly get to \eref{eq:Maximal-IR-Divergent}, the maximally IR divergent solution. The IRE of a vertex with $n_A$ diagonal gluon legs, $n_B$ off-diagonal gluon legs and $n_c$ ghost legs is then
\begin{align}
\ka(n_A,n_B,n_c)=\mhalfo(n_A-n_B-n_c)\ka_{MAG} \qquad \text{($n_A+n_B+n_c$ even)}.
\end{align}

However, it is not possible to construct an IR leading diagram purely from four-point functions for a vertex with an odd number of legs. The open question is the IR behavior of the three-point function for which we can only find upper and lower bounds from the bare vertices and the maximally IR divergent solution, respectively:
\begin{align}
 -\frac{1}{2}\ka_{MAG}\leq\ka_{AB^2}\leq0,&\qquad -\frac{1}{2}\ka_{MAG}\leq\ka_{Ac^2}\leq0,\nnnl
 -\frac{3}{2}\ka_{MAG}\leq\ka_{BBB}\leq0,&\qquad -\frac{3}{2}\ka_{MAG}\leq\ka_{Bcc}\leq0.
\end{align}
If the IRE of the three-point functions was known, the IREs of the complete tower of vertices with an odd number of legs could be derived. Without this knowledge, which cannot be obtained from a power counting analysis, we have to introduce a parameter $\eta$ to take the different possibilities into account. The IRE of an n-point function with n odd reads then
\begin{align}
\ka(n_A,n_B,n_c)=\mhalfo(n_A-n_B-n_c+\eta)\ka_{MAG} \qquad (\text{$n_A+n_B+n_c$ odd)}.
\end{align}
For $SU(2)$ $\eta$ can be either $0$ or $1$, while for $SU(N>2)$ it can also be $2$ and $3$.

\subsection{Aspects of the infrared solution and comparison to Landau gauge}
\label{ssec:MAGSolution}

The IR solution for the MAG can be summarized as follows. The scaling relation reads
\begin{align}
\setlength{\fboxsep}{3mm}
\fbox{$ \displaystyle \ka_{MAG}:=-\de_A=\de_B=\de_c\geq0$}
\end{align}
and the IREs of vertices are given by
\begin{align}
\setlength{\fboxsep}{3mm}
\fbox{$ \displaystyle \ka(n_A,n_B,n_c)=\mhalfo(n_A-n_B-n_c)\ka_{MAG} \qquad \text{($n_A+n_B+n_c$ even)}$}
\end{align}
and
\begin{align}
\setlength{\fboxsep}{3mm}
\fbox{$ \displaystyle \ka(n_A,n_B,n_c)=\mhalfo(n_A-n_B-n_c+\eta)\ka_{MAG} \qquad (\text{$n_A+n_B+n_c$ odd)}.$}
\end{align}

One can see directly from the scaling relation that this solution is qualitatively quite different from the one in Landau gauge, where the ghosts are IR enhanced, as here it is the diagonal gluon propagator that is the dominant degree of freedom in the IR. This can be interpreted as a variant of Abelian IR dominance. Also in agreement with this hypothesis is the IR suppression of off-diagonal propagators. One should note, however, that it is different from the indication of Abelian IR dominance seen in lattice calculations of the MAG propagators where all propagators show a massive behavior. At momenta below the momentum scale induced by this mass the propagators decouple. Since the propagator of the diagonal gluon is least IR suppressed the diagonal degrees of freedom are dominant in the IR. Here, however, we find an IR enhanced diagonal gluon propagator and it is clear that in the DSEs/FRGEs the diagrams with the most diagonal propagators dominate.

There is even a direct connection between ghost dominance in the Landau gauge and Abelian dominance in the MAG. One can show that upon transforming gauge field configurations from the MAG to the Landau gauge that the diagonal (MAG-dominant) configurations are transformed to configurations lying on the Landau gauge Gribov horizon \cite{Greensite:2004ke}, which give rise to the IR enhancement of the ghost propagator.
This establishes not only a link between the dominant configurations of the Landau gauge and the MAG, but also between two confinement scenarios of different origin \cite{Greensite:2004ke}.

Besides the obvious difference that the dominant fields are different, there is also another aspect that distinguishes the MAG and the Landau gauge. The scaling relation in the latter is $0\leq2\ka_{LG}:=\de_A=-2\de_c$. A lower bound for the IREs of propagators is the value when the Fourier transforms of the propagators are no longer well-defined \cite{Lerche:2002ep,Fischer:2009tn}, i.e., here $\de_c\geq-1$ and $\de_A\geq-1$ which corresponds to propagators with a $1/p^4$ behavior. Hence we get
\begin{align}
 0\leq \ka_{LG}\leq 1.
\end{align}
Note, however, that more divergent propagators, specifically going like $1/p^6$, have been found also and interpreted via an analytic continuation of the exponent beyond $-2$ \cite{Zwanziger:2009je}. These bounds on $\ka_{LG}$ allow that the gluon propagator vanishes at zero momentum, for which we need $\de_A>1$, i.e., $\ka_{LG}>1/2$. In the MAG the upper bound is $\ka_{MAG}\leq1$. The condition for an IR vanishing propagator, on the other hand, leads to $\ka_{MAG}>1$. Thus the IR suppressed propagators in the MAG cannot vanish. On the contrary, they are still IR divergent, but less than the tree-level propagator. It should be stressed that also such propagators can violate positivity and belong to confined fields.

The reason for different IR dominant field configurations in the MAG and the Landau gauge is likely related to the different shapes of their Gribov regions. While both are convex and contain the origin, only the latter is bounded in all directions. The former, however, is unbounded in the direction of diagonal fields \cite{Capri:2010an,Capri:2008vk}. In this case the entropy argument that in a high-dimensional space most volume is contained at the boundary and thus configurations with a small value of the lowest eigenvalue of the Faddeev-Popov determinant are dominant does not apply \cite{Zwanziger:1989mf}. The unboundedness of the Abelian direction in field configuration space may be directly responsible for the enhancement of the diagonal propagator and the dominance of Abelian configurations.

The discrepancy presently observed between the results from this thesis and from lattice calculations \cite{Mendes:2008ux,Mendes:2006kc} may have its explanation in the existence of an additional solution for the MAG of the decoupling type which is also found in the refined Gribov-Zwanziger scenario \cite{Capri:2010an,Capri:2008ak}. Indeed one can see from the DSEs that a decoupling solution in the MAG features three massive propagators. The reason is that there are four-point interactions between all three fields ($AABB$, $AAcc$, $BBcc$) which lead to tadpole diagrams in the DSEs. If the internal propagator of a tadpole diagram is massive, it does not vanish in dimensional regularization and can render the solution of the DSE massive. In Landau gauge there exists only a gluon four-point interaction and thus a massive gluon propagator does not influence the ghost propagator via a tadpole diagram. In the MAG, however, a massive propagator automatically leads to the massiveness of the other two propagators by the tadpoles stemming from the vertices $AABB$, $AAcc$ and $BBcc$.

The recent suggestion that the decoupling and the scaling solutions in the Landau gauge correspond to different Landau gauges that only deviate from each other in the IR regime \cite{Maas:2009se} can also be considered in the MAG. Indeed its DSEs offer a similar setting. In the Landau gauge the realization of one of the two solutions is directly related to the ghost propagator at zero momentum. Only if the boundary condition is chosen such in the renormalization process that the ghost dressing function diverges in the IR, the scaling solution is obtained. In the MAG the same condition is required for the diagonal gluon propagator. Thus the zero momentum value of its dressing function may serve as a parameter in analogy to the B-parameter of the Landau-B gauges \cite{Maas:2009se}.

\subsection{Calculation of the parameter $\ka_{MAG}$}

The first step in establishing the existence of a scaling solution in the MAG was to find the scaling relation \eref{eq:MAGScalingRelation}. The next step would be a numerical solution of the DSEs over the complete momentum range. However, this is complicated mainly by the fact that the IR leading terms are two-loop diagrams. For a numerical solution the DSEs have to be truncated properly. This can be done in a consistent way if the IR leading diagrams only have one loop, as a one-loop truncation then contains the IR and UV leading parts. For two loops, however, a direct UV/IR consistent truncation does not exist: As some two-loop terms are required for the IR, one has to include all diagrams up to two loops in order to be consistent in the UV. Such a calculation would not only be of a tremendous complexity due to the high number of terms in the MAG, but is also aggravated by the fact that a new method for renormalizing and calculating the two-loop diagrams and ans\"atze for all involved vertices are required.

What can be done is the calculation of the IRE $\ka_{MAG}$ if we only take into account the sunset diagrams. It is not possible to include the squint diagrams as in this case we do not have enough equations for the determination of the free parameters. The free parameters are the coefficients in the power laws of the dressing functions in the IR, $d_A$, $d_B$, $d_c$, and the IRE $\ka_{MAG}$. So in total we have four unknowns and three equations. However, in the DSEs truncated to the sunsets the coefficients appear only in the combinations $d_A^2 d_B^2$ and $d_A^2 d_c^2$ and we can solve this system. If we included the squints not only would we need to know the IR behavior of the three-point functions, but also other combinations of the coefficients would appear rendering the system of equations unsolvable. Thus we assume here that only the sunset diagrams are IR leading.

For the calculation it suffices to determine the integrands in the IR and the system of DSEs reduces to the following simpler form after suitable projections in Lorentz space:
\begin{align}
 d_A^{-1}&=-X^A_{AABB}(p^2,\ka_{MAG}) d_A d_B^2 - X^A_{AAcc}(p^2,\ka_{MAG})  d_A d_c^2,\\
 d_B^{-1}&=-X^B_{AABB}(p^2,\ka_{MAG})  d_A^2 d_B,\\
 d_c^{-1}&=-X^c_{AAcc}(p^2,\ka_{MAG})  d_A^2 d_c.
\end{align}
The $X(p^2,\ka_{MAG})$ denote the sunset integrals without the coefficients from the propagator power laws. The superscript gives the corresponding DSE and the subscript the bare vertex contained in the diagram. Using the invariant combinations
\begin{align}
 I_1:=&d_A^2 d_B^2,\\
 I_2:=&d_A^2 d_c^2
\end{align}
the three equations can be combined to
\begin{align}\label{eq:eqForKappa}
 1= \frac{X^A_{AABB}(p^2,\ka_{MAG})}{ X^B_{AABB}(p^2,\ka_{MAG})} + \frac{X^A_{AAcc}(p^2,\ka_{MAG})}{X^c_{AAcc}(p^2,\ka_{MAG})}.
\end{align}
If the $X(p^2,\ka_{MAG})$ are known, this equation yields the solution(s) for $\ka_{MAG}$.

For the calculation of the $X(p^2,\ka_{MAG})$ we need to decide what we use for the dressed four-point functions. As we cannot solve their DSEs and get exact expressions we can only make ans\"atze. The simplest choice is using bare vertices, but we could use any expression that stays constant when all external momenta go to zero. However, the number of possibilities for modifying the dressed four-point functions is tremendous. First of all four-point functions have a large number of possible tensors: For two Lorentz indices ($AAcc$ vertex) there are ten possible tensors and for four Lorentz indices there are 138 ($AABB$ vertex). Of these five and 76, respectively, are transverse with respect to the diagonal gluon legs. In color space the situation is a little bit better, as the fact that two indices are diagonal allows only a subset of all possible color tensors (see Section \ref{ssec:2PointPropagators} for details). One could choose, for example, the following basis for $SU(N)$:
\begin{align}
 t^{abij}_1=\de^{ab}\de^{ij}, \qquad t^{abij}_2=f^{aic}f^{bjc}, \qquad t^{abij}_3=d_A^{abij},
\end{align}
where $d_A^{abij}$ is the totally antisymmetric tensor
\begin{align}
 d^{abij}_A=\frac1{6}\text{Tr}\left(T^a T^{(b} T^i T^{j)} \right).
\end{align}
The Jacobi identity tells us that another possible tensor, $f^{ajc}f^{bic}$, is related to $t^{abij}_2$ and hence does not appear in this basis.
For the calculation of $\ka_{MAG}$ we restrict ourselves to bare vertices.

Another thing we need is an analytic solution for the scalar integral
\begin{align}
 I_{SS}&(a,b, c,e,f;p^2):=\int \ddotp{q} \ddotp{r} (q^2)^{a}[(r)^2]^{b} [(p-q-r)^2]^{c}(2\,p\, q)^e (2\,p\, r)^f
\end{align}
for $\lbrace e,\,f\rbrace \in \mathbb{N}$.
As I am not aware of any existing solution in the literature except for $e=0$ I provide it in Appendix \ref{chp:sunset}.

The integrals $X(p^2,\ka_{MAG})$ can be calculated with \textit{Mathematica}. First the DSEs are derived with the package $\textit{DoDSE}$ as outlined in Appendix \ref{sec:DoDSEMAG}. Using the Feynman rules given in Appendix \ref{chp:MAGFeynmanRules} the detailed expressions of the integrals are derived. They are then processed by contracting Lorentz and color indices and calculated using \eref{eq:sunsetSol}. The contraction of color indices was done in \textit{Mathematica} by simple replacement rules, as no available program could deal directly with the splitting of the color algebra. The whole process is necessarily automated, because the resulting expressions cover several pages even in the simple case adopted here. This also allows to test the effect of different tensors or employing different dressings.

\begin{figure}[t]
\begin{center}
\includegraphics[width=0.49\textwidth]{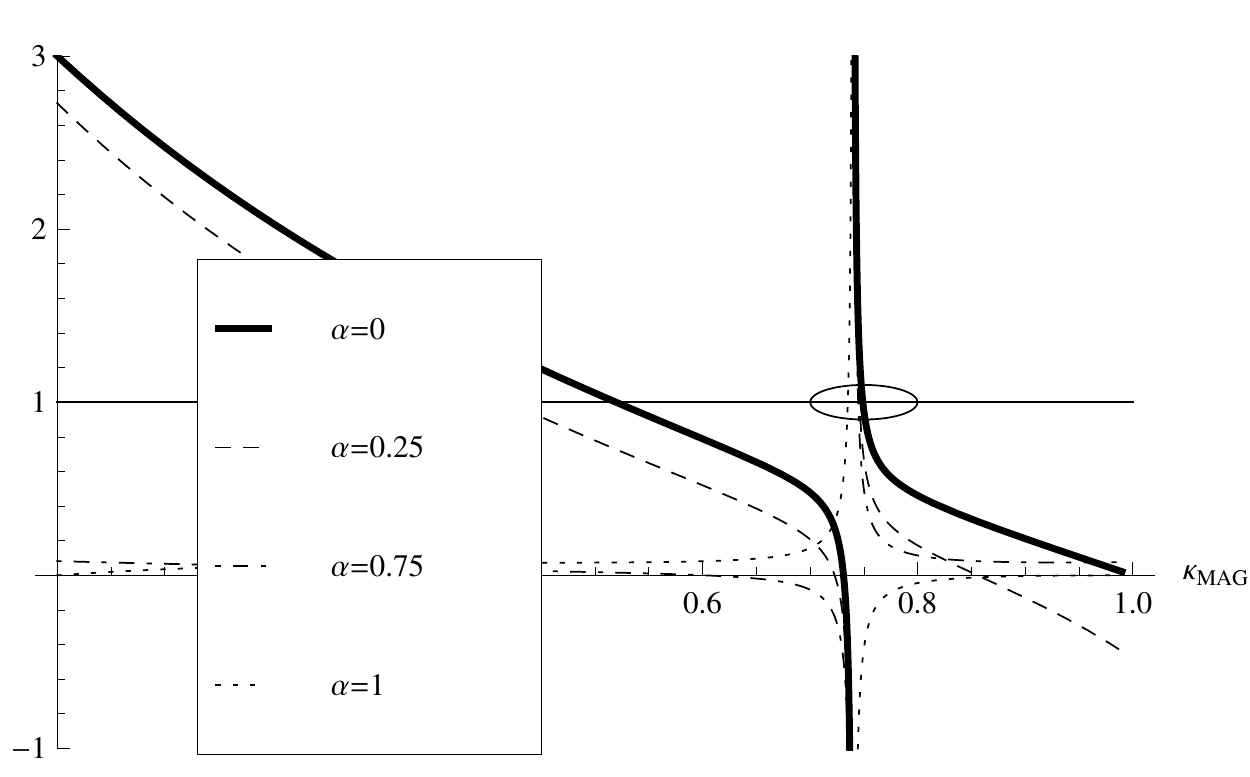}
\includegraphics[width=0.49\textwidth]{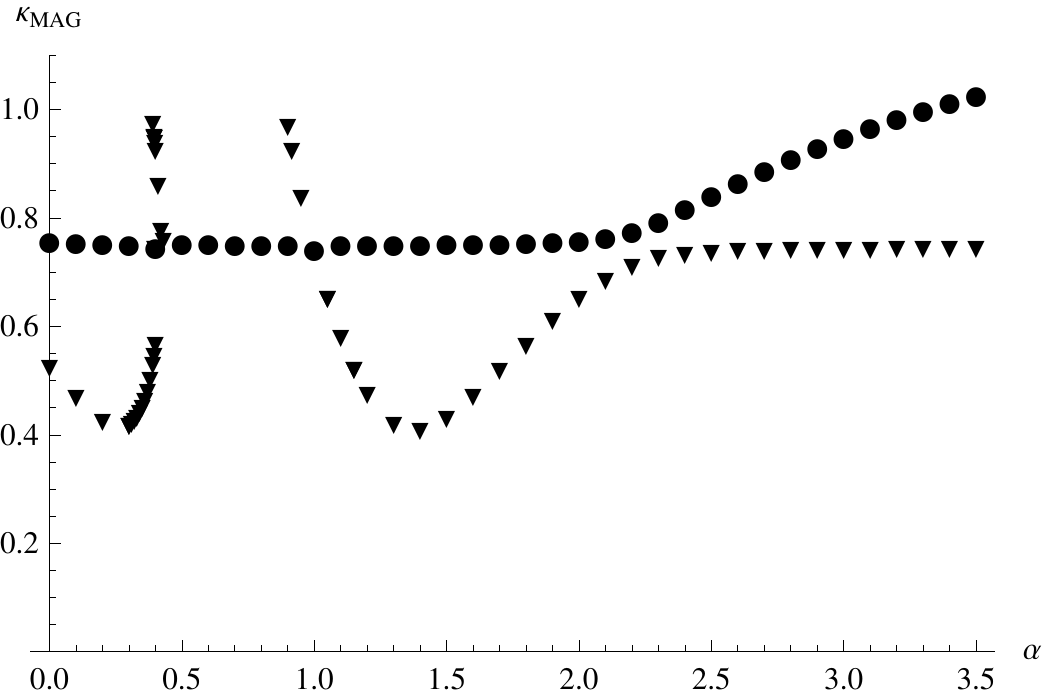}
\caption{\label{fig:kappaMAG}Left: Both sides of \eref{eq:eqForKappa} for several values of $\alpha$. Right: Dependence of $\ka_{MAG}$ on $\al$. The dots and triangles represent the first and second solution branches, respectively.}
\end{center}
\end{figure}

Having all analytic expressions for the sunset integrals $X(p^2,\ka_{MAG})$, we can plot the right-hand side of \eref{eq:eqForKappa} as a function of $\ka_{MAG}$. There is still one unfixed value which is the gauge fixing parameter $\al$. The obvious choice is $\al=0$, which corresponds to the pure MAG, but it is also interesting to check the dependence on $\al$. In \fref{fig:kappaMAG} on the left the two sides of \eref{eq:eqForKappa} are plotted for several values of $\al$. Solutions for $\ka_{MAG}$ can be read off where the straight line (left-hand side of \eref{eq:eqForKappa}) crosses the curves (right-hand side of \eref{eq:eqForKappa}). The singularity at $\ka_{MAG}=0.739908$ seems to be universal for all values of $\al$; this was checked up to $\al=100$. Consequently there is always one solution $\ka_{MAG}\approx 0.74$. However, as is depicted in \eref{eq:eqForKappa} on the right, there exists a second branch of solutions. For $\al$ between approximately $0.5$ and $0.9$ both coincide and when at $\al\approx 2.2$ the first branch starts to go away from $\ka_{MAG}\approx 0.74$, the first branch takes over. The first branch crosses $\ka_{MAG}=1$ at $\al=3.359$ and becomes thus unphysical.

As there is a solution for the parameter $\ka_{MAG}$ we have one more indication that the obtained IR solution really exists. The final proof would be a complete numerical calculation for all momenta.
\chapter{Dyson-Schwinger equations in the  Gribov region}
\label{chp:GZ}

In Chapter \ref{chp:IRYM} I explained how to fix the gauge according to the conventional Faddeev-Popov procedure and that for non-perturbative calculations this is not sufficient due to the appearance of further gauge copies. Now I will go into the details of the improved gauge fixing suggested by Gribov and how this leads to a local Lagrangian amenable to the usual tools of quantum field theory. The resulting action is called Gribov-Zwanziger action \cite{Gribov:1977wm,Zwanziger:1989mf} and constitutes the best option available so far to achieve a complete gauge fixing within the path integral. For lattice simulations the situation is slightly different and I will comment on this in more detail in Section \ref{ssec:GribovRegionOnLattice}. Of course the ultimate goal would be a complete gauge fixing as, for example, provided by restriction to the fundamental modular region. This, however, does not seem realizable directly within the path integral due to the non-trivial topology of this region.

One should note that the underlying idea of Gribov \cite{Gribov:1977wm} works for all gauges which are defined by a minimization problem like Landau gauge, Coulomb gauge or the maximally Abelian gauge. The first two were already considered in Gribov's original paper \cite{Gribov:1977wm}, while the last one was mainly investigated by Sorella and collaborators, see, e.g., refs. \cite{Capri:2005tj,Capri:2006cz,Capri:2010an,Capri:2008ak}. In this chapter, however, I only consider the Landau gauge.

The Gribov-Zwanziger action is described in Section \ref{sec:GZAction} and will be investigated in Section \ref{sec:IRGZ} with the tools developed in Chapter \ref{chp:scalingSolutions}. In this action fields mix at the level of two-point functions what leads to several complications that require a refinement of the method as described in Section \ref{sec:mixed2Point}. When the smoke clears two possible IR solutions remain, which, however, are very similar up to details. In particular, both feature the same qualitative behavior as the result obtained from the Faddeev-Popov theory. Interestingly one solution reduces in the IR exactly to the Faddeev-Popov theory. The results of this chapter have been published in ref. \cite{Huber:2009tx}.

\section{The Gribov-Zwanziger action}
\label{sec:GZAction}

This section contains a short overview of the derivation of the local formulation of the Gribov-Zwanziger action. It is named after N. Gribov, who was the first to suggest a solution to overcome the incomplete gauge fixing, and D. Zwanziger, who subsequently brought this idea into its final form in terms of a local action and derived several properties of the first Gribov region.

\subsection{Restriction to the first Gribov region}

As mentioned in Section \ref{chp:IRYM} the first Gribov region $\Omega$ is defined as the set of gauge field configurations for which the Faddeev-Popov operator is positive and which fulfill the gauge fixing condition, i.e., for Landau gauge
\begin{align}
 \Omega:= \{A\, | \, \partial_\mu A_\mu=0, \, M >0 \}
\end{align}
with
$M=-\partial_\mu D_\mu$ the Faddeev-Popov operator. The boundary of this region is called the (first) Gribov horizon. It is defined by those gauge field configurations for which the lowest non-trivial eigenvalue of the Faddeev-Popov operator vanishes. Consequently the second Gribov horizon is where the second eigenvalue vanishes and so on.
The conventional gauge fixing already restricts the field configurations to fields obeying the Landau gauge condition $\partial_\mu A_\mu=0$ and only the second condition, the positivity of the Faddeev-Popov operator, has to be implemented additionally.

The Faddeev-Popov operator is related to the ghost propagator $D^{ab}_{\bar{c}c}(k)$ by
\begin{align}
 D^{ab}_{\bar{c}c}(k)=-\frac{\delta^{ab}}{k^2}c_{\bar{c}c}(k)=-( M^{-1} )^{ab}.
\end{align}
Hence in order to be within the first Gribov region the ghost propagator dressing function $c_{\bar{c}c}(k)$ must be positive. Gribov parametrized the ghost propagator by \cite{Gribov:1977wm}
\begin{align}
 D^{ab}_{\bar{c}c}(k)=-\frac{\delta^{ab}}{k^2}\frac1{1-\sigma(k,A)}.
\end{align}
As $\sigma(k,A)$ increases with $k$ decreasing, it is sufficient to demand that $\sigma(0,A)<1$. This is known as the no-pole condition \cite{Gribov:1977wm}: When the ghost form factor $\sigma(0,A)$ becomes one, the Faddeev-Popov operator has a zero eigenvalue and one is directly on a horizon. $\sigma(0,A)$ is calculated as a series in $A$, which can be regarded as an external gluon field here. The no-pole condition is enforced in the path integral via a Heaviside functional:
\begin{align}
 \theta(1-\sigma(0,A))=\int_{-\infty}^{\infty} \frac{d\beta}{2\pi\,i}\frac{e^{i\,\beta(1-\sigma(0,A))}}{\beta-i\,\epsilon}.
\end{align}
The integral over $\beta$ can be evaluated with the saddle-point method. It yields a value for $\beta\rightarrow \gamma^4$ which takes the role of a new mass parameter. $\gamma$ is called the Gribov parameter.

The positivity of the Faddeev-Popov operator means that all its eigenvalues $\lambda_n$ are positive. Relaxing this condition to $tr(\lambda_n)>0$ allowed Zwanziger to derive a closed expression, the so-called horizon function, that restricts the integration to the Gribov region \cite{Zwanziger:1989mf}:
\begin{align}\label{eq:hor-func3}
 h(x)=\underset{\gamma(x)\rightarrow \gamma}{\lim}\int dy \big(D^{ac}_\mu(x)\gamma^2(x)\big)\left( M^{-1}\right )^{ab}(x,y) \big(D^{bc}_\mu(y) \gamma^2(y)\big).
\end{align}
The limit $\gamma(x)\rightarrow \gamma$ can only be taken after localizing the action. Details on this can be found in ref. \cite{Dudal:2010fq}.
The horizon function corresponds to a generalization of the ghost form factor $\sigma(0,A)$ to all orders. This has explicitly been worked out up to third order \cite{Gomez:2009tj}.

The condition for being within the first Gribov region is phrased as the horizon condition:
\begin{align}
 \int dx \, h(x) < d\, \gamma^4 (N^2-1) V,
\end{align}
where $d$, $N$ and $V$ are the number of dimensions, the number of colors and the space-time volume, respectively. Again this condition can be enforced via a Heaviside functional. Details on how to get from the step function to an additional term in the action can be found in ref. \cite{Zwanziger:2009je}. An interesting observation is that the resulting path integral has similarities with the partition function of a canonical ensemble which is equivalent to the microcanonical ensemble in the thermodynamic limit. Thus the Heaviside functional can be replaced by a Dirac delta functional here, i.e., the horizon condition becomes
\begin{align}
 \int dx \, h(x) = d\, \gamma^4 (N^2-1) V.
\end{align}
This amounts to restricting the integration to configurations lying directly at the Gribov horizon.

The resulting action that enforces the restriction to the first Gribov region is known as Gribov-Zwanziger action given by
\begin{align}
 S^{non-local}=S_{FP}+S_h,
\end{align}
where
\begin{align}
 S_{FP}&=S_{YM}+S_{gf},\\
 S_h&=\int dx\,(h(x)-\gamma^4 d (N^2-1)),\\
 S_{YM}&=\frac{1}{4} \int dx F_{\mu \nu }^rF^r_{\mu \nu},\\
 S_{gf}&=\int dx\,\left( i\,b^r (\partial_\mu A^r_\mu)+\bar{c}^r\,M^{rs}\,c^s\right).
\end{align}
The trivial term $\gamma^4 d (N^2-1)$ is introduced for later convenience.
Since $S_h$ is non-local the standard tools of quantum field theory can not be employed. However, it is possible to localize it as discussed in the next section.

\subsection{The local action} 
\label{ssec:GZAction}

In order to localize the horizon function we need two pairs of additional fields: $\vp_\mu^{ab}$ and $\vpb_\mu^{ab}$, which are complex conjugate to each other, and $\omega_\mu^{ab}$ and $\bar{\omega}_\mu^{ab}$, which are also complex conjugate to each other but Grassmann fields. These new fields are BRST doublets:
\begin{align}
 s\,\vp^{ab}_\mu=\omega^{ab}_\mu, \qquad s\,\omega^{ab}_\mu=0,\\
 s\,\bar{\omega}^{ab}_\mu=\bar{\vp}^{ab}_\mu, \qquad s\,\bar{\vp}^{ab}_\mu=0.
\end{align}
The three indices are one Lorentz and two color indices in the adjoint representation, i.e., $\mu=1,\ldots,d$ and $a,b=1,\ldots,N^2-1$.
The bosonic fields are used for localizing the non-local term $S_h$ and the fermionic fields for canceling the determinant arising from that procedure.
Using the Gaussian integration formulas
\begin{align}
 \int \mathcal{D}[\bar{\vp}\vp] e^{-\int dx\, \left(\bar{\vp}_\mu^{ac}M^{ab}\vp_\mu^{bc} + \gamma^2\, g\,f^{abc}A_\mu^a(\vp_\mu^{bc}-\vpb_\mu^{bc}) \right)}=&(\det\,M)^{-d(N^2-1)}e^{\left( -\gamma^4\,g^2\, f^{ace}A_\mu^a (M^{-1})^{cd}f^{bde}A_\mu^b \right) },\\
 \int \mathcal{D}[\bar{\omega}\omega] e^{ \int dx \,\bar{\omega}_\mu^{ac}M\omega_\mu^{bc}}=&(\det\,M)^{d(N^2-1)}
\end{align}
one can show
\begin{align}\label{eq:Sh-localized}
 e^{-S_h}=\int& \mathcal{D}[\bar{\vp}\vp\bar{\omega}\omega] e^{-\int dx\, \left(\bar{\vp}_\mu^{ac}M^{ab}\vp_\mu^{bc} + \gamma^2 \,g\,f^{abc}A_\mu^a(\vp_\mu^{bc}-\vpb_\mu^{bc})-\bar{\omega}_\mu^{ac}M\omega_\mu^{bc} -d\,\gamma^4\, (N^2-1) \right)}.
\end{align}
The local Gribov-Zwanziger action is then given by
\begin{align}\label{eq:SGZ-local}
 S^{local}=&S_{FP}+S_{GZ},\\
 S_{GZ}=&\int dx\, \left(\bar{\vp}_\mu^{ac}M^{ab}\vp_\mu^{bc} + \gamma^2 \,g\,f^{abc}A_\mu^a(\vp_\mu^{bc}-\vpb_\mu^{bc})-\bar{\omega}_\mu^{ac}M\omega_\mu^{bc} -d\,\gamma^4\, (N^2-1) \right) \label{eq:SGZ-local2}.
\end{align}

Note that a slightly different version is sometimes used that contains an additional term \cite{Zwanziger:1992qr}
\begin{align}
S_{\Delta_{BRST}}=\int dx\,g\,f^{abe}\bar{\omega}^{ac}_\mu \partial_\nu \left[ (D^{ed}_\nu c^d ) \vp^{bc}_\mu \right].
\end{align}
It can be obtained by a shift of the $\omega$ field,
\begin{align}
 \omega_\mu^{ab}(x)\rightarrow \omega_\mu^{ab}(x) - g \int dy\, (M^{-1})^{ac}(x,y)f^{cde}\partial_\nu\left[(D_\nu^{df}c^f(y))\vp_\mu^{eb}(y)\right],
\end{align}
and has the advantage that the remaining BRST symmetry is more manifest, since almost all terms in \eref{eq:SGZ-local2} are BRST exact:
\begin{align}
 s(\bar{\omega}^{ac}_\mu M^{ab} \vp^{bc}_\mu)=\bar{\vp}^{ac}_\mu M^{ab} \vp^{bc}_\mu-\bar{\omega}^{ac}_\mu M^{ab} \omega^{bc}_\mu+g\,f^{abe}\bar{\omega}^{ac}_\mu \partial_\nu \left[ (D^{ed}_\nu c^d ) \vp^{bc}_\mu \right].
\end{align}
This property is useful in the process of renormalization \cite{Zwanziger:1992qr}. The additional vertex derived from $S_{\Delta_{BRST}}$ does not influence any calculations as it features the fields $c$ and $\omega$, but there is no corresponding term with $\bar{c}$ and $\bar{\omega}$. Thus this vertex does not appear in any diagrams.

The only term in the action $S^{local}+S_{\Delta_{BRST}}$ which is not BRST invariant is the one proportional to $\gamma^2$:
\begin{align}
 s\left(\gamma^2 \,g\,f^{abc}A_\mu^a(\vp_\mu^{bc}-\vpb_\mu^{bc})\right)=\gamma^2 \,g\,f^{abc}\left((-D_\mu^{ad}c^d)(\vp_\mu^{bc}-\vpb_\mu^{bc})+A_\mu^a \omega^{bc}_\mu \right).
\end{align}
This is a soft breaking, i.e., only of mass dimension two.\footnote{The ghost fields $c$ and $\bar{c}$ have mass dimensions zero and two, respectively \cite{Zwanziger:1992qr}.} Hence one can show the renormalizability of the Gribov-Zwanziger action by embedding it into a larger BRST invariant action, for which the usual tools of algebraic renormalization apply \cite{Piguet:1995er}. The complete proof of renormalizability can be found in ref. \cite{Dudal:2010fq}. The original literature is refs. \cite{Zwanziger:1992qr,Maggiore:1993wq,Dudal:2005na}.

Another consequence of the BRST breaking is that physical operators can no longer be determined by their cohomology, because BRST exact quantities, with which a physical operator mixes in the renormalization process, can now influence expectation values. The complete form of the operator $F_{\mu\nu}^2$ under this mixing was determined in ref. \cite{Dudal:2009zh} for the Gribov-Zwanziger Lagrangian. Another approach was taken in ref. \cite{Baulieu:2009ha}, where operators with a physical cut structure were constructed by adding BRST exact terms. The precise meaning of the BRST breaking, however, remains to be clarified further.

In terms of the new fields the horizon condition becomes
\begin{align}
 \langle g\, f^{abc}A_\mu^a(\vp_\mu^{bc}-\vpb_\mu^{bc}) \rangle = 2\,d\,\gamma^2\, (N^2-1)
\end{align}
as can be inferred by differentiation of \eref{eq:Sh-localized} with respect to $\gamma^2$. Often this is phrased in terms of the gap equation
\begin{align}
 \frac{\partial \Gamma[0]}{\partial \gamma^2}=0,
\end{align}
where $\Gamma[0]$ is the vacuum energy defined by
\begin{align}
 e^{-\Gamma[0]}=\int \mathcal{D}[A\bar{c}c\vpb \vp \bar{\omega}\omega]e^{-S^{local}}.
\end{align}
Writing the horizon condition in this form a spurious solution for the value of the Gribov parameter arises, namely $\gamma=0$. As this corresponds to the starting action, it has to be discarded \cite{Dudal:2008sp}.

One point which has been stressed repeatedly in the literature is that the Gribov-Zwanziger action is tightly connected to the gap equation as was already pointed out by Gribov \cite{Gribov:1977wm}. Only if $\gamma$ takes the value as determined by the horizon condition the Gribov-Zwanziger Lagrangian corresponds to a gauge theory. It was pointed out in ref. \cite{Zwanziger:2010iz} that the gap equation breaks several symmetries of the original effective action $\Gamma$. It is argued that this symmetry breaking is spontaneous and the related Goldstone particles are the fields $\bar{c}$, $c$, $\bar{\omega}$, $\omega$ and parts of $\bar{\vp}$ and $\vp$. Their propagators are all found to be IR enhanced like $1/k^4$.

It is advantageous for the present task to rewrite the local Gribov-Zwanziger action given in \eref{eq:SGZ-local} by splitting the bosonic auxiliary fields into real and imaginary parts $U$ and $V$, respectively, as done in ref. \cite{Zwanziger:2009je}:
\begin{align}
S^{local}&=S_{FP}+S'_{GZ},\\
S'_{GZ}&=S_U+S_V+S_{UV}-\bar{\omega}M\omega,\\
S_U&=\frac1{2} \int dx \, U_\mu^{ac}\,M^{ab}\,U_\mu^{bc},\\
S_V&=\frac1{2} \int dx \,  V_\mu^{ac}\,M^{ab}\,V_\mu^{bc} +i\, g\,\gamma^2 \sqrt{2} f^{abc}\int dx \,  A_\mu^a V_\mu^{bc},\\
S_{UV}&=\frac1{2}i\,g f^{abc}\int dx \,  U_\mu^{ad} V_\mu^{bd} \partial_\nu A_\nu^c,
\end{align}
where $U$ and $V$ are defined by
\begin{align}
\varphi=\frac1{\sqrt{2}}\left( U+i\,V\right), \quad \bar{\varphi}=\frac1{\sqrt{2}}\left( U-i\,V\right).
\end{align}
When the Landau gauge condition $\partial_\mu A_\mu=0$ is enforced, $\mathcal{L}_{UV}$ vanishes and the only mixing on the level of two-point functions is between the gluon field $A$ and the imaginary part of the bosonic auxiliary field $V$, whereas the $U$ field does not mix. This splitting simplifies calculations, because we only have to deal with a two-by-two matrix instead a three-by-three matrix for the mixing. Perturbative calculations for the propagators derived from this Lagrangian can be found in ref. \cite{Gracey:2009mj}, where the $U$ and $V$ fields are denoted by $\rho$ and $\xi$, respectively.

A further simplification is achieved by combining the Faddeev-Popov ghosts $c$ and $\bar{c}$, the fermionic auxiliary fields $\omega$ and $\bar{\omega}$ and the real part of the bosonic auxiliary field $U$ into one single field. This is possible, since all of them only interact with the gluon field via the Faddeev-Popov operator and are quadratic in the action. Hence they can be integrated out in the path integral:
\begin{align}
\int \mathcal{D}[\bar{c}c] e^{\bar{c}\,M\,c}&=det\,M,\\
\int \mathcal{D}[\bar{\omega}\omega] e^{\bar{\omega}\,M\,\omega}&=(det\,M)^{d(N^2-1)},\\
\int \mathcal{D}[U] e^{-\frac1{2}U\,M \,U}&=(det\,M)^{-\frac{d}{2}(N^2-1)}.
\end{align}
The different exponents of the determinant of the Faddeev-Popov operator are due to the different numbers of degrees of freedom. For $\gamma=0$ also the $V$ field can be integrated out and all determinants from auxiliary fields cancel so that the original Faddeev-Popov Lagrangian is recovered as required. For the present purpose we can treat all these non-mixing fields as new fermionic fields $\eta$ and $\bar{\eta}$ with the appropriate number of degrees of freedom by localizing the resulting determinant again.
The field $V$ cannot be included due to its mixing with the gluon field and therefore the two fields $V$ and $\eta$ can have a different infrared behavior. An overview of the different fields is given in tab.~\ref{tab:ghosts}. The final action reads
\begin{align}\label{eq:final-GZ-action}
S&=\int dx \Bigg(\frac1{4}F^a_{\mu\nu}F^a_{\mu\nu}+\frac{1}{2\xi}(\partial_\mu A_\mu)^2-\bar{\eta}^{a}_c \,M^{ab}\, \eta^{b}_c+\nnnl
 &+\frac1{2} V_\mu^{ac}\,M^{ab}\,V_\mu^{bc} +i\, g\,\gamma^2 \sqrt{2} f^{abc} A_\mu^a V_\mu^{bc}+\frac1{2}i\,g f^{abc} U_\mu^{ad} V_\mu^{bd} \partial_\nu A_\nu^c\Bigg),
\end{align}
where the subscript index of the new ghost fields $\eta$ and $\bar{\eta}$ runs from $1$ to $\frac{d}{2}(N^2-1)+1$ and the superscript index is the usual color index, running from $1$ to $N^2-1$. Note that for odd dimensions and even $N$ this number is half-integer and therefore this transformation is not directly possible. However, we can consider only integer values and perform an analytic continuation to half-integer values, if necessary. Alternatively one keeps the original fields separated and will get the appropriate numerical factors in front of diagrams.

A further diagonalization of the Lagrangian would require a further splitting of the fields in color space due to the different number of color indices of the $A$ and $V$ fields. The new fields would mean a significant complication of the Lagrangian at the level of vertices although at the two-point level it becomes simpler. Such a diagonalization was performed in ref. \cite{Baulieu:2009ha} and in a slightly simpler form in ref. \cite{Sorella:2010fs} and leads to fields with complex masses.
Due to the resulting structure of the vertices of the diagonalized Lagrangian we continue with the expression given in \eref{eq:final-GZ-action}.

\begin{table}[t]
\begin{center}
\begin{tabular}{l|c|c}
Field & Number of degrees of freedom & Statistics \\
\hline
\hline
$c$, $\bar{c}$ & $1$ & fermionic \\
$\omega$, $\bar{\omega}$ & $d(N^2-1)$ & fermionic \\
$\phi$, $\bar{\phi}$ & $d(N^2-1)$ & bosonic \\
\hline
$U$ & $d/2(N^2-1)$ & bosonic \\
$V$ & $d/2(N^2-1)$ & bosonic \\
$\eta$, $\bar{\eta}$ & $d/2(N^2-1)+1$ & fermionic
\end{tabular}
\end{center}
\caption{\label{tab:ghosts} The numbers of degrees of freedom and the statistics of the Faddeev-Popov ghosts $c$ and $\bar{c}$, the original auxiliary fields $\omega$, $\bar{\omega}$, $\phi$ and $\bar{\phi}$, the bosonic auxiliary fields $U$ and $V$ and the fermionic fields $\eta$ and $\bar{\eta}$. A factor $N^2-1$ from the adjoint index common to all fields is not taken into account here.}
\end{table}

\subsection{The first Gribov region on the lattice}
\label{ssec:GribovRegionOnLattice}

The restriction to the first Gribov region works differently on the lattice. In fact, the usual gauge fixing algorithms automatically lead to the first Gribov region as they are based on the minimization of a functional corresponding to \eref{eq:R}. Consequently it is not possible to fix with such a method to another Gribov region as the first one. Configurations just beyond the Gribov horizon can be obtained due to limitations in the numerical accuracy as was demonstrated in ref. \cite{Pawlowski:2009iv}, where in addition also an alternative approach for gauge fixing, stochastic quantization, was employed.

Since there are several minima along a gauge orbit, it depends on the algorithm and its parameters which minima are chosen. This leads to the definition of several Landau gauges. The minimal Landau gauge takes the first minimum found by the minimization algorithm, see, e.g., refs. \cite{Zwanziger:1993dh,Cucchieri:1995pn}. The absolute Landau gauge is defined by taking those gauge field configurations that correspond to the global minimum of \eref{eq:R} and thus to the fundamental modular region. This is a hard computational problem similar to spin-glass problems and can only be realized approximately. Calculations where this was attempted can be found, for example, in refs. \cite{Cucchieri:1997dx,Maas:2008ri,Bogolubsky:2009dc,Bornyakov:2008yx}.

Of course this is not the only possible prescription to choose a unique representative of each gauge orbit as in principle every choice is allowed.
A class of recently suggested gauges are the Landau-B gauges \cite{Maas:2009se}. Instead of using the value of the minimizing functional, which is equivalent to the trace of the gluon propagator, as a criterion to decide which gauge copy is used, one takes the value of the ghost propagator at the lowest available lattice momentum. This value is denoted by B. Using B instead of the trace of the gluon propagator is in the spirit of functional equations, where the value of the ghost dressing function at vanishing momentum serves as a boundary condition for the system of equations. Two examples of Landau-B gauges are the min-B and the max-B gauges, where the lowest and highest values of B are chosen, respectively. It is possible that the max-B gauge corresponds to the scaling solution \cite{Maas:2009se}, but currently only data for small lattices in three dimensions \cite{Maas:2009se} and in the strong coupling limit \cite{Maas:2009ph} are available.

\subsection{The Dyson-Schwinger equations of the Gribov-Zwanziger action}

A manual derivation of the Dyson-Schwinger equations becomes quite tedious because the mixed propagator leads to many additional terms. For example, already at the perturbative level an $AAV$ vertex appears at one loop. For the $\eta$ field things stay simple as their number is conserved. This is due to invariance under the same scale transformation that guarantees ghost number conservation in the Faddeev-Popov theory \cite{Nakanishi:1990qm}:
\begin{align}
 \eta \rightarrow& \eta e^{\theta},\\
 \bar{\eta} \rightarrow& \bar{\eta}e^{-\theta}.
\end{align}
Thus in a sense the $\eta$ fields behave similar as the Faddeev-Popov ghosts. However, one should not forget that they couple with the $V$ field, e.g., via the (dressed) $V\bar{\eta}\eta$ vertex.

\begin{figure}
\begin{center}
\includegraphics[width=\textwidth]{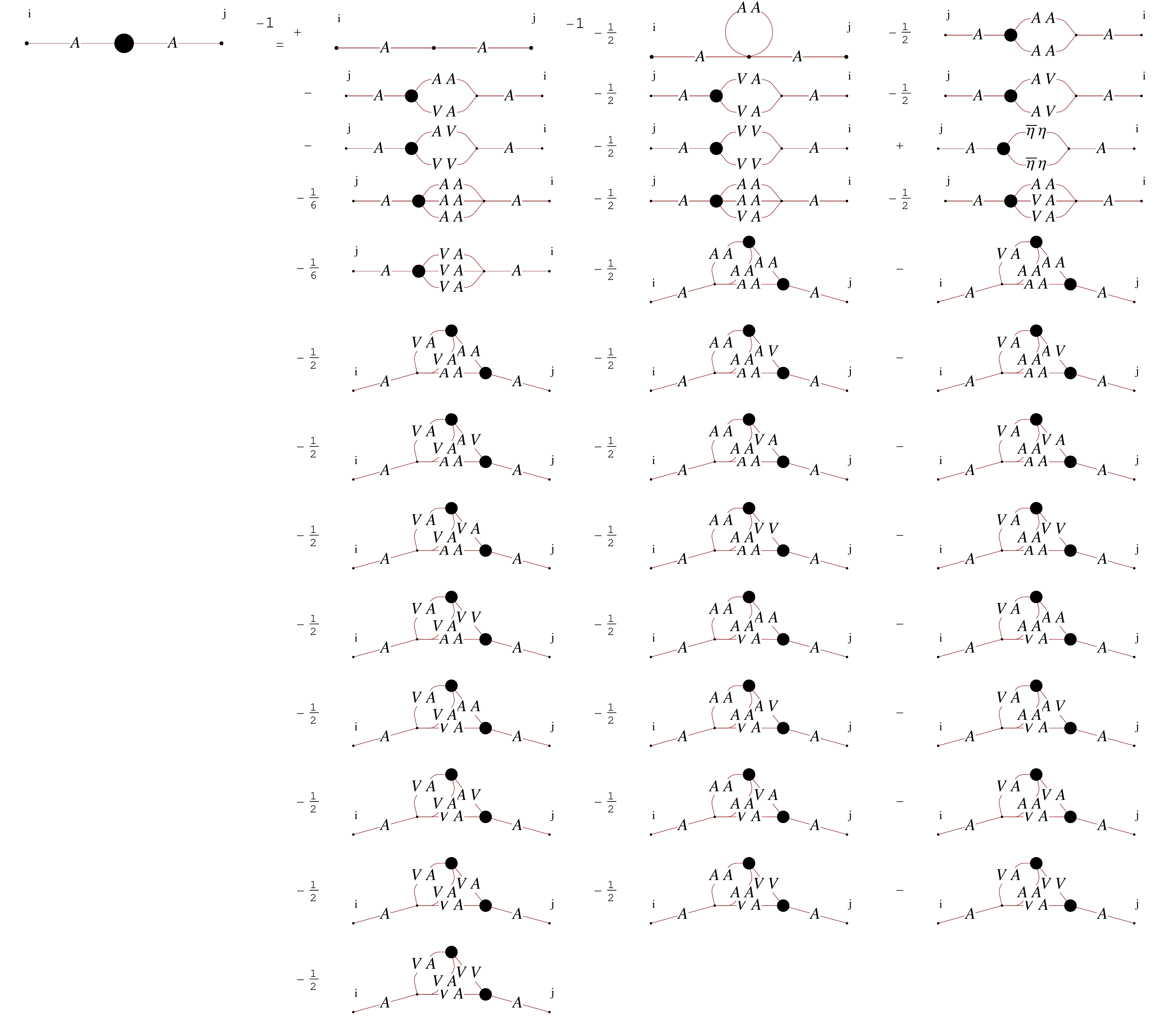}
\end{center}
\caption{\label{fig:AA-prop-DSEs}The DSE of the gluon two-point function. A propagator with the exponent $-1$ denotes the two-point function.
This convention, strictly speaking being mathematically incorrect, is chosen for the purpose of diagrammatic representation only. 
The propagators and vertices are labeled by the respective fields. The indices $i$ and $j$ denote the first and second fields of the depicted two-point function DSE. Internal propagators are all dressed. Bare and dressed n-point functions are denoted by thin and thick blobs, respectively.
}
\end{figure}

The non-trivial relation between propagators and two-point functions in the Gribov-Zwanziger theory is not an obstacle for the derivation of the DSEs, as dressed two-point functions only appear on the left-hand sides of their own DSEs and the bare counterparts on the right-hand sides.
It is again advantageous to employ the \textit{Mathematica} package \textit{DoDSE} which can also handle mixed propagators.
The resulting DSEs are given diagrammatically in figs.~\ref{fig:AA-prop-DSEs}, \ref{fig:VV-prop-DSEs}, \ref{fig:VA-prop-DSEs} and \ref{fig:etaeta-prop-DSEs}. Their derivation is described in Appendix \ref{sec:DoDSEGZ}.

\begin{figure}
\begin{center}
\includegraphics[width=\textwidth]{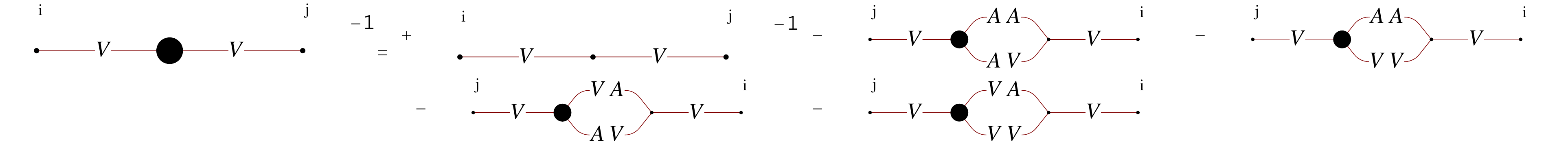}
\end{center}
\caption{\label{fig:VV-prop-DSEs}The DSE of the $V$ field two-point function.}
\end{figure}

\begin{figure}
\begin{center}
\includegraphics[width=\textwidth]{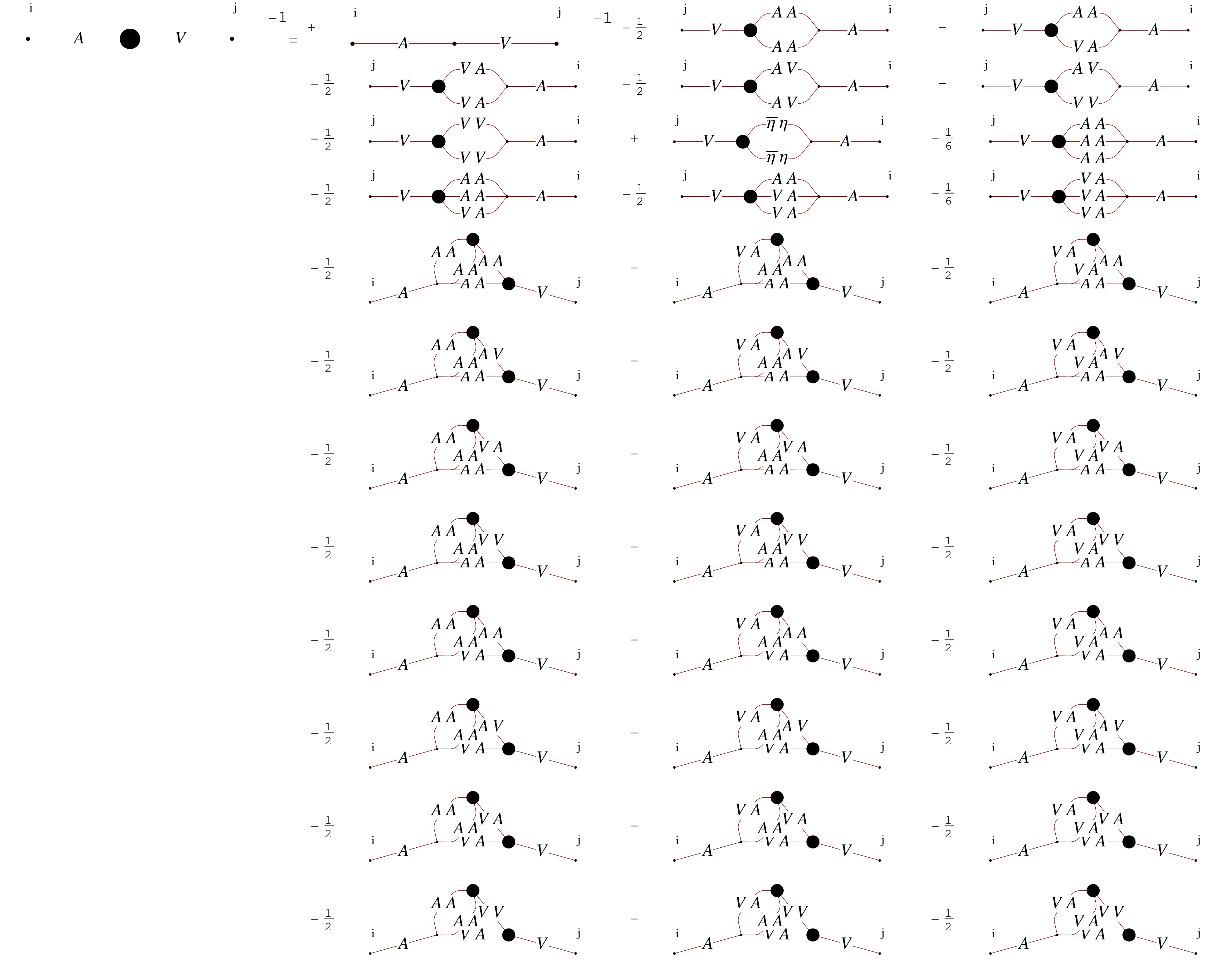}\\
\vskip10mm
\includegraphics[width=\textwidth]{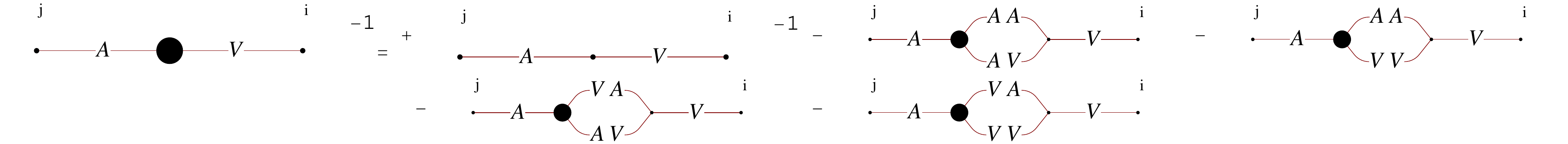}
\end{center}
\caption{\label{fig:VA-prop-DSEs}The DSEs of the $A$-$V$-mixed two-point function. The first is the $AV$ DSE and the second one the $VA$ DSE.}
\end{figure}

\begin{figure}
\begin{center}
\includegraphics[width=\textwidth]{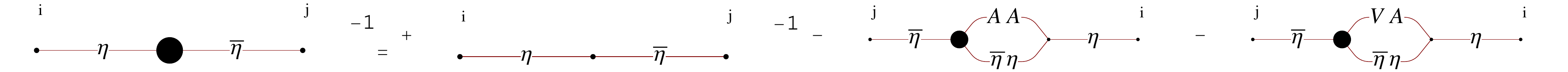}
\end{center}
\caption{\label{fig:etaeta-prop-DSEs}The DSE of the $\eta$ field two-point function.}
\end{figure}

\section{Infrared analysis of the Gribov-Zwanziger action}
\label{sec:IRGZ}

As already mentioned the matrix relation between propagators and two-point functions complicates the analysis as it yields non-trivial relations between the two sets of quantities. In this section the details of this relation are worked out. It is found that four possible scenarios for the IR behavior emerge which are discussed in Section \ref{ssec:IRPropagators}. Two of these scenarios can be discarded as they turn out to be inconsistent and the remaining two lead to the same qualitative IR behavior of the propagators as in the Faddeev-Popov theory.

\subsection{Two-point functions and propagators}
\label{ssec:2PointPropagators}

In addition to the complication due to the mixing at the two-point level the IR analysis is complicated by the tensor structure of the propagators/two-point functions that has to be considered explicitly as different dressing functions can have a different momentum dependence. The combination of mixing and several dressing functions leads to an even richer structure, as the matrix inversion counterintuitively can lead to terms where the determinant does not appear.

I start by discussing the color structure.
The gluon correlation function has only two color indices. Thus the only color tensor is $\de^{ab}$. I exclude here explicitly the occurrence of an antisymmetric tensor $\epsilon^{ab}$ as there is no indication of such a tensor neither in lattice simulations nor in other calculations. The $A$-$V$ mixed correlation function has three color indices. Excluding here the totally symmetric tensor $d^{abc}$, which cannot appear directly from any diagram, only the totally antisymmetric tensor $f^{abc}$ remains.
The $VV$ correlation function poses the greatest challenge as it has four color indices. This allows in the most general case six different color tensors \cite{vanRitbergen:1998pn,Cvitanovic:2008}:
\begin{align}
 {\de^{ab}\de^{cd},\, \de^{ac}\de^{bd}, \, \de^{ad}\de^{bc}, \, f^{abe}f^{cde}, \, f^{ace}f^{bde}, d^{abcd}_A}.
\end{align}
The last tensor is the totally antisymmetric tensor of rank four given by
\begin{align}
 d^{abcd}_A=\frac1{6}\text{Tr}\left(T^a T^{(b} T^c T^{d)} \right),
\end{align}
where $T^a$ is the generator of the group $SU(N)$.
Other tensors can be constructed, but they are not linearly independent; e.g., $f^{ade}f^{bce}$ is related to $f^{abe}f^{cde}$ and $f^{ace}f^{bde}$ via the Jacobi identity. Similar identities exist that allow to express tensors constructed from the totally symmetric tensor $d^{abc}$ with the six basis tensors given above.

The structure in Lorentz space is simpler: As both the $A$ and the $V$ field have one Lorentz index, all three correlation functions can have a transverse and a longitudinal tensor.

Finally there are the $\eta$ and $\bar{\eta}$ fields. They have one adjoint index and one index that runs from $1$ to $\frac{d}{2}(N^2-1)+1$. The behavior of the latter index is best explained in terms of the original fields $\bar{c}^a$, $c^a$, $\bar{\omega}^{ab}_\mu$, $\omega^{ab}_\mu$ and $U^{ab}_\mu$. The second color index and the Lorentz index of the last three fields were only introduced to obtain the correct number of degrees of freedom. They are in some sense static: In diagrams with external $\omega/\bar{\omega}/U$ fields and no $\omega/\bar{\omega}/U$ loops they do not contribute and in $\omega/\bar{\omega}/U$ loops they only lead to additional numerical coefficients. Thus they can be combined to the second index of the $\eta$ and $\bar{\eta}$ fields together with a trivial $1$ from the $c$ and $\bar{c}$ fields. In the calculations this was most easily implemented by using the Faddeev-Popov ghost instead of the $\eta$ field and supplementing the results afterwards with the correct coefficients.

In the present work not the full color tensor basis was used. Instead only the tensors appearing in the action were taken into account for the two-point functions. Nevertheless this truncation results in a non-trivial propagator for the $V$ field. The matrix of dressed two-point functions is defined by
\begin{align}
\Gamma^{\phi\phi}&=\begin{pmatrix}
\Gamma^{AA} & \Gamma^{AV}\\
\Gamma^{VA} & \Gamma^{VV}
\end{pmatrix},
\end{align}
where the individual two-point functions are given by
\begin{align}
\Gamma^{AA,ac}_{\mu\nu}&=\delta^{ac}p^2 c_A^\bot(p^2) P_{\mu\nu}+\delta^{ac}\frac{1}{\xi} c_A^\parallel(p^2) p_\mu p_\nu,\\
\Gamma^{VV,abcd}_{\mu\nu}&=\delta^{ac}\delta^{bd}p^2 c_V(p^2) g_{\mu\nu},\\
\Gamma^{AV,cab}_{\mu\nu}&=f^{cab} i\, p^2 c_{AV}(p^2) g_{\mu\nu}.
\end{align}
The functions $c_{ij}(p^2)$ are the dressing functions and $P_{\mu\nu}$ is the transverse projector, $P_{\mu\nu}=g_{\mu\nu}-p_\mu p_\nu/p^2$. The tree-level expressions are obtained by setting $c_A(p^2)=1$, $c_V(p^2)=1$ and $c_{AV}(p^2)=g\,\gamma^2/p^2$.

The longitudinal part has to be added to the gluon two-point function in order to be able to invert the matrix. Although it may seem that this might provide a direct generalization of the Gribov-Zwanziger Lagrangian from the Landau gauge to general linear covariant gauge, this is not so. The reason is that the Faddeev-Popov operator is no longer hermitian and can have complex eigenvalues \cite{Sobreiro:2005vn}. Consequently the usual definition of the Gribov region as the set of field configurations with a positive Faddeev-Popov operator does no longer make sense. Defining linear covariant gauges in the usual way via a Gaussian distribution over the gauge orbit would be the first idea. However, while in Landau gauge we are only dealing with a discrete set of Gribov copies inside the Gribov region, we have no idea what is changed by smearing out these Landau gauge Gribov copies over the gauge orbit. For example, if two Landau gauge Gribov copies are very close on the gauge orbit, the region between the two is in a sense preferred, as it gets contributions from both Gaussians. It can even be that there are not two maxima but only one, if the two Gribov copies are close enough. Such a preference does not agree with the original idea of a single Gaussian distribution and its significance is currently unknown.
A study of Gribov copies for small values of the gauge fixing parameter $\xi$ was done in ref. \cite{Sobreiro:2005vn}.

Normally one would expect that the inverse matrix is of the structure
\begin{align}
 (\Gamma^{\phi\phi})^{-1}=D^{\phi\phi}&=
\begin{pmatrix}
D^{AA} & D^{AV}\\
D^{V A} & D^{VV}
\end{pmatrix}=\frac{1}{\Gamma^{AA} \Gamma^{VV}-(\Gamma^{AV})^2}
\begin{pmatrix}
\Gamma^{VV} & -\Gamma^{AV}\\
-\Gamma^{VA} & \Gamma^{AA}
\end{pmatrix},
\end{align}
but because of the additional tensor structure in Lorentz and color space it is not that simple. Performing the inversion manually one finds  in the employed truncation the following propagators:
\begin{align}
D^{AA,ab}_{\mu\nu}&=\delta^{ab}\frac{1}{p^2} P_{\mu\nu} \frac{c_V(p^2)}{c_A^\bot(p^2) c_V(p^2)+2 N\, c^2_{AV}(p^2)},\\
D^{VV,abcd}_{\mu\nu}&=\frac{1}{p^2}\frac{1}{c_V(p^2)}\delta^{ac}\delta^{bd}g_{\mu\nu}-f^{abe}f^{cde}\frac1{p^2}P_{\mu\nu}\frac{2  c_{AV}^2(p^2)}{c_A^\bot(p^2) c_V^2(p^2)+2N\, c_{AV}^2(p^2) c_V(p^2)},\\
D^{AV,abc}&=-i\,f^{abc}\frac1{p^2}P_{\mu\nu}\frac{\sqrt{2} c_{AV}(p^2)}{c_A^\bot(p^2) c_V(p^2)+2N\,c_{AV}^2(p^2)}.
\end{align}
While, except for some coefficients, the $AA$ and $AV$ propagators indeed resemble the conjectured form, the $VV$ propagator comes with two tensors. Thus the employed truncation leads to a non-trivial structure of the propagators what will be important for the IR analysis.

For the fermionic ghost the standard relation is valid:
\begin{align}
D^{\eta\bar{\eta},ab}_{cd}=(\Gamma^{\eta\bar{\eta},ab}_{cd})^{-1}=-\de^{ab}\de_{cd}\frac{c_{\eta}(p^2)}{p^2}.
\end{align}
The minus was factored out in order to have a positive dressing function, which is required for being inside the first Gribov region.

\subsection{Infrared behavior of the propagators}
\label{ssec:IRPropagators}

In order to assess the behavior at low momenta we make as usual the ansatz that all dressing functions have a power law form in the infrared:
\begin{align}
c_{ij}(p^2)\overset{IR}{=}d_{ij}\,(p^2)^{\ka_{ij}}.
\end{align}
The asymptotic form of the propagators depends on the behavior of the determinant $c_A^\bot(p^2) c_V(p^2)+2N\,c_{AV}^2(p^2)$ in the IR. We can think of four possible cases:
\begin{center}
\begin{enumerate}
 \item[I:] $c_{AV}^2>c_A c_V \leftrightarrow \ka_A+\ka_V>2\ka_{AV}$
 \item[II:] $c_A c_V>c_{AV}^2 \leftrightarrow 2\ka_{AV}>\ka_A+\ka_V$
 \item[III:] $c_{AV}^2\sim c_A c_V \leftrightarrow \ka_A+\ka_V=2\ka_{AV}$, no cancelations
 \item[IV:] $c_{AV}^2\sim c_A c_V \leftrightarrow \ka_A+\ka_V=2\ka_{AV}$, cancelations
\end{enumerate}
\end{center}
In case I the dressing function of the mixed two-point function dominates the determinant, in case II the combination of gluon and $V$ field dressing functions and in cases III and IV all terms contribute equally. The difference between cases III and IV is that it is possible that the leading contributions cancel exactly and the determinant is less IR divergent. This possibility is considered as case IV. Details on this cancelation are given below in Section \ref{ssec:caseIV}.

In the following I will derive the IR behavior of the propagators in all four cases. This yields relations between the IREs of the two-point functions and the propagators. With this information the IR analysis will be performed based on the results of Chapter \ref{chp:scalingSolutions}.

\subsubsection{Case I: $\ka_A+\ka_V>2\ka_{AV}$}

By assumption the dressing function $c_{AV}$ dominates in the numerator. Therefore the propagators take the following form:
\begin{align}
D^{AA,ab}_{\mu\nu}&=\delta^{ab}\frac{1}{p^2} P_{\mu\nu} \frac{c_V(p^2)}{2 N\, c^2_{AV}(p^2)},\\
D^{VV,abcd}_{\mu\nu}&=\frac{1}{p^2}\frac{1}{c_V(p^2)}\left(\delta^{ac}\delta^{bd}g_{\mu\nu}-f^{abe}f^{cde}P_{\mu\nu}\frac{1}{N}\right),\\
D^{AV,abc}&=-i\,f^{abc}\frac1{p^2}P_{\mu\nu}\frac{1}{\sqrt{2}N\,c_{AV}(p^2)}.
\end{align}
Note that both tensors contribute to the $VV$ propagator, since they have the same IRE.
The relations between the IREs of propagators and two-point functions are
\begin{align}\label{eq:caseI-props-IRE}
\de_A&=\ka_V-2\ka_{AV},\\
\de_V&=-\ka_V,\\
\de_{AV}&=-\ka_{AV}.
\end{align}
These relations allow to calculate the additional term that appeared in the equations used in the IR analysis, $\Delta_{AV}$:
\begin{align}
\Delta_{AV}=2\de_{AV}-\de_A-\de_V=-2\ka_{AV}-\ka_V+2\ka_{AV}+\ka_V=0.
\end{align}
Consequently all equations become the same as for the non-mixing case and the analysis is straightforward.

I will demonstrate now that case I does not allow a consistent solution. For this we need the lower bound for IREs given by \eref{eq:maxIRDivSol-App},
\begin{align}
\ka_{v,max} = & \left(\frac{d}{2}-2\right)\left(1-\frac{1}{2}\sum_{i}m_{i}\right)-\frac{1}{2}\sum_{i}m_{i}\delta_{{i}},
\end{align}
and the inequality derived from the IR leading diagram, \eref{eq:leadingDiagramMod},
\begin{align}\label{eq:leadingDiagramModI}
\ka_i+\frac1{2}\sum_j \de_j m_j-n^b_{{i_1}\ldots {i_r}}\left(\left(\frac{d}{4}-1\right)(r-2)+\frac{1}{2}\sum_{j}k_{{j}}^{{i_1}\ldots {i_r}}\delta_{{i}}\right)\geq0.
\end{align}
The latter formula is used for the $VV$ DSE. Note that in this DSE only the bare $AVV$ vertex appears, and consequently \eref{eq:leadingDiagramModI} has only one possible realization:
\begin{align}\label{eq:caseI-AV}
\ka_{V}+\frac1{2}2\de_V-\frac1{2}(\de_A+2\de_V)-\frac1{2}\left(\dhalf-2\right) \geq0 \quad \Rightarrow \quad
2\ka_{V} -\de_A - \dhalf+ 2 \geq0.
\end{align}
Using the expression for $\de_A$, \eref{eq:caseI-props-IRE}, we get
\begin{align}
 2\ka_V-\ka_V+2\ka_{AV}-\dhalf+2 \geq0 \quad \Rightarrow \quad 2\ka_{AV}\geq -\ka_V+\dhalf-2.
\end{align}
This is the first condition we need for the inconsistency proof.

\begin{figure}
\begin{minipage}[c]{0.45\linewidth}
\includegraphics[width=\textwidth]{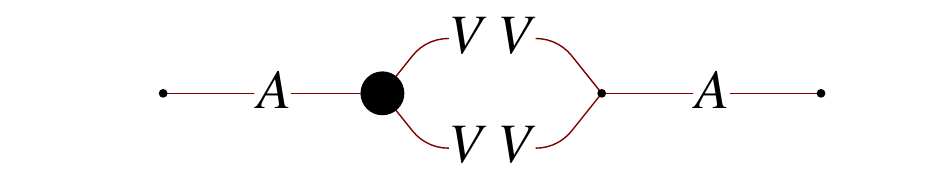}
\end{minipage}
\begin{minipage}[c]{0.45\linewidth}
\includegraphics[width=\textwidth]{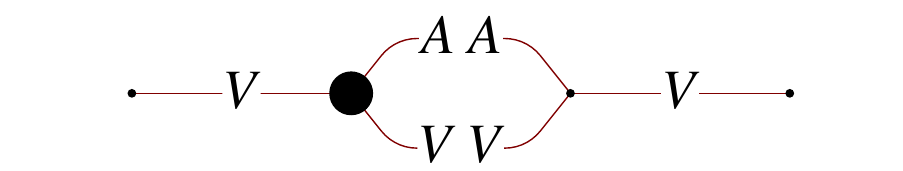}
\end{minipage}
\caption{\label{fig:AA-WW-loops}Diagrams of the $AA$ and $VV$ DSEs used explicitly in the text.}
\end{figure}

Next we consider the $AA$ DSE, to be more precise the $VV$ loop it contains as depicted in \fref{fig:AA-WW-loops} on the left. In the usual way we obtain the inequality
\begin{align}
\ka_A&\leq2\de_V+\frac{d}{2}-2+\ka_{AVV} \quad \Rightarrow \quad \ka_A+\ka_V\leq-\ka_V+\dhalf-2
\end{align}
by comparing the left- and right-hand side. We also used $\ka_{AVV}\leq0$ from the $AVV$ DSE. This inequality constitutes the second condition.

Now we insert both conditions into the defining inequality of case I:
\begin{align}
-\ka_V+\dhalf-2\geq\ka_A+\ka_V&>2\ka_{AV}\geq-\ka_V+\dhalf-2,\nnnl
0&>0.
\end{align}
As this inequality cannot be fulfilled, we conclude that case I is not consistent.

Note that in this proof we did not need the DSE of the $AV$ two-point function explicitly, i.e., we did not have to assume that any specific part in its DSE is IR leading. This point is important, because the analysis includes also the case that its tree-level part is IR leading. Thus there is no need to think about the cancelation of the tree-level part.

\subsubsection{Case II: $\ka_A+\ka_V<2\ka_{AV}$ (The simple scaling solution)}
\label{sssec:CaseII}

According to the definition the combination $c_A(p^2)c_V(p^2)$ dominates over the mixed dressing function in this case. The propagators take then the following form:
\begin{align}
D^{AA,ab}_{\mu\nu}&=\delta^{ab}\frac{1}{p^2} P_{\mu\nu} \frac{1}{c_A^\bot(p^2)},\\
D^{VV,abcd}_{\mu\nu}&=\frac{1}{p^2}\frac{1}{c_V(p^2)}\left(\delta^{ac}\delta^{bd}g_{\mu\nu}-f^{abe}f^{cde}P_{\mu\nu}\frac{2  c_{AV}^2(p^2)}{c_A^\bot(p^2) c_V(p^2)}\right)
\rightarrow \frac{1}{p^2}\frac{1}{c_V(p^2)}\delta^{ac}\delta^{bd}g_{\mu\nu},\\
D^{AV,abc}&=-i\,f^{abc}\frac1{p^2}P_{\mu\nu}\frac{\sqrt{2} c_{AV}(p^2)}{c_A^\bot(p^2) c_V(p^2)}.
\end{align}
At leading order only the first part of the $VV$ propagator contributes, as the second one is suppressed compared to the first due to the definition of case II. The relations between the IREs are
\begin{align}
\de_A&=-\ka_A,\label{eq:delta-kappaA-II}\\
\de_V&=-\ka_V,\label{eq:delta-kappaV-II}\\
\de_{AV}&=\ka_{AV}-\ka_A-\ka_V.\label{eq:delta-kappaAV-II}
\end{align}

Inserting these expressions into the additional term for mixed two-point functions, $\Delta_{AV}$, we get
\begin{align}\label{eq:caseII-addTerm}
\Delta_{AV}=2\de_{AV}-\de_A-\de_V=2\ka_{AV}-2\ka_A-2\ka_V+\ka_{A}+\ka_V=2\ka_{AV}-\ka_A-\ka_V>0.
\end{align}
The last step follows from the definition of case II. As a consequence all diagrams that feature a mixed propagator are additionally suppressed, as the coefficient of $\Delta_{AV}$ is the number of mixed propagators, see \eref{eq:IREArbitraryDiagram-mixed}. This also entails that vertices are additionally suppressed that require a mixed propagator, as, for example, the $AAV$ vertex. In the following these vertices are called $V$-odd vertices, as the number of their $V$ legs is odd. For the functional equations this means that we can neglect all diagrams containing $AV$ propagators and $V$-odd vertices in the IR. Additionally, as the $V$ propagator reduces to the same form as the $U$ and $\omega$ propagators, we can subsume all graphs containing $\eta$ and $V$ fields. The DSEs reduce then to the familiar DSEs of the Faddeev-Popov theory plus the DSEs of the $AV$ two-point function and $V$-odd vertices. 
A crucial point is that the Faddeev-Popov DSEs do not depend on the contributions from mixed propagators and can thus be solved self-consistently. The mixed correlation function, on the other hand, depends on the gluon and $V$ field propagators. Hence, having their solutions, one can also calculate them.

First I will give the results for the diagonal correlation functions and then for the mixed one. As we have recovered the standard Faddeev-Popov theory, we can directly use the known results \cite{vonSmekal:1997vx,Lerche:2002ep,Zwanziger:2001kw,Alkofer:2004it,Huber:2007kc,Fischer:2009tn}. The propagators of the Faddeev-Popov ghosts and all auxiliary fields are described by the same IRE $\ka:=\ka_c=\ka_\omega=\kappa_U=\kappa_V$. In the following I denote all these fields as ghost field. The unique scaling relation is given by \cite{Lerche:2002ep,Zwanziger:2001kw}
\begin{align}\label{eq:caseII-scalRel}
 2\ka_A+\ka=\dhalf-2
\end{align}
and the result for an n-point function with $2n$ ghost- and $m$ gluon-legs by \cite{Alkofer:2004it,Huber:2007kc}
\begin{align}
\ka_{2n,m}=(n-m)\ka+(1-n)\left(\mhalf{d}-2\right).
\end{align}
This equation is obtained as described in Section \ref{ssec:skeletonExpansion}: One starts with an appropriate two-point function, takes the IR leading diagram and plugs in as many ghost-gluon vertices as required. The value of $\ka$ can be calculated from the leading diagrams in the gluon and ghost DSEs using bare vertices as $0.595\,353$ \cite{Lerche:2002ep,Zwanziger:2001kw}. This value was reproduced from the equations of the Gribov-Zwanziger theory in order check the code employed in other calculations of this thesis. Everything boils down to solve the equation
\begin{align}\label{eq:kappa}
 -\frac{(1+\kappa ) (2+\kappa )}{12 (3+4 (-2+\kappa ) \kappa )}=1,
\end{align}
which results in $\kappa=\frac{1}{98} \left(93-\sqrt{1201}\right)=0.595\,353$ as the only solution with $\ka<1$.

\begin{figure}
\begin{center}
\includegraphics[width=0.45\textwidth]{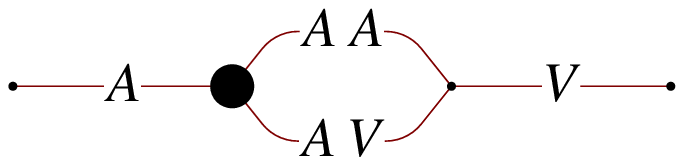}
\includegraphics[width=0.45\textwidth]{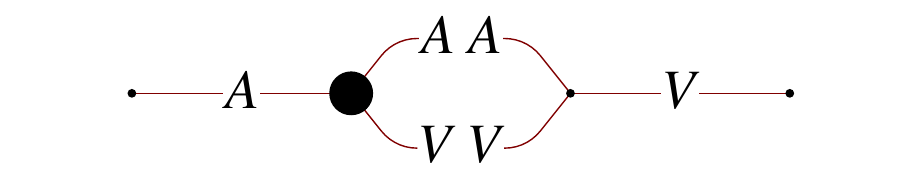}\\
\vskip10mm
\includegraphics[width=0.45\textwidth]{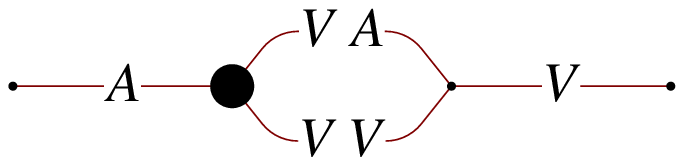}
\end{center}
\caption{\label{fig:VA-loops}Diagrams of the $VA$ DSE used explicitly in the text.}
\end{figure}

Next the value of the IRE of the mixed two-point function is evaluated. The diagrams that require at least one $AV$ propagator are depicted in \fref{fig:VA-loops}. There is one other diagram which has two $AV$ propagators and is thus IR suppressed, see \fref{fig:VA-prop-DSEs}. The diagrams of \fref{fig:VA-loops} have the following IREs:
\begin{align}\label{eq:VA-diagrams1}
\de_A+\de_{AV}+\ka_{AAA},
 \quad \de_A+\de_V+\ka_{AAV}, 
 \quad \de_V+\de_{AV}+\ka_{AVV}.
\end{align}
$\ka_{AVV}$ is zero due to the scaling relation and the IREs of $\ka_{AAA}$ and $\ka_{AAV}$ can be determined from their DSEs as
\begin{align}
\ka_{AAA}&=3\de_V,\\
\ka_{AAV}&=\de_{AV}+2\de_V.
\end{align}

Plugging these into \eref{eq:VA-diagrams1} and using eqs. (\ref{eq:delta-kappaA-II}) - (\ref{eq:delta-kappaAV-II}) as well as the scaling relation \eref{eq:caseII-scalRel}, we self-consistently get $\ka_{AV}$ for all three diagrams. In the numerical calculation of $\ka_{AV}$ we have several unknowns: The coefficients of the dressing functions ($d_{AV}$, $d_{V}$, $d_A$) and the IREs $\ka_V$ and $\ka_{AV}$. However, $d_{AV}$ always drops out of the equations and $d_V$ and $d_A$ appear in the combination $g^2/ d_A d_V^2$, which can be calculated from the Faddeev-Popov part as $0.0267784$. That leaves only $\ka_{AV}$.

For a practical calculation a further truncation is required and we only take into account the $AV$-$VV$ loop. This is in a sense a one-loop truncation and also employed for the Faddeev-Popov theory. There the ghost-gluon vertex is taken as bare, but in principle also other diagrams in its DSE contribute at leading IR order. However, if one inserted all those IR leading diagrams into the two-point function DSE, one would obtain two-loop diagrams. Due to the same argument we only consider diagrams in the truncation with an $AVV$ vertex, since vertices like $AAV$ effectively lead to two-loop diagrams. In other words we take only into account the order $d_A d_V^2$ and neglect everything of order $(d_A d_V^2)^2$ and higher. However, one should keep in mind that such truncations are only needed to calculate a numeric value for the IREs and that the qualitative features and the scaling relations themselves are derived for the full system without truncations.

\begin{figure}
\begin{center}
\includegraphics{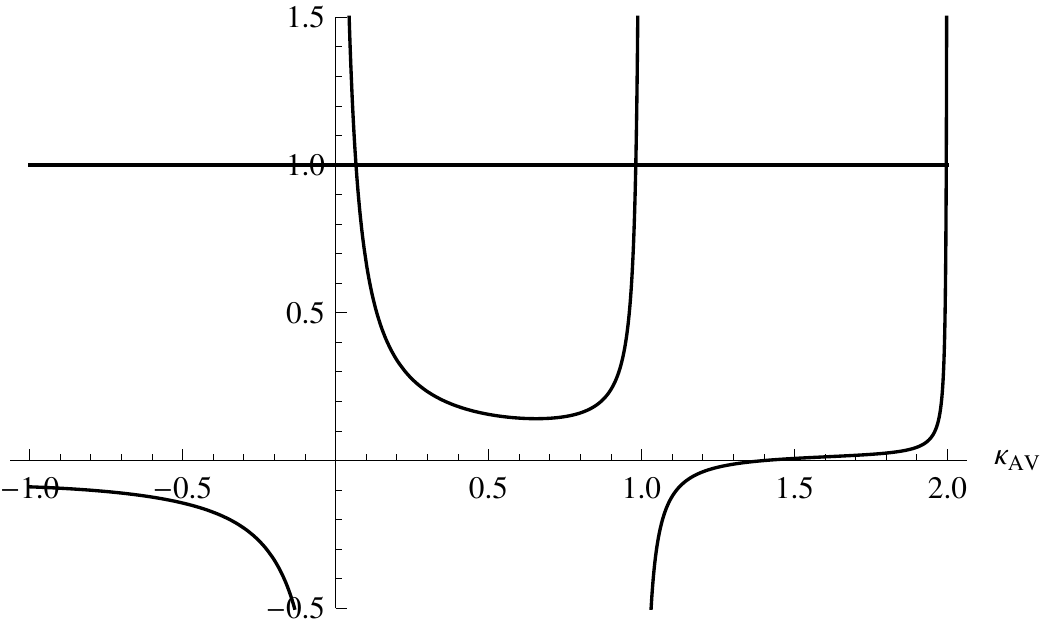}
\caption{\label{fig:kappaAV}The plot to determine solutions for $\ka_{AV}$. On the 
$y$-axis the left- and right-hand sides of \eref{eq:kappaAV} are plotted.}
\end{center}
\end{figure}

The equation to determine $\ka_{AV}$ is obtained by projecting the $AV$-$VV$ loop transversely in Lorentz space and by $f^{abc}$ in color space:
\begin{align}
(d-1)(p^2)^{\ka_{AV}+1} d_{AV}=- g^2\,N\int \frac{d^dq}{(2\pi)^d} \frac{d_{AV}  ((p+q)^2)^{-1+\de_V} (q^2)^{-1+\de_{AV}} \left(p^2 q^2-(p\,q)^2\right)}{\sqrt{2} p^2 d_A d_V^2}.
\end{align}
The coefficient $d_{AV}$ drops out of the equation. The final equation that has to be solved numerically is
\begin{align}\label{eq:kappaAV}
1&=\frac{0.0513093 \Gamma(1-\ka_{AV}) \Gamma(0.595353+\ka_{AV})}{\Gamma(1.40465-\ka_{AV}) \Gamma(1+\ka_{AV})}-\frac{0.0370825 \Gamma(-\ka_{AV}) \Gamma(0.595353+\ka_{AV})}{\Gamma(1.40465-\ka_{AV}) \Gamma(2+\ka_{AV})}-\nnnl
&-\frac{0.0513093 \Gamma(-\ka_{AV}) \Gamma(1.59535+\ka_{AV})}{\Gamma(0.404647-\ka_{AV}) \Gamma(2+\ka_{AV})}-\frac{0.0210772 \Gamma(-1-\ka_{AV}) \Gamma(0.595353+\ka_{AV})}{\Gamma(1.40465-\ka_{AV}) \Gamma(3+\ka_{AV})}-\nnnl
&-\frac{0.0247217 \Gamma(-1-\ka_{AV}) \Gamma(1.59535+\ka_{AV})}{\Gamma(0.404647-\ka_{AV}) \Gamma(3+\ka_{AV})}-\frac{0.0513093 \Gamma(-1-\ka_{AV}) \Gamma(2.59535+\ka_{AV})}{\Gamma(-0.595353-\ka_{AV}) \Gamma(3+\ka_{AV})}-\nnnl
&-\frac{0.0237307 \Gamma(-2-\ka_{AV}) \Gamma(0.595353+\ka_{AV})}{\Gamma(1.40465-\ka_{AV}) \Gamma(4+\ka_{AV})}-\frac{0.0210772 \Gamma(-2-\ka_{AV}) \Gamma(1.59535+\ka_{AV})}{\Gamma(0.404647-\ka_{AV}) \Gamma(4+\ka_{AV})}-\nnnl
&-\frac{0.0370825 \Gamma(-2-\ka_{AV}) \Gamma(2.59535+\ka_{AV})}{\Gamma(-0.595353-\ka_{AV}) \Gamma(4+\ka_{AV})}+\frac{0.0513093 \Gamma(-2-\ka_{AV}) \Gamma(3.59535+\ka_{AV})}{\Gamma(-1.59535-\ka_{AV}) \Gamma(4+\ka_{AV})}.
\end{align}
The left- and the right-hand sides of this equation are plotted in \fref{fig:kappaAV}. As one can see, there are several solutions. A constraint for $\ka_{AV}$ is given by the definition of case II: $2\ka_{AV}>\ka_V+\ka_A=-\ka_V+d/2-2$.
The smallest solutions fulfilling this constraint are $0.0668776$ and $0.981386$. Further solutions always just a little bit below integer numbers follow. At this point we cannot say which one is correct. In our truncation all of them are admissible.

An interesting observation is the fact that the $AV$-$VV$ loop (see \fref{fig:VA-loops}) is automatically at leading order:
\begin{align}
\de_V+\de_{AV}+\ka_{AVV}=-\ka_V+\ka_{AV}-\ka_A-\ka_V=\ka_{AV}.
\end{align}
If the tree-level part is kept in the $AV$ DSE and it should be subdominant, $\ka_{AV}$ has to be lesser than $-1$. However, the constraint on $\ka_{AV}$ is $2\ka_{AV}>-\ka_V+d/2-2$ and thus $\ka_V>\dhalf$. Since this is not true for the value determined for $\ka_V$, there has to be some mechanism that cancels the tree-level term in order to allow this solution. Although the details of such a cancelation are not known, it can be expected that it is connected to the horizon condition, which is responsible for the cancelation in the Faddeev-Popov ghost DSE, see e.g., \cite{Zwanziger:2009je}. For the $\omega$, $U$ and $V$ two-point functions the same argument as for the Faddeev-Popov ghost can be employed without problems.

\subsubsection{Case III: $\ka_A+\ka_V=2\ka_{AV}$, no cancelations (The strict scaling solution)}
\label{sssec:CaseIII}

Here both terms contribute equally in the determinant. Abbreviating it as $det\, C=  c_A c_V +  2 N c_{AV}^2= (p^2)^{2\ka_{AV}} det\,D= (p^2)^{\ka_A+\ka_V} det\,D$,  we get 
\begin{align}\label{eq:propsIII}
D^{AA,ab}_{\mu\nu}&=\delta^{ab}\frac{1}{p^2} P_{\mu\nu} \frac{d_V}{(p^2)^{\ka_A}}\frac{1}{det\,D},\\
D^{VV,abcd}_{\mu\nu}&=\frac{1}{p^2}\frac{1}{c_V(p^2)}\left(\delta^{ac}\delta^{bd}g_{\mu\nu}-f^{abe}f^{cde}P_{\mu\nu}\frac{2  d_{AV}^2}{det\,D}\right),\\
D^{AV,abc}&=-i\,f^{abc}\frac1{p^2}P_{\mu\nu}\frac{d_{AV}}{(p^2)^{\ka_{AV}}}\frac{\sqrt{2}}{det\,D}.
\end{align}
By definition no cancelations in the determinant occur and $det \, D$ is just a momentum independent constant.

The IREs of the propagators are
\begin{align}
\de_A=-\ka_A,\\
\de_V=-\ka_V,\\
\de_{AV}=-\ka_{AV}.
\end{align}
The additional term in the formulas for the IREs, eqs. (\ref{eq:maxIRDivSol-mixing-App}) and (\ref{eq:leadingDiagramMod}), becomes
\begin{align}
\Delta_{AV}=2\de_{AV}-\de_A-\de_V=-2\ka_{AV}+\ka_{A}+\ka_V=0.
\end{align}
Thus again the additional terms vanish, i.e., we can directly get the scaling relation from the vertices in the Lagrangian, if we assume that the tree-level two-point function in the $VV$ DSE gets canceled in a similar way as that for the other IR enhanced fields. Employing the usual technique, we consider possible scaling solutions as obtained from the interactions in the Lagrangian. The only non-trivial solution arises from the $A\bar{\eta}\eta$  and $AVV$ vertices:
\begin{align}
\ka:=\ka_V=\ka_\eta, \quad \ka_A+2\ka=\dhalf-2,\quad \ka_{AVV}=\ka_{A\eta\eta}=0.
\end{align}
The IRE of the mixed two-point function can be calculated from the defining assumption of case III as
\begin{align}
\ka_{AV}=-\frac{\ka}{2}+\frac{d}{4}-1.
\end{align}
In all DSEs the diagrams with a bare $A\bar{\eta}\eta$ or $AVV$ vertex are IR leading and the formula for the IRE of a vertex is
\begin{align}
\ka_{2n,m}=(n-m)\ka+(1-n)\left(\mhalf{d}-2\right),
\end{align}
where $m$ is the number of gluon legs and $n$ the number of legs of ghosts or auxiliary fields.

Thus the qualitative behavior is the same as in case II, i.e., the propagators of the Faddeev-Popov ghost and the auxiliary fields are IR enhanced and the gluon propagator is IR suppressed, as is the mixed propagator. The difference between case II and III is that the mixed propagator is more IR suppressed with respect to $\ka$ in case II, since there it holds that $\de_{AV}>\ka/2$, whereas in case III we have $\de_{AV}=\ka/2$.

Having found a scaling relation does, however, not yet mean that we can be sure there really exists a corresponding scaling solution. A first indication of its existence is a solution for the IRE $\ka$. This is an intricate task as the system of equalities that has to be solved involves the coefficients of the power laws, the $d_{ij}$, in a non-linear way. In case II only the IRE $\ka$ appears non-linearly and the remaining equation could be rewritten into a simple form, see \eref{eq:kappa}. For the present system, however, this is not possible.

For case III it is necessary that the tree-level term in the $AV$ DSE gets canceled by some mechanism that may be related to the horizon condition. If this is not the case the $AV$-$VV$ loop can only be leading for $\ka\geq2$ or - if the tree-level term leads - we get $\ka=2$.

\subsubsection{Case IV: $\ka_A+\ka_V=2\ka_{AV}$, cancelations}
\label{ssec:caseIV}

The final case that remains to be investigated has the same condition on the IREs of the two-point functions as case III, i.e., both terms in the determinant scale equally. However, even the coefficients of the power laws are the same so that the leading terms of each expression cancel each other:
\begin{align}
 d_A d_V=-2\,N\,d_{AV}^2.
\end{align}
The IRE of the term that takes over is undetermined, so we introduce an additional IRE $\ka_D$ which gives the correction of the IRE of the determinant:
\begin{align}
 det\,C=&c_A c_V + 2 N c_{AV}^2=\nnnl
 =&(p^2)^{\ka_A+\ka_V}(d_A d_V+2N d_{AV}^2 + d_D (p^2)^{\ka_D}+\ldots )=\nnnl
 =&(p^2)^{\ka_A+\ka_V}( d_D (p^2)^{\ka_D}+\ldots).
\end{align}
Consequently $\ka_D$ is non-negative, $\ka_D>0$.

The propagators are then almost the same as in case III, except for the $VV$ propagator, where now the first term is suppressed because there is no determinant:
\begin{align}
D^{AA,ab}_{\mu\nu}&=\delta^{ab}\frac{1}{p^2} P_{\mu\nu} \frac{d_V}{(p^2)^{\ka_A}}\frac{1}{det\,D},\\
D^{VV,abcd}_{\mu\nu}&=\frac{1}{p^2}\frac{1}{c_V(p^2)}\left(\delta^{ac}\delta^{bd}g_{\mu\nu}-f^{abe}f^{cde}P_{\mu\nu}\frac{2  d_{AV}^2}{det\,D}\right)
\rightarrow \frac{1}{p^2}\frac{1}{c_V(p^2)}\left(-f^{abe}f^{cde}P_{\mu\nu}\frac{2  d_{AV}^2}{det\,D}\right),\\
D^{AV,abc}&=-i\,f^{abc}\frac1{p^2}P_{\mu\nu}\frac{d_{AV}}{(p^2)^{\ka_{AV}}}\frac{\sqrt{2}}{det\,D}.
\end{align}
The IREs of the propagators are in this case
\begin{align}
\de_A=-\ka_A-\ka_D,\label{eq:caseIV-IREsA-relations}\\
\de_V=-\ka_V-\ka_D,\label{eq:caseIV-IREsV-relations}\\
\de_{AV}=-\ka_{AV}-\ka_D.\label{eq:caseIV-IREsAV-relations}
\end{align}
Hence we have again $\Delta_{AV}=2\de_{AV}-\de_A-\de_V=0$.

The appearance of the new IRE $\ka_D$ does not influence the general analysis as described in Section \ref{sec:mixed2Point}, because it only appears in the relation between the IREs of propagators and two-point functions, eqs. (\ref{eq:caseIV-IREsA-relations}) - (\ref{eq:caseIV-IREsAV-relations}). So we can employ in the usual way \eref{eq:leadingDiagramMod} for the $VV$ two-point function, where only a bare $AVV$ vertex can appear:
\begin{align}
\ka_V+\de_V-\frac1{2}\left( \de_A+2\de_V\right)\geq&0,\nnnl
\ka_A+2\ka_V+\ka_D\geq&0.
\end{align}
On the other hand, from counting the $VV$-$AA$ loop in the $VV$ DSE, see \fref{fig:AA-WW-loops} on the right, we get another bound:
\begin{align}
\ka_V\leq&\de_V+\de_A+\ka_{AVV}\, ,\nnnl
\ka_A+2\ka_V+2\ka_D\leq&0,
\end{align}
where we used $\ka_{AVV}\leq0$ from the DSE of the $AVV$ vertex. Note that these two inequalities look alike except that the coefficients of $\ka_D$ are different. Combining them yields therefore additional information on $\ka_D$:
\begin{align}
-\ka_D\leq\ka_A+2\ka_V&\leq-2\ka_D,\nnnl
\ka_D\geq2\ka_D.
\end{align}
Since $\ka_D$ is non-negative this inequality yields $\ka_D=0$. This corresponds to case III and thus case IV does not yield a solution.

\subsection{The infrared behavior of the propagators in the Gribov-Zwanziger theory}

Although the analysis of the Gribov-Zwanziger action was complicated by the mixing of the gluon field $A$ and the auxiliary field $V$, it was possible to obtain a scaling relation. The most important observation is certainly that the qualitative picture for the behavior of the gluon and ghost propagators is not changed, as the scaling relation $\de_A=-2\de_c$ is still valid, i.e., the gluon propagator is IR suppressed and the Faddeev-Popov ghost propagator IR enhanced. Furthermore, the IREs of the propagators of the auxiliary fields $\omega$, $U$ and $V$ are the same as for the Faddeev-Popov ghost propagator. The two solutions obtained differ mainly in the behavior of the mixed propagator: Though IR suppressed in both solutions, this suppression is more pronounced in case II, where $\de_{AV}>\ka/2$. In case III the IRE $\de_{AV}$ is determined by the scaling relation as $\de_{AV}=\ka/2$.

One of the solutions found was not accessible to a numerical solution for $\ka$, so it remains unclear, if it really exists. For the other solution, case II, the calculation of $\ka$ was unexpectedly easy as the system of DSEs reduces in the IR to the system of the Faddeev-Popov theory. Due to this one can directly employ all the results known from there. The mixed propagator does not influence the Faddeev-Popov system and its IRE can be calculated with input from there. Several allowed solutions were found.

The obtained results confirm the conjecture by Zwanziger that the cutoff of the integration at the Gribov horizon does not influence the results from DSEs, but it has to be taken into account via choosing appropriate boundary conditions \cite{Zwanziger:2001kw}. The boundary condition adopted here is derived from the horizon condition and allowed the IR enhancement of the propagators of the ghosts and the auxiliary fields. It is expected that the use of an alternative boundary condition will lead to a decoupling type of solution as in the Faddeev-Popov theory \cite{Fischer:2008uz}. It is possible to arrive at such an alternative condition by adding dimension two condensates to the Gribov-Zwanziger action \cite{Dudal:2008sp,Dudal:2007cw,Dudal:2008rm}. Their effect is that no longer the Gribov copies directly at the horizon dominate in the path integral but gauge field configurations lying within the Gribov region. This should be in direct relation to the boundary condition imposed on the DSEs of the two-point functions of the ghost and the auxiliary fields.
\chapter{Conclusions}
\label{chp:conclusions}

The central topic of this thesis has been the IR behavior of correlation functions in different gauges. The IR regime is directly linked to the property of confinement of the elementary fields. Prominent confinement scenarios are the Kugo-Ojima or the Gribov-Zwanziger scenarios, which are based on completely different arguments but lead to the same conclusions in the Landau gauge. Also the hypothesis of Abelian IR dominance, motivated by the dual superconductor picture of confinement, is related to IR properties. Consequently it is of interest to determine the behavior of propagators and vertices at low momenta and test the different scenarios where applicable. Furthermore, knowing the relations between different gauges further elucidates the overall picture of the non-perturbative regime of Yang-Mills theory.

A significant obstacle in the investigation of other gauges than the Landau gauge is the complication of the structure of the interactions between the fields. This makes their IR analysis a very tedious task with the conventional techniques. In Chapter \ref{chp:scalingSolutions} I described an extension of these methods that allows to assess the existence of different IR solutions. Although the proof is rather technical, the main statement is rather simple: One can infer possible IR solutions directly from the interactions of the Lagrangian. This is a remarkable result insofar as one can derive the IR behavior of the whole infinite tower of functional equations. The derivation of the known Landau gauge results reduces with this method to a few lines. Further gauges shortly investigated in Chapter \ref{chp:scalingSolutions} are linear covariant and ghost anti-ghost symmetric gauges.

Another complication of non-Landau gauges is that the mere derivation of the DSEs becomes very time consuming. The natural consequence was to do this task with the help of a computer algebra system and led to the development of the \textit{Mathematica} package \textit{DoDSE}. Its use is not restricted to the purposes of this thesis, but it can derive the DSEs for arbitrary actions. The underlying algorithm is also suited for manual calculations and described in Chapter \ref{chp:FunctionalEquations}. Having the DSEs directly available on the computer alleviates many task as, for example, the calculation of IREs. The length of the resulting expressions makes clear that manual calculations are extremely difficult when going beyond the Landau gauge.

The maximally Abelian gauge was investigated in Chapter \ref{chp:MAG}. Due to the complexity of this gauge the improved method for the assessment of the IR behavior of correlation functions was of great help here. The main results were an IR enhanced diagonal gluon propagator and IR suppressed off-diagonal propagators. This very nicely supports the hypothesis of Abelian IR dominance according to which the off-diagonal degrees of freedom should be suppressed compared to the diagonal ones in the IR. It has to be mentioned, however, that this result is not in direct agreement with current lattice simulations. The reason for this disagreement is most probably the same as in the Landau gauge, i.e., two solutions exist which depend on the boundary conditions imposed on the system of functional equations. For the maximally Abelian gauge this would be implemented by the renormalization condition for the diagonal gluon propagator. The scaling relation of the maximally Abelian gauge constitutes the first consistent solution of the system of infrared exponents that was found beyond the Landau gauge.
Furthermore, the connection between ghost dominance in the Landau gauge and Abelian dominance in the maximally Abelian gauge is further elucidated by these results.
Also on the influence of the shape of the Gribov region, which is different for the Landau and the maximally Abelian gauges, on the IR behavior of Green functions was speculated.
Finally, it should be mentioned that this work was the first one to investigate the maximally Abelian gauge for the physical gauge group $SU(3)$. As the action is different in $SU(2)$ and $SU(N>2)$, it is rather non-trivial that the obtained IR behavior of propagators and vertices is the same in both cases.

Chapter \ref{chp:GZ} concerned the non-perturbative gauge fixing in Landau gauge. A common way to overcome difficulties due to Gribov copies to a certain extent is to use the Gribov-Zwanziger action. Its IR behavior was determined with the methods developed in Chapter \ref{chp:scalingSolutions}. The semi-perturbative results of an IR suppressed gluon and an IR enhanced ghost propagator were confirmed with this non-perturbative analysis. At the same time the qualitative behavior is the same as obtained from functional equations when the standard Landau gauge fixing is employed. This corroborates the conjecture by Zwanziger that only the proper choice of the boundary condition is relevant for the functional equations, as one can formally cut the integration at the Gribov horizon.

The methods developed in this thesis allow first steps in the investigation of Yang-Mills theory beyond the Landau gauge. Due to the inherent complications the following two tools are of great help: An automated derivation of the functional equations and a method for a qualitative determination of the IR behavior capable of dealing with many interaction terms. Their formulations are not specific to the problems investigated in this thesis but have an even greater area of applicability. Thus they form a sound basis for further investigations to improve our understanding of the strong interaction.

\chapter*{Acknowledgments}

The last three years have been an exciting, fascinating and enthralling journey.
On these pages I want to thank those people who contributed in one or the other way to its success and made sure I would arrive.

I would like to thank my advisor Reinhard Alkofer who offered me the opportunity to explore the land of Yang-Mills theory. On my way through this interesting landscape he supported me in many ways and gave me the freedom to explore also some paths besides the main road what led, for example, to the development of \textit{DoDSE}. I am very grateful for his support, advice and encouragement during my journey.

A part of my way I was accompanied by Silvio Sorella and we passed together through the so-called Gribov region. Although we found it to be a dark and foggy place, we managed to keep true and steer through the mist on the search for operators with physical properties. And we indeed found some first promising traces of such an operator, called i-particles, which are, however, not covered in this thesis. We also came across evidence of the scaling solution in the Gribov region. I am very grateful for his good guidance in this dark region and his warm hospitality.

My companion on the first half of the trip was Kai Schwenzer. Together we sailed the deep waters of scaling solutions in different gauges and when we came to the island of the maximally Abelian gauge, we spent some time there, looking for indications of confinement. On our way we also circumnavigated the cliffs of kinematic divergences successfully. Thanks to Kai for helping me cross these dangerous seas.

I also had many stimulating exchanges with Axel Maas. I would like to thank him for sharing his insights about the Gribov region gained on his own journey and for conveying the latest reports from distant parts of the land of Yang-Mills theory to me.

At several points on my journey my path crossed with that of Daniel Zwanziger with whom I had many long and stimulating discussions about the landscape of the Gribov region. They helped me in finding my way and I would like to thank him for sharing his insights.

There are some other travelers I met: David Dudal, Christian S. Fischer, Leonard Fister, Marcelo S. Guimar\~aes, Florian Hebenstreit, Klaus Lichtenegger, Mario Mitter, Jan M. Pawlowski, Lorenz von Smekal and Nele Vandersickel. With all of them I enjoyed enlightening discussions and they helped me to continue on my way.

I also would like to thank all the members of the doctorate college "Hadrons in vacuum, nuclei and stars": The students, who struggled along with me through the world of QCD, and the professors, especially Christian Lang, who was my co-advisor.


A journey without money is troublesome. I am glad I was supported by the Karl-Franzens-University Graz, the Graz Advanced School of Science and the FWF under contract W1203-N08. Furthermore, I acknowledge the help of the KUWI program of the Karl-Franzens-University Graz to explore the Brazilian parts of the Gribov region at the Universidade do Estado do Rio de Janeiro.

Finally, I thank my partner Birgit Plauder, who never questioned my urge to advance deeper and deeper into the mysterious land of Yang-Mills theory and always had a place to come home to so I could replenish my strength.


\appendix
\chapter{Details on the formulae for the scaling analysis}
\label{chp:details-scaling}

I present in this appendix the derivations of several equations of Chapter \ref{chp:scalingSolutions}. All calculations are done in $d$ dimensions. Another generalization is that mixed propagators are taken into account wherever necessary, but the case of a diagonal two-point matrix can be directly inferred from the results.

\section{Inequalities from DSEs and FRGEs in $d$ dimensions}

For completeness I shortly summarize the derivation of inequality (\ref{eq:verts-props-inequal1}) in $d$ space-time dimensions which was already given in four dimensions in Section \ref{subsec:ineqs}.

We start with three-point functions. From the FRGE of a generic three-point function, as depicted in \fref{eq:RG-3-point-App}, one can extract the inequality
\begin{align}\label{eq:RG-3-point-App}
\ka_{ABC}\leq 3\ka_{ABC}+\delta_A+\delta_B+\delta_C+\frac{d}{2}-2 \quad \Rightarrow \quad \ka_{ABC}+\mhalfo(\delta_A+\delta_B+\delta_C)+\frac{d}{4}-1\geq0.
\end{align}

\fig{th}{fig:RG-3-point-App}{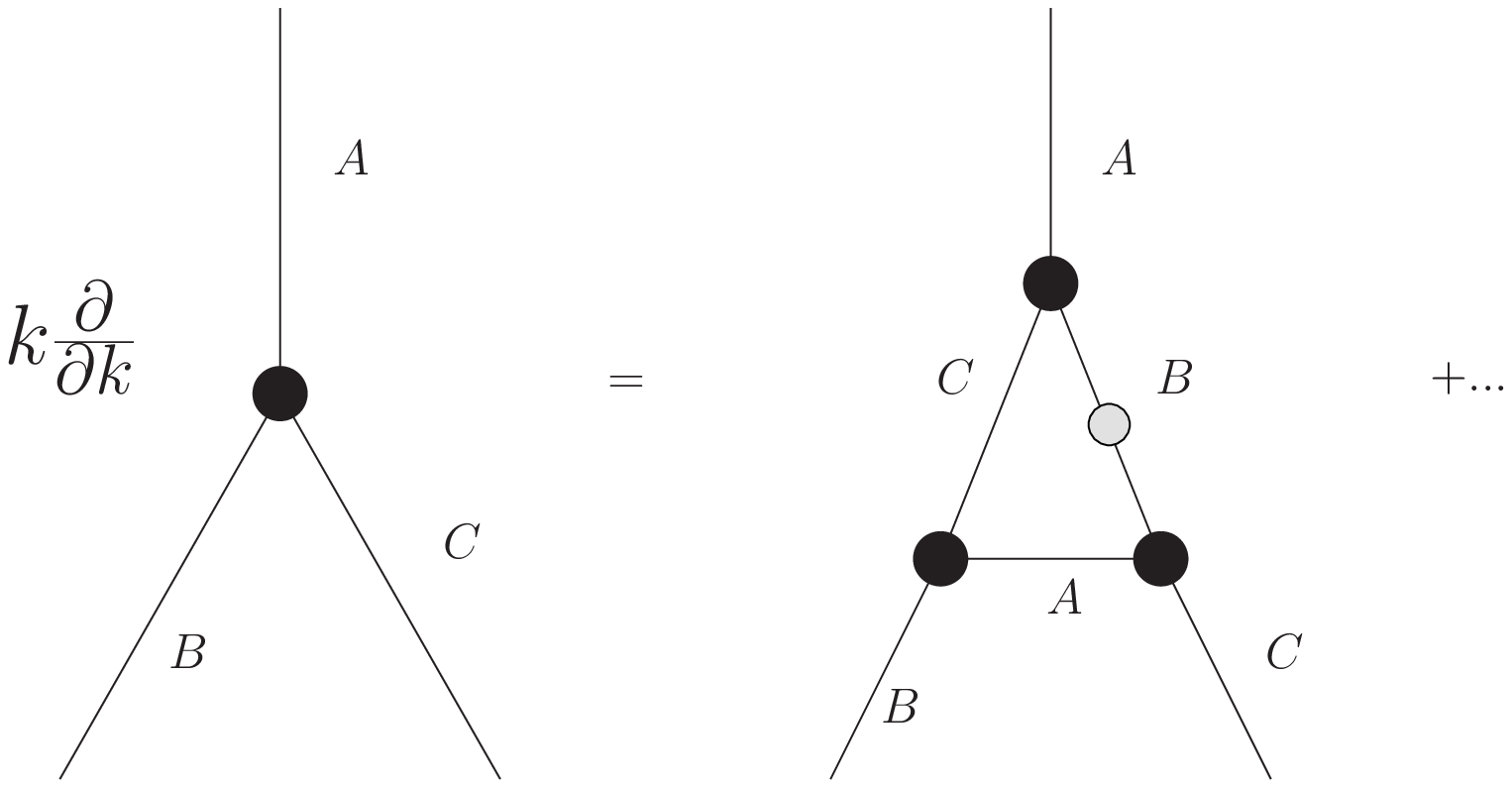}{One specific diagram in the FRGE of a generic three-point function. Internal lines represent dressed propagators, black blobs dressed vertices. The grey blob is a regulator insertion.}{width=0.48\textwidth}

\begin{figure}[b]
\begin{center}
 a) \includegraphics[width=0.43\textwidth]{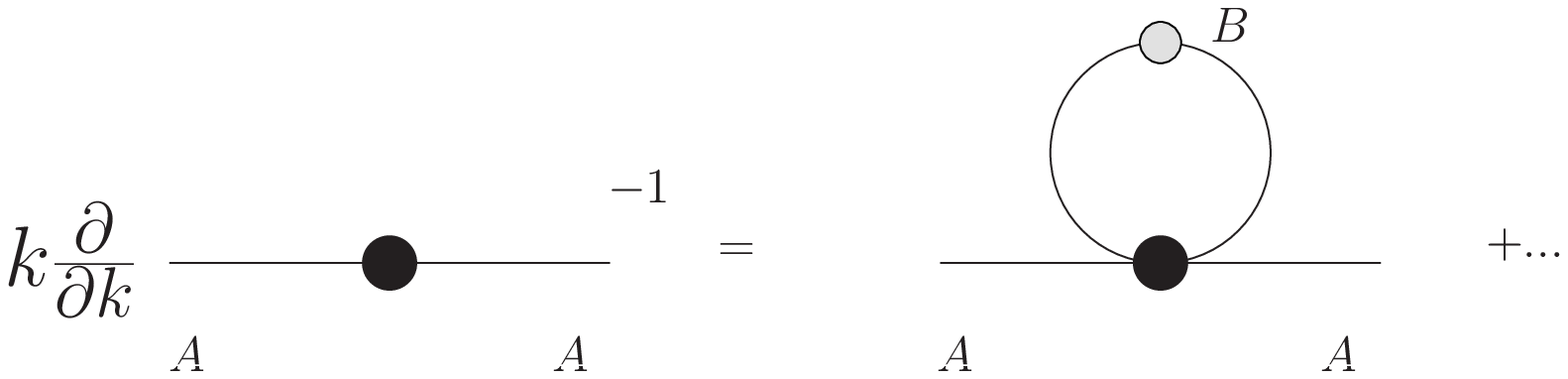}\hskip1cm
 b) \includegraphics[width=0.43\textwidth]{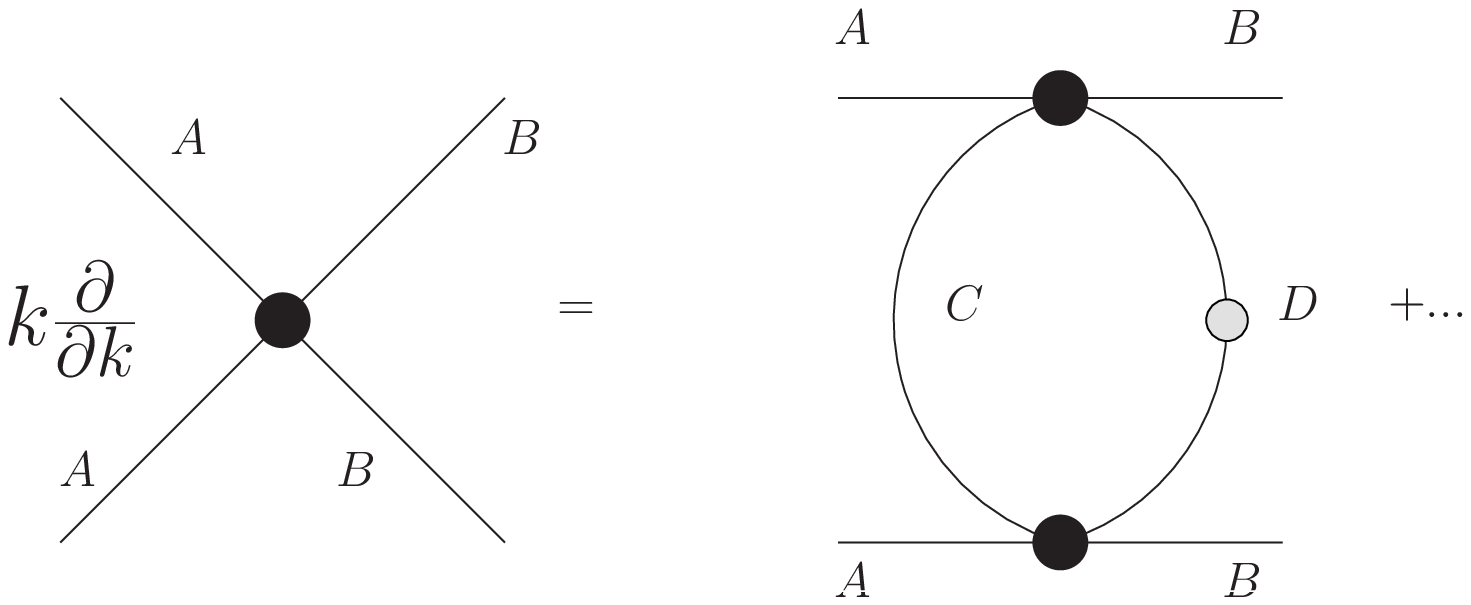}
\caption{\label{fig:RG-4-point-App} Parts of the FRGEs of generic four-point functions.}
\end{center}
\end{figure}

For four-point functions we get from \fref{fig:RG-4-point-App} the following two inequalities:
\begin{align}
 -\delta_A-\delta_B -\left(\frac{d}{2}-2\right) &\leq \ka_{AABB},\label{eq:4Point1}\\
 \ka_{AABB}&\leq 2\ka_{ABCD}+\delta_{C}+\delta_{D}+\frac{d}{2}-2 \label{eq:4Point2}.
\end{align}
Combining them yields
\begin{align}
 \ka_{ABCD}+\frac{1}{2} \left(\delta_{A}+\delta_{B}+\delta_{C}+\delta_{D}\right) +2\left(\frac{d}{4}-1\right)\geq0.
\end{align}

The inequalities for the three- and four-point functions fulfill
\begin{align}\label{eq:IR-ineqs1-App}
\left(\frac{d}{4}-1\right)(r-2)+\ka_{{i_1}\ldots {i_r}}+\frac{1}{2}\sum_{j}k_{{j}}^{{i_1}\ldots {i_r}}\delta_{{i}}\geq0,
\end{align}
where $r=3,4$ denotes the number of legs the n-point function has.
Higher n-point function also obey this inequality as can be shown by induction.

\begin{figure}[t]
\begin{center}
 a) \includegraphics[width=0.43\textwidth]{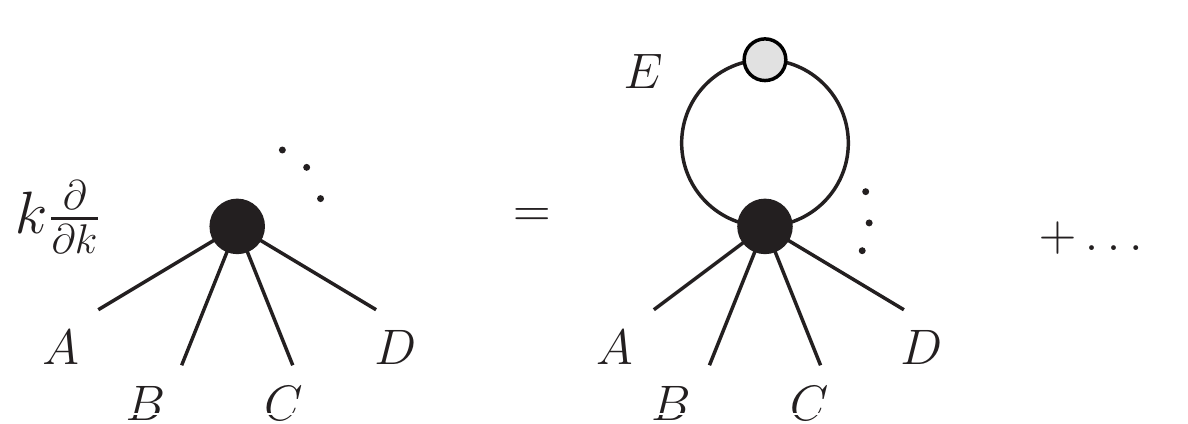}\hskip1cm
 b) \includegraphics[width=0.43\textwidth]{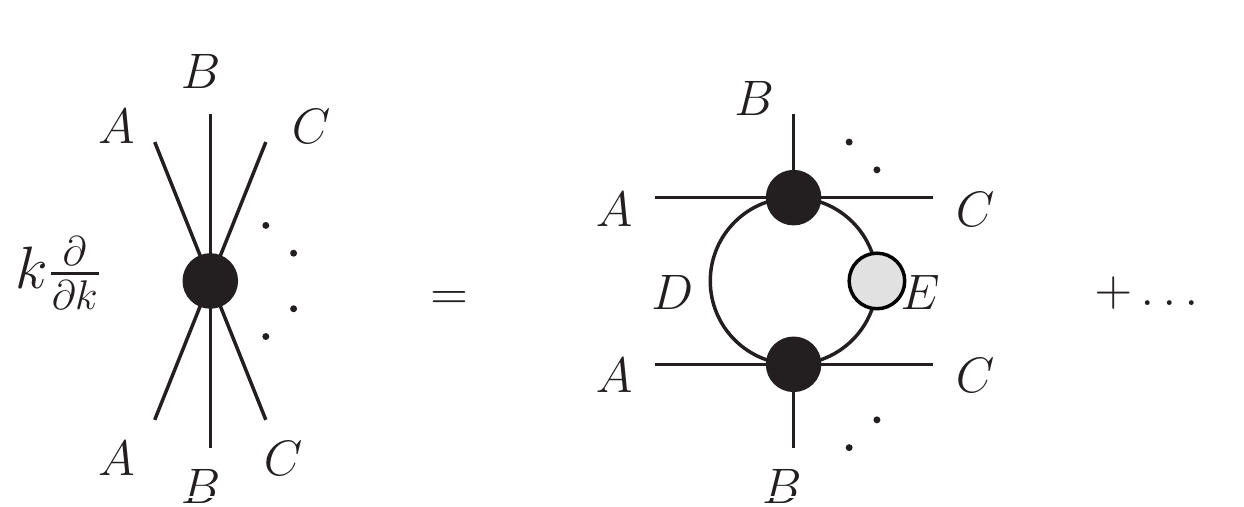}
\caption{\label{fig:RG-3+6+-app} Parts of the FRGEs of generic n-point functions.}
\end{center}
\end{figure}

For the proof we need the following two inequalities which can be inferred from \fref{fig:RG-3+6+-app}:
\begin{align}
 \ka_{ABCD\cdots}&\leq \ka_{ABCDEE\cdots}+\de_E+\frac{d}{2}-2,\label{eq:nPoint1-App}\\
 \ka_{AABBCC\cdots}&\leq 2\ka_{ABCDE\cdots}+\de_D+\de_E+\frac{d}{2}-2.
\end{align}
The dots represent further legs as indicated in  \fref{fig:RG-3+6+-app}. Note that these inequalities are generalizations of eqs. (\ref{eq:4Point1}) and (\ref{eq:4Point2}).
The first inequality can be used to write down an equation for the vertex $AABBCC$ as appearing in the second inequality:
\begin{align}
 \ka_{AABB\cdots}-\de_{C}-\left(\frac{d}{2}-2\right)\leq\ka_{AABBCC\cdots}&\leq 2\ka_{ABCDE\cdots}+\de_D+\de_E+\frac{d}{2}-2,\\
 \frac{1}{2}\left(\ka_{AABB\cdots}-\de_C-\de_D-\de_E\right)-\left(\frac{d}{2}-2\right) &\leq \ka_{ABCDE\cdots}\,\label{eq:nPoint2-App}.
\end{align}
This inequality connects an n-point function with a (2n-6)-point function. The goal is to rewrite the equation such that only propagator IREs and the IRE of the n-point function remain. For this one can successively use \eref{eq:nPoint1-App} to replace the IRE of the remaining other vertex and arrives at
\begin{align}
 -\frac{1}{2}\left(\de_A+\de_B+\de_C+\de_D+\de_E +\dots \right)-\left(\frac{d}{4}-1\right)(r-2) &\leq \ka_{ABCDE\cdots}\,,
\end{align}
where $r$ gives the number of legs of the n-point function. The coefficient of $d/4-1$ is obtained as follows: One gets $2(d/4-1)$ in \eref{eq:nPoint2-App} and $d/4-1$ from every further application of \eref{eq:nPoint1-App}. The latter has to be done until $2n-6-2x=2$, where $x$ is the number of iterations. Hence the total coefficient of $d/4-1$ is $2+x=n-2$.  This establishes \eref{eq:IR-ineqs1-App} for all n-point functions.

For the second group of inequalities the non-positivity of the IREs of interactions appearing in the Lagrangian, \eref{eq:primDivVerts}, is used in \eref{eq:IR-ineqs1-App}:
\begin{align}\label{eq:IR-ineqs2-App}
\left(\frac{d}{4}-1\right)(r-2)+\frac{1}{2}\sum_{j}k_{{j}}^{{i_1}\ldots {i_r}}\delta_{{i}}\geq0 \qquad \forall \text{ primitively divergent vertices}.
\end{align}

\section{Infrared exponent for an arbitrary diagram}
\label{sec:details-scaling-App}

It is possible to express the IRE $\ka_v$ of an arbitrary diagram $v$ by counting the IREs of all its propagators and dressed vertices. For $d$ dimensions also the canonical dimensions have to be taken into account:
\begin{align}\label{eq:deltav-start}
\ka_{v}= & l\frac{d}{2}+\sum_{i}n_i(\delta_{{i}}-1)+\sum_{vertices,r\geq3}n^{d}_{{i_1}\ldots {i_r}}(\ka_{{i_1}\ldots {i_r}}+c_{{i_1}\ldots {i_r}})+\nnnl
 & +\sum_{vertices,r\geq3}n^{b}_{{i_1}\ldots {i_r}}c_{{i_1}\ldots {i_r}}-c_{v}.
\end{align}
$n_i$ are the number of internal propagators with IR exponents $\delta_{{i}}$, whereas the numbers of vertices $\phi_{i_1}\cdots \phi_{i_r}$ are $n_{{i_1}\ldots {i_r}}$. Superscripts $d$ and $b$ stand for dressed and bare, respectively. In case none is given, I refer to both. The sums $\sum_{vertices,r\geq3}$ extend over all vertices with $r$ legs. $m_{i}$ is the number of external legs of field type $\phi_i$ and $l$ is the number of loops. The canonical dimensions of the vertex $\phi_{i_1}\cdots \phi_{i_r}$ are given by $c_{{i_1}\ldots {i_r}}$.

Without mixed propagators it is possible to use topological relations to get completely rid of the internal propagators. In the more general case, however, when mixed propagators appear, the dependence on them will remain. To avoid too cumbersome notation we restrict ourselves to the case of two fields only. We denote them explicitly by $A$ and $V$, as these are the fields for which the mixing occurs in the Gribov-Zwanziger Lagrangian; see Section \ref{ssec:GZAction} for details. The case of no mixing fields can be obtained by setting the number of mixing propagators $n_{AV}$ to zero and extending the sums over all fields in the formulae to come. Moreover, we have $\de_i=-\ka_i$ in this case.

The first topological relation we need gives the number of loops in terms of the numbers of propagators and vertices:
\begin{align}\label{eq:numberOfLoops}
l=\sum_{i=A,V}n_i+n_{AV}+1-\sum_{vertices,r\geq3}n_{{i_1}\ldots {i_r}}.
\end{align}
Furthermore, we express the number of internal $A$- and $V$-propagators by
\begin{align}\label{eq:relation-verts-props}
n_i=\frac{1}{2}\left(\sum_{vertices,r\geq3}k_{{i}}^{{i_1}\ldots {i_r}}n_{{i_1}\ldots {i_r}}-m_{i}-n_{AV}\right), \quad \quad i=A,V,
\end{align}
where $k_{{i}}^{{i_1}\ldots {i_r}}$ denotes the number of times the field $\phi_{i}$ appears in the vertex $\phi_{i_1}\cdots \phi_{i_r}$.
Plugging these expressions into \eref{eq:deltav-start}, we get
\begin{align}
\ka_{v} = & \left(\sum_{i=A,V}\frac{1}{2}\left(\sum_{vertices,r\geq3}k_{{i}}^{{i_1}\ldots {i_r}}n_{{i_1}\ldots {i_r}}-m_{i}-n_{AV}\right)+1+n_{AV}-\sum_{vertices,r\geq3}n_{{i_1}\ldots {i_r}}\right)\frac{d}{2}+\nnnl
 & +n_{AV}(\de_{AV}-1)+\sum_{i=A,V}\frac{1}{2}\left(\sum_{vertices,r\geq3}k_{{i}}^{{i_1}\ldots {i_r}}n_{{i_1}\ldots {i_r}}-m_{i}-n_{AV}\right)(\delta_{{i}}-1)+\nnnl
 & +\sum_{vertices,r\geq3}n^{d}_{{i_1}\ldots {i_r}}\left(\delta_{{i_1}\ldots {i_r}}+2-\frac{r}{2}\right)+\sum_{vertices,r\geq3}n^{b}_{{i_1}\ldots {i_r}}\left(2-\frac{r}{2}\right)-2+\frac{1}{2}\sum_{i=A,V}m_{i}=\nnnl
= & \left(\frac{d}{2}-2\right)\left(1-\frac{1}{2}\sum_{i=A,V}m_{i}\right)-\frac{1}{2}\sum_{i=A,V}m_{i}\delta_{{i}}+\sum_{vertices,r\geq3}n^{b}_{{i_1}\ldots {i_r}}\left(-\frac{d}{2}+2-\frac{r}{2}\right)+\nnnl
 & +\sum_{vertices,r\geq3}n^{d}_{{i_1}\ldots {i_r}}\left(-\frac{d}{2}+\delta_{{i_1}\ldots {i_r}}+2-\frac{r}{2}\right)+n_{AV}\left((\delta_{AV}-1)-\frac1{2}\sum_{i=A,V}(\de_{i}-1)\right)+\nnnl
 & +\sum_{i=A,V}\frac{1}{2}\left(\sum_{vertices,r\geq3}k_{{i}}^{{i_1}\ldots {i_r}}n_{{i_1}\ldots {i_r}}(\frac{d}{2}+\delta_{{i}}-1)\right),
\end{align}
where it was used that the canonical dimension of a vertex is given by $(4-r)/2$ with $r$ being the number of external legs.
Reordering terms yields
\begin{align}\label{eq:IREArbitraryDiagram-mixing-App}
\ka_{v} = & \left(\frac{d}{2}-2\right)\left(1-\frac{1}{2}\sum_{i=A,V}m_{i}\right)-\frac{1}{2}\sum_{i=A,V}m_{i}\delta_{{i}}+\frac1{2}n_{AV}(2\de_{AV}-\de_A-\de_V)+\nnnl
 & +\sum_{vertices,r\geq3}n^{d}_{{i_1}\ldots {i_r}}\left(\left(\frac{d}{4}-1\right)(r-2)+\ka_{{i_1}\ldots {i_r}}+\frac{1}{2}\sum_{i=A,V}k_{{i}}^{{i_1}\ldots {i_r}}\delta_{{i}}\right)+\nnnl
 & +\sum_{vertices,r\geq3}n^{b}_{{i_1}\ldots {i_r}}\left(\left(\frac{d}{4}-1\right)(r-2)+\frac{1}{2}\sum_{i=A,V}k_{{i}}^{{i_1}\ldots {i_r}}\delta_{{i}}\right).
\end{align}
This is the formula for the IRE of the diagram $v$ with $m_i$ external legs of the field $\phi_i$ and the number and type of vertices given by the $n^b$ and $n^d$. The dependence on the internal propagators is given via the term $n_{AV}(2\de_{AV}-\de_A-\de_V)/2$.
The maximally IR divergent solution in the $d$-dimensional case is
\begin{align}\label{eq:maxIRDivSol-mixing-App}
\ka_{v,max} = & \left(\frac{d}{2}-2\right)\left(1-\frac{1}{2}\sum_{i=A,V}m_{i}\right)-\frac{1}{2}\sum_{i=A,V}m_{i}\delta_{{i}}+\frac1{2}n_{AV}(2\de_{AV}-\de_A-\de_V).
\end{align}
It is derived from \eref{eq:IREArbitraryDiagram-mixing-App} and the non-negativity of the last two terms, see eqs. (\ref{eq:IR-ineqs1-App}) and (\ref{eq:IR-ineqs2-App}).

The corresponding formulae in the case of no mixing fields are obtained by setting $n_{AV}=0$:
\begin{align}\label{eq:DSE-2-point-App}
\ka_{v} = & \left(\frac{d}{2}-2\right)\left(1-\frac{1}{2}\sum_{i}m_{i}\right)-\frac{1}{2}\sum_{i}m_{i}\delta_{{i}}+\nnnl
 & +\sum_{vertices,r\geq3}n^{d}_{{i_1}\ldots {i_r}}\left(\left(\frac{d}{4}-1\right)(r-2)+\ka_{{i_1}\ldots {i_r}}+\frac{1}{2}\sum_{i}k_{{i}}^{{i_1}\ldots {i_r}}\delta_{{i}}\right)+\nnnl
 & +\sum_{vertices,r\geq3}n^{b}_{{i_1}\ldots {i_r}}\left(\left(\frac{d}{4}-1\right)(r-2)+\frac{1}{2}\sum_{i}k_{{i}}^{{i_1}\ldots {i_r}}\delta_{{i}}\right)
\end{align}
and
\begin{align}\label{eq:maxIRDivSol-App}
\ka_{v,max}=\left(\frac{d}{2}-2\right)\left(1-\frac{1}{2}\sum_{i}m_{i}\right)-\frac{1}{2}\sum_{i}m_{i}\delta_{{i}}
\end{align}
for the maximally IR divergent solution.

\section{Inequality from the leading diagram}
\label{sec:Prop-Eqs}

Under the assumption that a certain diagram is leading in a two-point DSE one can derive an additional inequality constraining the system. For this one starts with the general expression of a two-point function IRE, obtained from \eref{eq:IREArbitraryDiagram-mixing-App}:
\begin{align}\label{eq:leading-diagram}
\ka_{i} = & \left(\frac{d}{2}-2\right)\left(1-\frac{1}{2}\sum_{j}m_{j}\right)-\frac{1}{2}\sum_{j}m_{j}\delta_{{j}}+\frac1{2}n_{AV}(2\de_{AV}-\de_A-\de_V)+\nnnl
 & +\sum_{\substack{dressed\\vertices}}n^{d}_{{i_1}\ldots {i_r}}\ka_{{i_1}\ldots {i_r}}
  +\sum_{\substack{all\\vertices}}n_{{i_1}\ldots {i_r}}\left(\left(\frac{d}{4}-1\right)(r-2)+\frac{1}{2}\sum_{j}k_{{j}}^{{i_1}\ldots {i_r}}\delta_{{i}}\right).
\end{align}
Here $\ka_i$ can be any two-point IRE and $\de_j$ is restricted to IREs of non-mixing propagators.
Using $\sum_j m_j=2$ in the case of propagators, this can be written as
\begin{align}
\ka_{i}& +\frac{1}{2}\sum_{j}m_{j}\delta_{{i}}-\frac1{2}n_{AV}(2\de_{AV}-\de_A-\de_V)-\nnnl
 &-\sum_{\substack{all\\vertices}}n_{{i_1}\ldots {i_r}}\left(\left(\frac{d}{4}-1\right)(r-2)+\frac{1}{2}\sum_{j}k_{{j}}^{{i_1}\ldots {i_r}}\delta_{{i}}\right)=\sum_{\substack{dressed\\vertices}}n^{d}_{{i_1}\ldots {i_r}}\ka_{{i_1}\ldots {i_r}}.
\end{align}
We can use the lower bound for the IREs of the vertices on the right-hand side given by the maximally IR divergent solution, \eref{eq:maxIRDivSol-mixing-App}, to get a new inequality:
\begin{align}
\ka_i&+\frac1{2}\sum_j \de_j m_j-\frac{1}{2}n_{AV}(2\de_{AV}-\de_A-\de_V)-\nnnl
 &\quad-\sum_{\substack{all\\vertices}}n_{{i_1}\ldots {i_r}}\left(\left(\frac{d}{4}-1\right)(r-2)+\frac{1}{2}\sum_{j}k_{{j}}^{{i_1}\ldots {i_r}}\delta_{{i}}\right) \geq \nnnl
\geq& \sum_{\substack{dressed\\vertices}} n^d_{{i_1} \ldots {i_r}}  \Bigg(\left(\frac{d}{2}-2\right)\left(1-\frac{1}{2}\sum_{j}k_{j}^{{i_1} \ldots {i_r}}\right)-\nnnl
 & \quad - \frac1{2} \sum_j \de_{j} k_{j}^{{i_1} \ldots {i_r}} +\frac1{2} \bar{k}^{{i_1} \ldots {i_r}}_{AV}(2\de_{AV}-\de_A-\de_V) \Bigg) .
\end{align}
Here $\bar{k}^{{i_1} \ldots {i_r}}_{AV}$ indicates the number of times a mixed propagator is contained in the diagram that determines the IRE of the vertex $\phi_{i_1} \cdots \phi_{i_r}$. Note that $\bar{k}^{{i_1} \ldots {i_r}}_{AV}$ only is different from zero for vertices that necessarily contain an $AV$-propagator like the $AAV$-vertex.
The right-hand side depends on dressed vertices only, indicated by the superscript $d$ of $n$. On the other hand, the left-hand side sums over dressed and bare vertices, so that in total only the bare vertex remains in the sums over vertices:
\begin{align}\label{eq:leadingDiagramMod}
\ka_i&+\frac1{2}\sum_j \de_j m_j-n^b_{{i_1}\ldots {i_r}}\left(\left(\frac{d}{4}-1\right)(r-2)+\frac{1}{2}\sum_{j}k_{{j}}^{{i_1}\ldots {i_r}}\delta_{{i}}\right)  -\nnnl
&-\frac1{2}(2\de_{AV}-\de_A-\de_V)\left( n_{AV}+\sum_{\substack{dressed\\vertices}} n^d_{{i_1} \ldots {i_r}} \bar{k}^{{i_1} \ldots {i_r}}_{AV} \right) \geq0.
\end{align}
Here $\sum_i k_i^{i_1 \ldots i_r}=r$ was used.

Again the case of a Lagrangian with only diagonal two-point functions is obtained by setting $n_{AV}$ and consequently also $\bar{k}^{{i_1} \ldots {i_r}}_{AV}$ to zero. Additionally we have $\ka_i=-\de_i$ so that the first two terms cancel:
\begin{align}\label{eq:leadingDiagramSimple}
-n^b_{{i_1}\ldots {i_r}}\left(\left(\frac{d}{4}-1\right)(r-2)+\frac{1}{2}\sum_{j}k_{{j}}^{{i_1}\ldots {i_r}}\delta_{{i}}\right) \geq0.
\end{align}
Combining it with \eref{eq:IR-ineqs2-App} we obtain the equation
\begin{align}\label{eq:leadingDiagramSimpleEq}
\hat{n}^b_{{i_1}\ldots {i_r}}\left(\left(\frac{d}{4}-1\right)(r-2)+\frac{1}{2}\sum_{j}k_{{j}}^{{i_1}\ldots {i_r}}\delta_{{i}}\right) =0,
\end{align}
where the hat indicates that it is only valid for a specific n-point function.
\chapter{Grassmann fields}
\label{chp:GrassmannFields}

In many quantum field theories not only commuting, but also anti-commuting fields appear, be they fundamental particles like electrons or quarks or mathematical constructs like ghosts. These Grassmann fields entail additional complications as one has to carefully take into account their anti-commutativity. For their treatment one has to choose one of several conventions concerning the direction of derivatives and the ordering of the fields. It is also possible to introduce a metric, see, for example, ref. \cite{Pawlowski:2005xe}. This appendix summarizes the convention employed in this thesis, which has also been adopted for the program \textit{DoDSE}. In the following Grassmann fields are denoted by $\psi$ and their anti-fields, here called anti-Grassmann fields, by $\bar{\psi}$. The corresponding sources are $\bar{\eta}$ and $\eta$, respectively.

Derivatives with respect to anti-commuting fields or sources are defined as acting from the right and left for Grassmann and anti-Grassmann quantities, respectively:
\begin{align}
& \frac{\delta}{\delta \psi} := \frac{\overset{\leftarrow}{\delta}}{\delta \psi}, & \frac{\delta}{\delta \bar{\psi}} := \frac{\overset{\rightarrow}{\delta}}{\delta \bar{\psi}}.
\end{align}
Hence by definition all derivatives are written at the left side of an expression but act from the correct side.
Consequently quantities should always be ordered such that Grassmann fields are right of anti-Grassmann fields. This is valid for the products of fields and sources appearing in the path integral as well as for derivatives. So the path integral for a theory with the two Grassmann fields $\psi$ and $\bar{\psi}$ reads
\begin{align}
 Z[\eta,\bar{\eta}]=\int \mathcal{D}[\bar{\psi}\psi] e^{-S[\bar{\psi},\psi]+\bar{\eta}\psi + \bar{\psi}\eta}
\end{align}
and a quartic Grassmann interaction has the form
\begin{align}
 \Gamma_{ijkl}=-\frac{\partial \Gamma}{\partial \bar{\psi}_i \partial \bar{\psi}_j \partial \psi_k \partial \psi_l}\Big|_{\bar{\psi}=\psi=0}.
\end{align}
For easier readability (and also in correspondence with \textit{DoDSE}) the indices of $\Gamma_{ijkl}$ do not reflect the order of how the derivatives are performed, but rather have the order in which the derivatives appear, i.e., the differentiation with respect to $\bar{\psi}_j$ has to be performed before that with respect to $\bar{\psi}_i$, but $\psi_k$ comes before $\psi_l$.

The replacements of the fields to obtain the generating DSE for 1PI functions, see \eref{eq:DSE-master}, are given by
\begin{align}
 \bar{\psi}_r \rightarrow \bar{\psi}_r+ D^{J,\psi\Phi}_{rt}  \frac{\delta}{\delta \Phi_t},\\
 \psi_r \rightarrow \psi_r+ D^{J,\bar{\psi}\Phi}_{rt}  \frac{\delta}{\delta \Phi_t}
\end{align}
for anti-commuting fields.
Furthermore, eqs. (\ref{eq:derivatives}) have to be amended by
\begin{subequations}\label{eq:derivsGrassmann}
\begin{align}
\label{eq:deriv-propagator-anti-Grassmann}\frac{\delta}{\delta\bar{\psi}_{i}}D^J_{jk} & = D^J_{jm}\Gamma^{J,\bar{\psi}\Phi\Phi}_{imn} D^J_{nk},\\
\label{eq:deriv-propagator-Grassmann}\frac{\delta}{\delta\psi_{i}} D^J_{jk} & = D^J_{jm} \Gamma^{J,\Phi\Phi \psi}_{mni} D^J_{nk},\\
\label{eq:deriv-vertex-anti-Grassmann}\frac{\delta}{\delta\bar{\psi}_{i}} \Gamma^J_{j_{1}\cdots j_{n}} & =\Gamma^J_{ij_{1}\cdots j_{n}},\\
\label{eq:deriv-vertex-Grassmann}\frac{\delta}{\delta \psi_{i}} \Gamma^J_{j_{1}\cdots j_{n}} & =\Gamma^J_{j_{1}\cdots j_{n}i},
\end{align}
\end{subequations}
where the fields were added as superscripts for easier identification.
Care has to be taken for the internal super fields $\Phi$, which can be commuting or anti-commuting. As long as it is not determined if they are Grassmann or anti-Grassmann, i.e., if the derivatives are right- or left-derivatives, respectively, they should be considered as floating. Only at the end, when the sources are set to zero and the super fields become physical fields, one places them at the left or the right side, but one still has to obey the order in which the derivatives have been applied. This is an important point as the required ordering yields the signs expected normally for Feynman diagrams with fermion loops.

An example of how to order the fields should help to illustrate this point. We consider the quark respectively ghost loop in the gluon DSE of Landau gauge where $\psi=\{q,c\}$. After performing the first derivative with respect to the gluon field $A_i$, we have to replace the fields as indicated in \eref{eq:DSE-master}. For the anti-Grassmann field this is $\bar{\psi}_r \rightarrow \bar{\psi}_r+ D^{J,\psi\bar{\psi}}_{rt}  \frac{\delta}{\delta \psi_t}$:
\begin{align}
-S^{A\bar{\psi}\psi}_{irs} \bar{\psi}_r \psi_s \rightarrow -S^{A\bar{\psi}\psi}_{irs} \left(\bar{\psi}_r \psi_s + D^{J,\psi\bar{\psi}}_{rs} \right).
\end{align}
Differentiating once more with respect to $A_j$ yields
\begin{align}
\Gamma^{AA}_{ij}=
 - S^{A\bar{\psi}\psi}_{irs} D^{\psi\bar{\psi}}_{rr'} D^{\psi\bar{\psi}}_{s's} \frac{\delta \Gamma}{\delta A_j \delta \psi_{r'} \delta \bar{\psi}_{s'}}\Bigg|_{J=0}+\text{gluonic terms},
\end{align}
where the external sources have already been set to zero.
Ordering the derivatives to get the canonical order changes the sign of the expression and leads to the expected relative minus sign of closed fermion loops:
\begin{align}
\Gamma^{AA}_{ij}=
  S^{A\bar{\psi}\psi}_{irs} D^{\bar{\psi}\psi}_{r'r} D^{\bar{\psi}\psi}_{ss'} \Gamma^{A\bar{\psi}\psi}_{js'r'}+\text{gluonic terms}.
\end{align}

Finally, we consider the expansion of the action when Grassmann fields are involved. First, by definition all anti-Grassmann fields have to be left of the Grassmann fields. Second, the expansion coefficients are antisymmetric in the indices belonging to anti-commuting fields. This entails that we can differentiate with respect to these fields as usual, e.g.,
\begin{align}
\frac{\delta}{\delta \bar{\psi}_i}S^{\bar{\psi}\bar{\psi}\psi\psi}_{rstu} \bar{\psi}_r \bar{\psi}_s \psi_t \psi_u=&S^{\bar{\psi}\bar{\psi}\psi\psi}_{istu} \bar{\psi}_s \psi_t \psi_u - S^{\bar{\psi}\bar{\psi}\psi\psi}_{ritu} \bar{\psi}_r \psi_t \psi_u=\nnnl
 =& S^{\bar{\psi}\bar{\psi}\psi\psi}_{istu} \bar{\psi}_s \psi_t \psi_u + S^{\bar{\psi}\bar{\psi}\psi\psi}_{irtu} \bar{\psi}_r \psi_t \psi_u= 2 S^{\bar{\psi}\bar{\psi}\psi\psi}_{istu} \bar{\psi}_s \psi_t \psi_u.
\end{align}

The additional rules for Grassmann fields described in this section allow their inclusion in the derivation of DSEs, but also for FRGEs this convention is adequate.
\chapter{Feynman rules of the maximally Abelian gauge}
\label{chp:MAGFeynmanRules}

I collect here the Feynman rules of the MAG. The action is given in \eref{eq:L-MAG} with the interpolating gauge fixing parameter chosen as $\zeta=1$ and the gauge fixing parameters $\al$ and $\xi$ unfixed. The indices $i$, $j$, $k$ are diagonal and $a$, $b$, $c$, $d$, $e$,  off-diagonal; $r$ stands for both. The momenta are always chosen as ingoing, except for anti-ghosts. The momentum convention for the Fourier transformation to momentum space is defined with a positive sign for ingoing momenta, e.g.,
\begin{align}
 \Gamma_{A\bar{c}c,\mu}^{iab,(0)}&(p_3,p_1,p_2)=\int dx \,dy \, \Gamma(z,x,y) e^{i(z\,p_3+y\, p_2 - x\,p_1 )}
\end{align}
for a ghost-gluon vertex, where $p_2$ and $p_3$ are incoming and $p_1$ is outgoing.
The order of indices corresponds to the order of fields given as subscripts indicating the vertex.

The bare propagators are
\begin{align}
 D_A^{ij,(0)}(p^2)&=\de^{ij}\frac1{p^2}\left(g_{\mu\nu}-(1-\xi)\frac{p_\mu p_\nu}{p^2}\right),\\
 D_B^{ab,(0)}(p^2)&=\de^{ab}\frac1{p^2}\left(g_{\mu\nu}-(1-\al)\frac{p_\mu p_\nu}{p^2}\right),\\
 D_c^{ab,(0)}(p^2)&=-\de^{ab}\frac1{p^2}.
\end{align}

The bare three-point vertices are
\begin{align}
 \Gamma_{ABB,\mu\nu\rho}^{iab,(0)}&(p_3,p_1,p_2)=g\,i\,f^{abi}(2\pi)^4\de(p_1+p_2+p_3) \times\nnnl
 &\times\left( g_{\mu\nu}(p_2-p_1)_\rho+g_{\mu\rho}(p_1-p_3)_\nu+g_{\nu\rho}(p_3-p_2)_\mu+\frac{1}{\al}(g_{\nu\rho}p_{1_\mu}-g_{\mu\rho}p_{2_\nu}) \right),\\
 \Gamma_{BBB,\mu\nu\rho}^{abc,(0)}&(p_1,p_2,p_3)=g\,i\,f^{abc}(2\pi)^4\de(p_1+p_2+p_3) \times\nnnl
 &\times\left( g_{\mu\nu}(p_2-p_1)_\rho+g_{\mu\rho}(p_1-p_3)_\nu+g_{\nu\rho}(p_3-p_2)_\mu \right),\\
 \Gamma_{A\bar{c}c,\mu}^{iab,(0)}&(p_3,p_1,p_2)=g\,i\,f^{abi}(2\pi)^4\de(-p_1+p_2+p_3) (p_2+p_1)_\mu,\\
 \Gamma_{B\bar{c}c,\mu}^{cab,(0)}&(p_3,p_1,p_2)=g\,i\,f^{abc}(2\pi)^4\de(-p_1+p_2+p_3) \left(\frac{1}{2}p_3+p_2\right)_\mu.
\end{align}

Finally, the bare four-point vertices are
\begin{align}
 &\Gamma_{AABB,\mu\nu\rho\sigma}^{ijab,(0)}(p_3,p_4,p_1,p_2)=g^2(2\pi)^4\de(p_1+p_2+p_3+p_4) f^{aie}f^{bje}\times\nnnl
 &\times\left(2g_{\mu\nu}g_{\rho\sigma}-\left(1-\frac{1}{\al}\right) g_{\mu\sigma}g_{\nu\rho}-\left(1-\frac{1}{\al}\right)g_{\mu\rho}g_{\nu\sigma} \right),\\
&\Gamma_{ABBB,\mu\nu\rho\sigma}^{iabc,(0)}(p_4,p_1,p_2,p_3)=g^2(2\pi)^4\de(p_1+p_2+p_3+p_4) \times\nnnl
 &\times\left( f^{abe}f^{cie} \left(g_{\mu\rho}g_{\nu\sigma}-g_{\mu\sigma}g_{\nu\rho}\right)+f^{bce}f^{aie}\left(g_{\mu\nu}g_{\rho\sigma}-g_{\mu\rho}g_{\nu\sigma}\right) +f^{cae}f^{bie}\left(g_{\mu\sigma}g_{\nu\rho}-g_{\mu\nu}g_{\rho\sigma}\right) \right),\\
&\Gamma_{BBBB,\mu\nu\rho\sigma}^{abcd,(0)}(p_1,p_2,p_3,p_4)=g^2(2\pi)^4\de(p_1+p_2+p_3+p_4) \times\nnnl
 &\times\left( f^{abr}f^{cdr} \left(g_{\mu\rho}g_{\nu\sigma}-g_{\mu\sigma}g_{\nu\rho}\right)+f^{bcr}f^{adr}\left(g_{\mu\nu}g_{\rho\sigma}-g_{\mu\rho}g_{\nu\sigma}\right) +f^{car}f^{bdr}\left(g_{\mu\sigma}g_{\nu\rho}-g_{\mu\nu}g_{\rho\sigma}\right) \right),\\
&\Gamma_{AB\bar{c}c,\mu\nu}^{icab,(0)}(p_4,p_3,p_1,p_2)=g^2(2\pi)^4\de(-p_1+p_2+p_3+p_4) g_{\mu\nu} \frac1{2}\left(2f^{cbe}f^{aie}-f^{abe}f^{cie}\right),\\
&\Gamma_{AA\bar{c}c,\mu\nu}^{ijab,(0)}(p_3,p_4,p_1,p_2)=g^2(2\pi)^4\de(-p_1+p_2+p_3+p_4) 2g_{\mu\nu}f^{aie}f^{bje},\\
&\Gamma_{BB\bar{c}c,\mu\nu}^{cdab,(0)}(p_3,p_4,p_1,p_2)=-g^2(2\pi)^4\de(-p_1+p_2+p_3+p_4) g_{\mu\nu} \left(f^{adk}f^{cbk}+f^{ack}f^{dbk}\right),\\
&\Gamma_{\bar{c}\bar{c}cc,\mu\nu}^{cdab,(0)}(p_1,p_2,p_3,p_4)=-g^2(2\pi)^4\de(-p_1-p_2+p_3+p_4) \times\nnnl
 &\times\frac{\al}{4} \left(4f^{abk}f^{cdk}+2f^{abe}f^{cde}+f^{ade}f^{bce}-f^{ace}f^{bde}\right).
\end{align}

\chapter{Calculation of the sunset diagram}
\label{chp:sunset}

The solution of the sunset diagram is
\begin{align}
 I_{SS}(a,b, c;p^2):=&\int \ddotp{q} \ddotp{r} (q^2)^{a}[(r)^2]^{b} [(p-q-r)^2]^{c}=\nnnl
 =&(4\pi)^{-d} (p^2)^{d+a+b+c} \frac{\Gamma(a+d/2)\Gamma(b+d/2)\Gamma(c+d/2)\Gamma(-a-b-c-d)}{\Gamma(-a)\Gamma(-b)\Gamma(-c)\Gamma(a+b+c+3d/2)}.
\end{align}
It can be derived, for example, with NDIM or by using the solution for the one-loop two-point function, \eref{eq:2-point}. Raising the number of loops by adding further propagators between the two vertices one gets the so-called water melon diagram for which the solution is also known, see e.g., ref. \cite{Berends:1993ee} for the massive case. Introductions to NDIM can be found, for example, in refs. \cite{Suzuki:1997wv,Anastasiou:1999ui,Huber:2007da}.

The main complication is that we need not only this integral but also the case where factors of $p\, q$, $p\, r$ and $q\, r$ appear. In contrast to the one-loop case it is here not possible to express the scalar products by the invariants appearing in the integral. Following an idea by Suzuki and Schmidt \cite{Suzuki:2000us}, who calculated the integral when arbitrary powers of $(2\,p\, q)$ appear, one can use NDIM for the calculation of
\begin{align}
 I_{SS}(a,b, c,e,f;p^2):=&\int \ddotp{q} \ddotp{r} (q^2)^{a}[(r)^2]^{b} [(p-q-r)^2]^{c}(2\,p\, q)^e (2\,p\, r)^f.
\end{align}
This indeed suffices as the scalar product between the two loop momenta can be expressed via the other scalar products:
\begin{align}
 q\, r=\frac{1}{2}\left((p-q-r)^2-p^2-q^2-r^2+2p \, q +2p\, r \right).
\end{align}

Following the NDIM procedure one obtains 276 solutions in terms of five-dimensional hypergeometric series, which should all be equivalent. We exploit this huge number by choosing one solution where the hypergeometric series terminates:
\begin{align}\label{eq:sunsetSol}
 I_{SS}&(a,b, c,e,f;p^2)=\nnnl 
 &=\pi^d (p^2)^{a+b+c+d+e+f}2^{e+f}\frac{(-a,2a+e+d/2)(-b,2b+f+d/2)(-c,2c+d/2)}{-a-b-c-d,2a+2b+2c+e+f+5/2)}\times\nnnl
 &\times \sum_{n_3,n_4,n_6,n_7,n_8=0}^{\max[e,f]} \frac{(-1)^{-n_4-n_8}2^{n_6}(c+d/2,n_3+n_7)}{n_3!n_4!n_6!n_7!n_8!(1-a-d/2-e,n_3+n_4-n_8)}\times\nnnl
 &\times \frac{P(e,n_3+n_4,n_6)P(f,n_7+n_8,n_6)}{(1-b-d/2-f,n_7+n_8-n_4)(1+a+b+c+d,n_3+n_4+n_6+n_7+n_8)}.
\end{align}
The symbol $P$ is defined as
\begin{align}
 P(a,b,c):=(-a/2,b+c/2)(1/2-a/2,b+c/2)
\end{align}
and
\begin{align}
 (a,b):=\frac{\Gamma(a+b)}{\Gamma(a)}
\end{align}
is the Pochhammer symbol.
Although a five-fold series is normally quite a nasty object to deal with, this is not a real problem here. The reason is that the series is truncated as we can see from the definition of the symbol $P$. Depending on if the first argument is even or odd, $P$ vanishes from certain values of the summation variables on, for example,
\begin{align}
 P(e,n_3+n_4,n_6)=0 \quad \text{for} \quad e<2n_3+2n_4+n_6 \quad \text{if} \quad e/2\in \mathbb{I} \wedge n_6/2 \in \mathbb{I}.
\end{align}
This can most easily be taken into account by restricting the sum to $\max[e,f]$. As $e$ and $f$ are typically quite small this is sufficiently fast in a numerical implementation.

\chapter{Deriving Dyson-Schwinger equations with \textit{DoDSE}}
\label{chp:DoDSE}

In the course of this thesis a \textit{Mathematica} package \cite{Wolfram:1999} was programmed to automate the derivation of DSEs. As we turn to more and more complicated gauges such a tool is indispensable, because the numbers of diagrams become very large. \textit{DoDSE}, what is short for "Derivation of Dyson-Schwinger equations", can handle complicated Lagrangians easily by resorting to a symbolic notation. This alleviates the process of the derivation considerably. After the DSEs have been derived in this symbolic notation the individual terms can be transformed into the proper algebraic expressions.

Available documentation for \textit{DoDSE} include ref. \cite{Alkofer:2008nt} and a \textit{Mathematica} notebook distributed together with the package, which, among other places, can be obtained from \href{http://cpc.cs.qub.ac.uk/summaries/AECT_v1_0.html}{http://cpc.cs.qub.ac.uk/summaries/AECT\_v1\_0.html}. Furthermore, the syntax and examples for all commands are listed via the \textit{Mathematica} command \verb|?|, e.g., \verb|?doDSE|. The present Appendix describes some general aspects of the program and provides examples used in the calculations of Chapters \ref{chp:MAG} and \ref{chp:GZ}. The final section contains a list with all functions of \textit{DoDSE}.

\section{Overview over \textit{DoDSE}}

The main challenge in doing functional calculations with a symbolic programming language is to implement the properties of fields, propagators and vertices properly. A very convenient way for this is provided by the index notation introduced in Sec. \ref{chp:FunctionalEquations}. Thereby all indices, e.g., color, Lorentz, Dirac, and the space or momentum dependence are described by only one index. This suffices as long as all quantities like color or momentum "flow" through the diagram together. The evident advantage is the reduction of redundancy in the notation.

\begin{figure}[t]
 \begin{center}
  \includegraphics[width=0.5\textwidth]{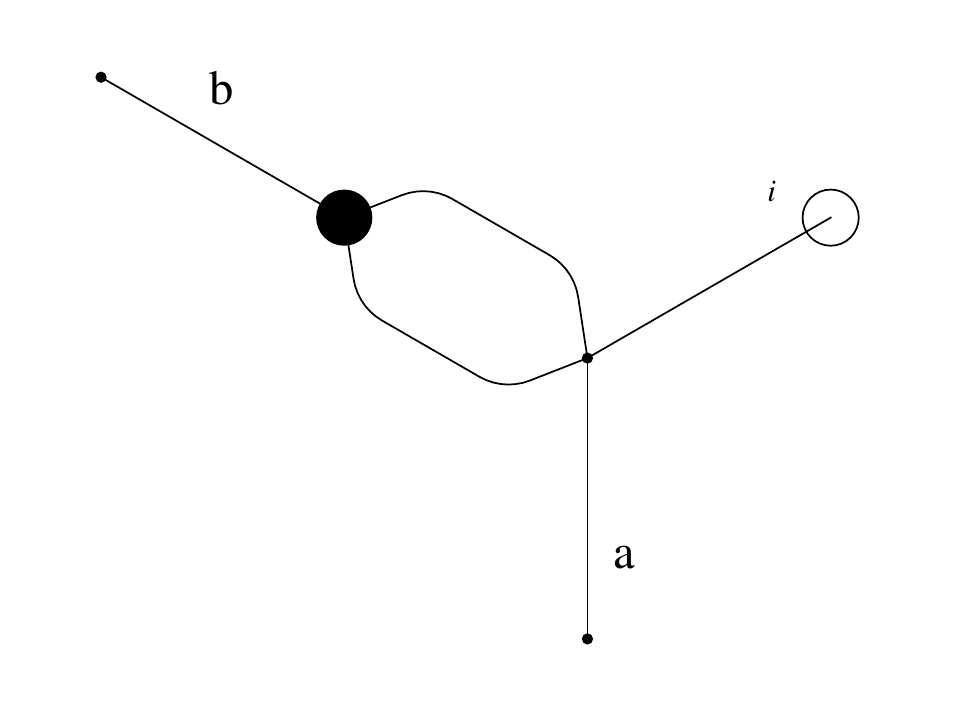}
  \caption{\label{fig:DoDSE-Ex1}The graphical representation of the example provided in the text. The circle denotes the external field.}
 \end{center}
\end{figure}

Instead of using the standard multiplication operator \textit{DoDSE} employs its own function called \verb|op|. The main reason is that the order of quantities is not changed unpredictably by \textit{Mathematica} within a calculation and also the anticommutativity property of Grassmann fields can be taken into account. The arguments of an \verb|op| function can be fields, propagators or vertices. Fields are defined as a list with two entries: The first determines the name of the field and the second gives its index, e.g., \verb|{A,i}|. If a field appears as an argument of an \verb|op| function it is an external field. n-point functions exist in bare and dressed form, \verb|S| and \verb|V|, respectively. Dressed propagators are denoted by \verb|P|. Note that in the derivation of DSEs there is no need for a bare propagator but only for a bare two-point function, which is denoted by \verb|S|. The arguments of \verb|V|, \verb|P| and \verb|S| are fields. A simple example containing an external field, dressed propagators and bare and dressed vertices is the expression
\begin{verbatim}
op[{A,i},
 P[{A,j1},{A,j2}], P[{A,k1},{A,k2}],
 S[{A,i},{A,j1},{A,k1},{A,a}],
 V[{A,j2},{A,k2},{A,b}]]
\end{verbatim}
It is depicted in \fref{fig:DoDSE-Ex1}. All internal indices are summed and the free indices \verb|a| and \verb|b| correspond to the external legs of the graph.

For using \textit{DoDSE} the package file \verb|DoDSE.m| has to be loaded. The standard way is
\begin{verbatim}
 <<DoDSE`
\end{verbatim}
if the package resides in the subdirectory \verb|DoDSE| of \verb|$UserAddOnsDirectory|.

The next step is the definition of an action. For the simplest cases \textit{DoDSE} only needs its basic structure given by lists of fields. The action of Yang-Mills theory, for example, is defined as
\begin{verbatim}
actionYM = {{A, A}, {c, cb}, {A, A, A}, {A, cb, c}, {A, A, A, A}};
\end{verbatim}
From this definition \textit{DoDSE} will automatically infer that the ghost and anti-ghost fields \verb|c| and \verb|cb|, respectively, are Grassmann fields as they appear in a pair. If this assumption is unwanted, because the fields are not anti-commuting, one can provide the option \verb|specificFieldDefinitions| to declare fermions and bosons specifically. An example to this is given in Section \ref{ssec:derivingDSEsGZ}.
A DSE is then derived from the action with the function \verb|doDSE|. Its two required arguments are the action and the list of fields with respect to which the derivatives are performed. For example, the two-point DSEs in Landau gauge are obtained by
\begin{verbatim}
AADSE = doDSE[actionYM, {A, A}];
ccDSE = doDSE[actionYM, {cb, c}];
\end{verbatim}

The result of \verb|doDSE| is a (sum of) \verb|op| function(s) each representing a single diagram. There are basically two ways of getting this into a useful representation. One can us the function \verb|shortExpression| (or \verb|sE|) to get the result in a shorthand notation where $\Delta$ represents a propagator, $\Gamma$ a dressed vertex and \texttt{S} a bare vertex. For the gluon two-point DSE this looks like
\begin{verbatim}
sE[AADSE]
\end{verbatim}
\begin{align}
 &S_{\text{i1 i2}}^{\text{A A}}-\frac{1}{2} \left(S_{\text{i1 i2 r1 s1}}^{\text{A A A A}} \Delta _{\text{r1 s1}}^{\text{A A}}\right)-\frac{1}{2} \left(S_{\text{i1 r1 s1}}^{\text{A A A}} \Gamma _{\text{i2 t1 u1}}^{\text{A A A}} \Delta _{\text{r1 t1}}^{\text{A A}} \Delta _{\text{s1 u1}}^{\text{A A}}\right) -\nnnl
 &\quad -\frac{1}{6} \left(S_{\text{i1 r1 r2 s1}}^{\text{A A A A}} \Gamma _{\text{i2 s2 t2 u2}}^{\text{A A A A}} \Delta _{\text{r1 s2}}^{\text{A A}} \Delta _{\text{r2 t2}}^{\text{A A}} \Delta _{\text{s1 u2}}^{\text{A A}}\right)+S_{\text{i1 r1 s1}}^{\text{A cb c}} \Gamma _{\text{i2 t1 u1}}^{\text{A cb c}} \Delta _{\text{s1 t1}}^{\text{c cb}} \Delta _{\text{u1 r1}}^{\text{c cb}}-\nnnl
 &\quad-\frac{1}{2} \left(S_{\text{i1 r1 r2 s1}}^{\text{A A A A}} \Gamma _{\text{i2 s2 t1}}^{\text{A A A}} \Gamma _{\text{u1 v2 w1}}^{\text{A A A}} \Delta _{\text{r1 s2}}^{\text{A A}} \Delta _{\text{r2 v2}}^{\text{A A}} \Delta _{\text{s1 w1}}^{\text{A A}} \Delta _{\text{u1 t1}}^{\text{A A}}\right) \nonumber
\end{align}
The superscripts denote the fields and the subscripts the corresponding indices.

It is also possible to use \textit{DoDSE} for drawing Feynman graphs. This is done with the function \verb|DSEPlot|:
\begin{verbatim}
DSEPlot[AADSE, actionYM, {{A, Red}, {c, Green}}]
\end{verbatim}
As second argument one has to give the action. The third argument is optional to give some graphics primitives for the different fields.

I would like to make a short comment on the graphical representation of DSEs using \textit{DoDSE}. \texttt{DSEPlot} employs the \textit{Mathematica} function \texttt{GraphPlot} that originally is intended for drawing graphs in graph theory. However, it is possible to abuse it also for drawing Feynman diagrams, but with a few limitations: It is not possible to use wiggly lines and the alignment of some graphs may be arbitrary. And when it comes to non-planar diagrams the usefulness of \texttt{GraphPlot} comes to an end, as the result cannot be distinguished from a planar diagram. Thus the ability to represent DSEs graphically is limited, but with a little bit of effort it can produce presentable figures.

In order to get the algebraic expressions with all the indices and the integrals one uses the function \verb|getAlg|. Before it can be used one has to specify the Feynman rules. These are defined globally by overloading the propagator and vertex functions. For example, the gluon propagator, the ghost propagator and the ghost-gluon vertex in the Landau gauge are defined as
\begin{verbatim}
S[c[c1_, p1_], cb[c2_, p2_], explicit -> True] := 
 -1/SPD[p1] SD[c1, c2];
P[c[c1_, p1_], cb[c2_, p2_], explicit -> True] :=
 -SPD[p1]^(\[Delta]c - 1) SD[c1, c2];
S[A[c1_, \[Mu]_, p1_], A[c2_, \[Nu]_, p2_], explicit -> True] := 
 SD[c1, c2] (MTD[\[Mu], \[Nu]] - 
  FVD[p1, \[Mu]] FVD[p1, \[Nu]]/SPD[p1])/SPD[p1];
P[A[c1_, \[Mu]_, p1_], A[c2_, \[Nu]_, p2_], explicit -> True] := 
 SD[c1, c2] (MTD[\[Mu], \[Nu]] -
  FVD[p1, \[Mu]] FVD[p1, \[Nu]]/SPD[p1]) SPD[p1]^(\[Delta]A - 1);
S[A[c1_, \[Mu]_, p1_], cb[c2_, p2_], c[c3_, p3_], explicit -> True] :=
 I g SUNF[c1, c2, c3] FVD[p2, \[Mu]];
V[A[c1_, \[Mu]_, p1_], cb[c2_, p2_], c[c3_, p3_], explicit -> True] := 
 S[A[c1,\[Mu],p1], cb[c2, p2], c[c3,p3]];
\end{verbatim}
Here \verb|SPD|, \verb|SD[a,b]| and \verb|FVD| represent scalar products, the color tensor $\de^{ab}$ and momenta. The choice of convention is up to the user and depends on the further processing of the expressions.
The propagators were dressed with a power law and the dressed ghost-gluon vertex was taken bare. These functions have to be defined as above so \textit{DoDSE} can use them. Their arguments are fields and the arguments of those are their indices and their momentum argument. The task of \verb|getAlg| is to replace all the propagators and vertices in the result of a \verb|doDSE| calculation by expressions with the correct indices and the correct flow of the momenta through the integral. After this it sets the option \verb|explicit| to \verb|True| and the algebraic expressions are plugged in.

Let me illustrate this with the diagrams of the two-point functions in the Landau gauge containing ghosts. For the ghost two-point function this is
\begin{verbatim}
IGhDSELoop =  getAlg[ccDSE[[2]],
 {{A, adj, lor}, {c, adj}},
 {A, {c, cb}},
 {p1, -p1},
 {LorentzContract, SUNContract}];
\end{verbatim}
and for the gluon two-point function
\begin{verbatim}
IGluonDSELoop = getAlg[AADSE[[4]],
 {{A, adj, lor}, {c, adj}},
 {A, {c, cb}},
 {p1, -p1},
 {LorentzContract,  SUNContract}];
\end{verbatim}
The first argument of \verb|getAlg| is the expression for the integrals we got from \verb|doDSE|. The second tells which indices the fields have and the third is a list of the fields. The fourth list contains the external momenta of the integral. The final argument is a list of functions applied to the output. \verb|LorentzContract| and \verb|SUNContract| are user-defined functions for the contraction of the indices.

With the resulting expressions one can calculate the IRE $\ka$ of Landau gauge \cite{Lerche:2002ep,Zwanziger:2001kw}: The gluon DSE is projected transversely and the integrals are performed with the help of \eref{eq:2-point}. The two equations are then combined to solve for the IRE $\ka$.

\section{DSEs of the maximally Abelian gauge}
\label{sec:DoDSEMAG}

The derivation of the two-point DSEs in the MAG is presented in this section and the way of discarding unphysical vertices is explained.

\subsection{Derivation of the DSEs in the maximally Abelian gauge}

First we define the action of the MAG based on \eref{eq:L-MAG}:
\begin{verbatim}
IAMAG={{A, A}, {B, B}, {cb, c}, {db, d}, 
 {A, B, B}, {A, cb, c}, {A, A, B, B}, {A, A, cb, c},
 {B, B, B, B}, {B, B, cb, c}, {cb, cb, c, c},{B, B, B},
 {B, cb, c}, {A, B, cb, c}, {A, B, B, B}};
\end{verbatim}
The fields \verb|A|, \verb|B|, \verb|c| and \verb|d| are the diagonal gluon, the off-diagonal gluon, the off-diagonal ghost and the diagonal ghost, respectively. \verb|cb| and \verb|db| are the corresponding anti-fields.
\textit{DoDSE} does not know anything about diagonal or off-diagonal indices and allows all combinations of fields. However, some structure functions vanish and only certain vertices exist, see Section \ref{ssec:abelianPart}. Hence we have to devise some test function that discards non-existent vertices. The function \verb|vertexTest| only allows three-point functions with no or one diagonal field and four-point functions with at most two:
\begin{verbatim}
Clear@vertexTest;
vertexTest[a_V]:= 
 Not@(Length@a == 3 && MatchQ[Length@Cases[a, A, \[Infinity]], 2 | 3]) &&  
 Not@(Length@a == 4 && MatchQ[Length@Cases[a, A, \[Infinity]], 3 | 4])
\end{verbatim}
The derivation of the two-point DSEs can now be done. It is important to provide the function \verb|vertexTest| as argument to discard unphysical vertices. Note that the indices are only given together with the fields to respect the convention of the MAG described in Section \ref{ssec:abelianPart}. Otherwise \textit{DoDSE} would automatically take \verb|i1| and \verb|i2| as indices:
\begin{verbatim}
AA = doDSE[IAMAG, {{A, i}, {A, j}}, vertexTest];
BB = Delete[doDSE[IAMAG, {{B, a}, {B, b}}, vertexTest], 23];
cc = Delete[doDSE[IAMAG, {{c, a}, {cb, b}}, vertexTest], 15];
dd = doDSE[IAMAG, {{d, i}, {db, j}}, vertexTest];
\end{verbatim}
With hindsight those diagrams are discarded manually in the off-diagonal equations which vanish due to their color structure, see Section \ref{ssec:MAGDSEs}.

For plotting the DSEs we can define some graphics primitives. The black and white definition as used in this thesis and a color definition which is better suited for presentations are
\begin{verbatim}
fieldPlotRulesBW = {{A, Thickness[0.01]}, {B, Thickness[0.01], 
 Dashing[0.05]}, {c, Dotted}, {d, Dotted, Thickness[0.03]}};
fieldPlotRulesColor = {{A, Red, Thickness[0.01]}, {B, Purple, 
 Thickness[0.01]}, {c, Darker@Darker@Green, Thickness[0.01]},
  {d, LightGreen, Thickness[0.01]}};
\end{verbatim}
When plotting the DSEs we use the argument \verb|factorStyle| to adjust the size and style of the numeric coefficients and set the number of diagrams per row to four:
\begin{verbatim}
DSEPlot[#, IAMAGSU2, fieldPlotRulesBW, 4,
 factorStyle :> {FontSize :> 20, FontWeight :> Bold}]& /@{AA, BB, cc, dd}
\end{verbatim}
This yields the two-point DSEs of the MAG as shown in figs.~\ref{fig:MAG-DSEs-AA}, \ref{fig:MAG-DSEs-BB} and \ref{fig:MAG-DSEs-cc}.

\subsection{Obtaining and calculating the integrals}
\label{ssec:MAGDoDSEIntegrals}

The integrals of the two-point DSEs are obtained with the function \verb|getAlg|. The Feynman rules of bare vertices are defined according to Appendix \ref{chp:MAGFeynmanRules}. Full propagators are dressed with a power law. Required arguments are a list with the fields and their indices, the list of fields and a list with the external momenta of the expression. For example, the sunset of the diagonal gluon two-point DSE with a bare $AABB$ vertex is obtained by
\begin{verbatim}
getAlg[AA[[6]],
 {{A, adj, lor}, {B, adj, lor}, {c, adj}},
 {A, B, {c, cb}},
 {p, -p},
 {LorentzContract, SUNMagContract}];
\end{verbatim}
The integral can be calculated with the solution for the sunset diagram given in Appendix \ref{chp:sunset}, \eref{eq:sunsetSol}, after a projection in Lorentz space to get a scalar expression.

\section{DSEs of the Gribov-Zwanziger action}
\label{sec:DoDSEGZ}

The derivation of the two-point DSEs of the Gribov-Zwanziger action is presented in this Section. As in this action fields mix at the two-point level one needs the option \verb|specificFieldDefinitions|. Furthermore, this mixing requires an additional step to identify equal diagrams at the end.

\subsection{Deriving the DSEs of the Gribov-Zwanziger action}
\label{ssec:derivingDSEsGZ}

The Gribov-Zwanziger action, see \eref{eq:final-GZ-action}, is defined for \textit{DoDSE} as follows:
\begin{verbatim}
IAFP = {{A, A}, {A, A, A}, {A, A, A, A}};
IAGZ = {{n, nb}, {A, nb, n}, {A, W, W}, {A, W}, {W, W}};
IA = Join[IAGZ, IAFP];
\end{verbatim}
The action contains the gluon field \verb|A|, the pair of Grassmann fields \verb|n| and \verb|nb| and the auxiliary field \verb|W|. They correspond to the fields $A$, $\eta$, $\bar{\eta}$ and $V$ of Chapter \ref{chp:GZ}, respectively.\footnote{The symbol \texttt{W} was chosen, as \texttt{V} represents in \textit{DoDSE} a dressed vertex.}
From the action we derive the individual propagator equations: 
\begin{verbatim}
AADSE = doDSE[IA, {A, A}, {{A, A}, {n, nb}, {W, W}, {A, W}}, 
 specificFieldDefinitions -> {A, W, {n, nb}}];
nnbDSE = doDSE[IA, {n, nb}, {{A, A}, {n, nb}, {W, W}, {A, W}},
 specificFieldDefinitions -> {A, W, {n, nb}}];
WWDSE = doDSE[IA, {W, W}, {{A, A}, {n, nb}, {W, W}, {A, W}}, 
 specificFieldDefinitions -> {A, W, {n, nb}}];
AWDSE = doDSE[IA, {A, W}, {{A, A}, {n, nb}, {W, W}, {A, W}}, 
 specificFieldDefinitions -> {A, W, {n, nb}}];
WADSE = doDSE[IA, {W, A}, {{A, A}, {n, nb}, {W, W}, {A, W}}, 
 specificFieldDefinitions -> {A, W, {n, nb}}];
\end{verbatim}
Note that the option \verb|specificFieldDefinitions| is required to define \verb|W| as a bosonic field. Otherwise \textit{DoDSE} would assume automatically from the definition of the action that it is a fermion field.

The number of terms in the DSEs can be determined with \verb|countTerms|:
\begin{verbatim}
countTerms /@ {AADSE, nnbDSE, WWDSE, AWDSE, WADSE}

 {51, 3, 5, 50, 5}
\end{verbatim}
However, this is not the final number. Because of the mixed propagators not all diagrams were yet identified properly and this has to be done in a separate step:
\begin{verbatim}
{AADSEId, nnbDSEId, WWDSEId, AWDSEId, WADSEId} = 
  identifyGraphs[#, compareFunction :> compareGraphs2] & /@
   {AADSE, nnbDSE, WWDSE, AWDSE, WADSE};
\end{verbatim}
Now the number of terms decreases considerably for DSEs, where the first derivative was done with respect to the gluon field:
\begin{verbatim}
countTerms /@ {AADSEId, nnbDSEId, WWDSEId, AWDSEId, WADSEId}

{37, 3, 5, 36, 5}
\end{verbatim}

Finally, we can plot the diagrams
\begin{verbatim}
pAADSE = DSEPlot[AADSEId, IA, 4, factorStyle -> {FontSize -> 20}]
pnnbDSE = DSEPlot[nnbDSEId, IA, 4, factorStyle -> {FontSize -> 20}]
pWWDSE = DSEPlot[WWDSEId, IA, 4, factorStyle -> {FontSize -> 20}]
pAWDSE = DSEPlot[AWDSEId, IA, 4, factorStyle -> {FontSize -> 20}]
pWADSE = DSEPlot[WADSEId, IA, 4, factorStyle -> {FontSize -> 20}]
\end{verbatim}
to obtain figs. \ref{fig:AA-prop-DSEs}, \ref{fig:VV-prop-DSEs}, \ref{fig:VA-prop-DSEs} and \ref{fig:etaeta-prop-DSEs}. As we do not give any graphics primitives for drawing the propagators, \textit{DoDSE} tags them by their field content.

Finally, the full expressions are obtained with the function \verb|getAlg|. After suitable projections in color and Lorentz space, the integrals can be performed using \eref{eq:2-point}.

\section{Tables of \textit{DoDSE} functions}

In the following I provide lists with all public functions of \textit{DoDSE}.

\bcommandlist
\multicolumn{2}{c}{\textbf{Main functions}}\\
 \hline
  Command & Description\\
\hline
\texttt{doDSE[ilist, clist]} & Derives the DSE for the correlation function \texttt{clist} for a theory with interactions \texttt{ilist}.\\
\texttt{doDSE[ilist, clist [, props, vertexTest, opts]]} & \texttt{vertexTest} is a function for determining if a vertex respects the symmetries of the Lagrangian. \texttt{props} is a list of allowed propagators given in the form \texttt{\{\{field1a, field1b\}, \{field2a, field2b\}, ...\}}. \texttt{doDSE} accepts the options \texttt{specificFieldDefinitions} and \texttt{sourcesZero} (prevents the replacement of super-field propagators and vertices when set to \texttt{False}).\\
\texttt{shortExpression[expr, opts]} \quad \texttt{sE[expr, opts]} & Rewrites a \textit{DoDSE} expression into a shorter form using \texttt{\$bareVertexSymbol}, \texttt{\$vertexSymbol} and \texttt{\$propagatorSymbol} for representation. Options of the internal \textit{Mathematica} function \texttt{Style} can be given.\\
\texttt{DSEPlot[expr, ilist~[,fRules,len,opts]]} & Plots the full DSE in graphical form. \texttt{expr} is an expression containing \texttt{op} functions, \texttt{ilist} the list of interactions and \texttt{fRules} a list of options for plotting individual fields. \texttt{len} determines how many graphs are shown in one line. If \texttt{fRules} is not given, the lines are named according to the fields. Possible options are: \texttt{output->List}, to get the result in list form, and \texttt{indexStyle} and \texttt{factorStyle} to change the style of the indices and the prefactors (e.g. font size or color).\\
\texttt{getAlg[exp, ilist, flist, mlist, funclist} & Derives the algebraic expression from \texttt{exp}. \texttt{ilist} contains the lists of fields and all their indices, \texttt{flist} the list of fields, \texttt{mlist} the list of external momenta and \texttt{funclist} a list of further functions to be applied on the result.
\ecommandlist

\bcommandlist
\multicolumn{2}{c}{\textbf{Functions for the individual computation steps}}\\
 \hline
  Command & Description\\
 \hline
  \texttt{generateAction[ilist[,flist]]} & Generates the action in internal representation from the interactions of the theory given in \texttt{ilist}. For mixed propagators \texttt{flist} specifies explicitly the type of fields in the form \texttt{\{boson1, boson2, ..., \{fermion1, antifermion1\}, \{fermion2, antifermion2\}, ...\}}.\\ 
  \texttt{deriv[expr,dlists]} & Differentiate \texttt{expr} with respect to the fields in \texttt{dlists}.\\
  \texttt{replaceFields[expr]} & Replaces the fields in \texttt{expr} by the corresponding expressions after 
the first differentiation is done to change from full to 1PI Green functions.\\
  \texttt{identifyGraphs[expr[, compareGraphs->cfunc]]} & Adds up equivalent graphs in \texttt{expr}. \texttt{cfunc} can be \texttt{compareGraphs} (standard) or \texttt{compareGraphs2}, the latter being necessary for mixed propagators but taking longer.\\
  \texttt{setSourcesZero[expr, flist [, props, vertexTest]]} & Sets the external fields in \texttt{flist} to zero, i.e. only physical propagators and vertices are left. \texttt{vertexTest} is a function for determining if a vertex respects the symmetries of the Lagrangian. \texttt{props} is a list of allowed propagators given in the form \texttt{\{\{field1a, field1b\}, \{field2a, field2b\}, ...\}}.\\
  \texttt{orderFermions[expr]} & Orders derivatives with respect to Grassmann fields such that the anti-fields are left of the fields thereby possibly giving a minus sign. \texttt{expr} is an \texttt{op}-function or a sum of those. Bare vertices are not affected by the ordering.
\ecommandlist

\bcommandlist
\multicolumn{2}{c}{\textbf{Functions for checks and tools}}\\
 \hline
  Command & Description\\
 \hline
  \texttt{countTerms[expr]} & Counts the number of terms appearing in the expression.\\
  \texttt{fieldQ[f]} & Determines if expression \texttt{f} is defined as a field.\\
  \texttt{bosonQ[f]} & Determines if expression \texttt{f} is defined as a bosonic field.\\
  \texttt{fermionQ[f]} & Determines if expression \texttt{f} is defined as a fermionic field.\\
  \texttt{antiFermionQ[f]} & Determines if expression \texttt{f} is defined as an anti-field to a fermionic field.\\
  \texttt{checkFields[expr]} & Checks if all fields in the expression are defined\\
  \texttt{checkIndices[expr]} & Checks if an index appears more often than twice.\\
  \texttt{checkSyntax[expr]} & Checks if \texttt{expr} has the correct syntax, i.e. \texttt{op} functions only contain propagators, vertices and fields.\\
  \texttt{checkAction[expr]} & Checks if all indices appear exactly twice, the syntax is ok and all fields are defined.\\
  \texttt{checkAll[expr]} & Performs a series of checks on \texttt{expr} (\texttt{checkIndices}, \texttt{checkSyntax}, \texttt{checkFields}).\\
  \texttt{defineFields[flist]} & Defines the fields of the action that are given in \texttt{flist} as single entries for bosons and grouped by braces for fermions.\\
     \texttt{\$vertexSymbol} & Symbol representing a vertex in \texttt{shortExpression}. Standard value: \texttt{$\Gamma$}.\\
   \texttt{\$bareVertexSymbol} & Symbol representing a bare vertex in \texttt{shortExpression}. Standard value: \texttt{S}.\\
   \texttt{\$PropagatorSymbol} & Symbol representing a propagator in \texttt{shortExpression}. Standard value: \texttt{$\Delta$}.
\ecommandlist

\setstretch{0.99} 
\bibliographystyle{utphys_thesis}
\bibliography{literature_thesis}

\end{document}